\documentclass[11pt]{article}

\usepackage{xr-hyper}
\usepackage{blmkt} 
\bibinput{mybib}

\makeatletter
\newcommand*{\addFileDependency}[1]{
  \typeout{(#1)}
  \@addtofilelist{#1}
  \IfFileExists{#1}{}{\typeout{No file #1.}}
}
\makeatother

\newcommand*{\myexternaldocument}[1]{%
    \externaldocument{aux-files/#1}%
}
\myexternaldocument{02_appendix}

\begin{document}

\begin{titlepage}
\title{The Black Market for Beijing License Plates\thanks{Warm thanks to Shanjun Li for comments and data. Thanks to John Barrios, Stéphane Bonhomme, Dan Black, Jean-Pierre Dub\'e, Alfred
Galichon, Chris Hansen, Ryan Kellogg, Carl Mela, Luxi Shen, Avner Strulov-Shlain, Baohong Sun, Chad Syverson, and Thomas
Wollmann for comments. Thanks to Charles Ahlstrom, Xinyao Kong, Cheryl Liu, Chun Pong Lau, and Omkar Katta for research assistance.

The R package implementing the methods developed herein, as well as pedagogical documentents for instructors, are available on the corresponding author's website.}}
\author{{\O}ystein Daljord\thanks{Chicago Booth School of Business, University of Chicago, 5807 South Woodlawn Avenue, Chicago, IL 60637,
USA. E-mail: \href{mailto:Oeystein.Daljord@chicagobooth.edu}{Oystein.Daljord@chicagobooth.edu}. Web: \url{faculty.chicagobooth.edu/oystein.daljord}} \and Guillaume Pouliot\thanks{Harris School of Public Policy, University of Chicago, 1155 E 60th St, Chicago, IL 60637 USA. E-mail: \href{mailto: guillaumepouliot@uchicago.edu}{guillaumepouliot@uchicago.edu}. Web: \url{sites.google.com/site/guillaumeallairepouliot}.  Corresponding author.}   \and Junji Xiao\thanks{
Department of Economics, Lingnan University. 8 Castle Peak Rd, Tuen Mun, Hong Kong. E-mail: \href{mailto: junji.xiao@gmail.com}{junji.xiao@gmail.com}. Web: \url{https://jjxiao.weebly.com/}} \and Mandy Hu\thanks{CUHK, Department of Marketing, Room 1105, 11/F, Cheng Yu Tung Building, 12 Chak Cheung Street Shatin, N.T., Hong Kong. E-mail: \href{mailto: mandyhu@baf.cuhk.edu.hk}{mandyhu@baf.cuhk.edu.hk}. Web: \url{bschool.cuhk.edu.hk/staff/hu-mandy-mantian}}}
\date{\today}
\maketitle
\begin{abstract}
\singlespacing
\noindent Black markets can reduce the effects of distortionary regulations by reallocating scarce resources toward consumers who value them most. The illegal nature of black markets, however, creates transaction costs that reduce the gains from trade. We take a partial identification approach to infer gains from trade and transaction costs in the black market for Beijing car license plates, which emerged following their recent rationing. 
We find that at least 
$11\%$
of emitted license plates are illegally traded. 
The estimated transaction costs suggest severe market frictions: between \numberblank{61\%} and \numberblank{82\%} of the realized gains from trade are lost to transaction costs.
\\
\noindent\textbf{Keywords:} informal economy/underground economy, optimal transport, partial identification, semiparametric and nonparametric methods\\
\noindent\textbf{JEL Codes:} E26, D450, P230, C140\\

\bigskip
\end{abstract}
\setcounter{page}{0}
\thispagestyle{empty}
\end{titlepage}
\pagebreak \newpage

\doublespacing 

\section{Introduction} \label{sec:introduction}
The informal economy, the trade in goods and services that goes undetected in official statistics, 
makes up an estimated one-sixth of the GDP in the world economy (\citealp{schneider10}). An important segment of the informal economy is black markets, where
goods and services are traded illegally. Black markets emerge in response to restrictions
on trade that create gains from trade. 
Such restrictions include prohibition, taxation, and rationing.
Even though black markets are usually considered undesirable \textit{per se}, it has been suggested that they can improve
welfare by reducing the impact of regulatory distortions (\citealp{davidson07}). Nevertheless, the
illicit nature of black markets creates transaction costs, such as potential legal
liabilities, search costs, or the difficulties of enforcing any kind of contract, which reduce the said gains from trade.  The question of their welfare impact is therefore case-specific and largely empirical.  It is furthermore challenging. 
Indeed, due to their being illegal, black market transactions are typically  unobserved, and little is known about how such markets perform. 
\vspace{0.2in}\\
\noindent In this paper, we ask how
much of the gains from trade are lost to transaction costs in a black market, which
is informative about unintended effects of trade restrictions and the efficacy of enforcement.
\vspace{0.2in}\\
\noindent We address this question in the context of the black market for license plates in
Beijing. In 2011, the Beijing government restricted driving within the city limits to all
but those who have Beijing license plates in an effort to tackle increasing pollution and
congestion. The license plates were rationed and the quota was allocated by lottery.
While Shanghai has used auctions to allocate license plates since the 1990s, the auction
format has found limited public support. \cite{li18} reports survey evidence which shows
that, even though a majority of Beijing residents recognized the need for rationing, less
than 10\% preferred an auction, and about 40\% preferred a hybrid lottery and auction
mechanism.\footnote{See footnote 13 of \cite{li18}.} At the time the Beijing lottery was introduced, there were growing political
concerns over the increasing auction prices in Shanghai, which by 2010 had reached
the level of the price of a typical new car. Having considered an auction format politically
unviable, Beijing chose a lottery out of fairness concerns (\citealp{huang19}).
\vspace{0.2in}\\
\noindent Allocating license plates by lottery creates gains from trade: some lottery winners may prefer to trade their license plates rather than use the plates themselves. In Section \ref{sec:lottery}, we report anecdotal evidence of a black market for license plates that emerged soon after the introduction of the lottery. News reports note that the market largely took place online, while car dealerships seem to have acted as occasional intermediaries. The market was supplied by both corrupt officials and lottery winners. Enforcement of the non-transferability was seemingly lax, according to anecdotal evidence. Both rentals, without formal ownership, and purchases, with or without formal ownership, were offered in the market. The reported transaction prices are of a similar magnitude to the Shanghai auction prices.
\vspace{0.2in}\\
\noindent We find evidence of black market trade in comprehensive car sales data. We argue that if the license plates were allocated to a random subset of the population of car buyers, and there was no black market trade, then fewer cars would be sold, but the sales distribution would not shift following the rationing. If instead black market trade allocated license plates towards wealthier households who empirically purchase more expensive models (\citealp{li18, xiao17}), the average prices would likely shift upwards. We use a standard difference-in-differences approach to document such an upward shift in the average Beijing car prices following the rationing. We find no such price shifts in comparable nearby cities, such as Tianjin and Shijiazhuang, that did not ration license plates.
\vspace{0.2in}\\
\noindent We do find evidence of a clear upward shift in the average prices in Tianjin after
it introduced a rationing mechanism, which distributed about 40\% of the license plates
by auction and the rest by lottery. By design, an auction allocates licenses to a non-random
subset of the population of car buyers just as a black market would reallocate
license plates towards those who are willing to pay, and leads to similar shifts in the sales
distribution. We do not find that the price jump in Beijing was caused by car dealerships
adapting their pricing to the rationing. Instead, we find evidence that vertical price restraints, which were widespread in the industry at the time, precluded car dealerships from
responding to the rationing by adapting their prices.
\vspace{0.2in}\\
\noindent In addition to being of intrinsic interest, the volume of trade on the black market is essential to assessing the gains from trade.  Though the difference-in-differences estimates are
consistent with the existence of black market trade, they carry little information about
the volume of trade. Our main empirical challenge is that the black market transactions
are unobserved. To overcome this challenge, we develop a novel, transparent, and intuitive empirical approach based
on optimal transport methods to identify a lower bound for the gains from trade. We
exploit the fact that a lottery, which randomly samples car buyers from the population, will generate a different sales distribution than a black market that allocates licenses
to a selected subset of the population. We show how a distance between the two
sales distributions is informative about the volume of black market trade. To address the potential issue of time trends in the sales distributions, we develop
an analogue to difference-in-differences for distributions which uses shifts in the sales
distribution in the nearby city of Tianjin, where there was no rationing at the time, to
control for common trends. We find that at least $11\%$ of the quota is traded on the black
market.
\vspace{0.2in}\\
\noindent In the final step, we combine information from a variety of sources, including news reports, to infer transaction costs in a market equilibrium model. To further bound the
volume of trade, we combine estimates of the willingness-to-pay for Beijing license plates from the recent literature (\citealp{li18}) with anecdotal evidence.  We thereby produce, in Table \ref{tab:seven}, decreasingly nested estimates of identified intervals for the volume of trade and share of transaction costs that correspond to increasingly nested sets of assumptions. By a Coase theorem type argument, we infer transaction costs as the wedge
between demand and supply that is necessary to rationalize the estimated volume of trade.
\vspace{0.2in}\\
\noindent Though some of the anecdotal evidence points to lenient enforcement of the non-transferability of the license plates (see Section \ref{sec:lottery}), our estimates suggest otherwise. Firstly, we find that sizable gains from trade are left unrealized. While the market would realize \numberblank{RMB 18.8 billion} in gains from trade in the absence of transaction costs, we find that the realized net gains from trade lie in a plausible range from \numberblank{RMB 1.3 billion to RMB 5.3 billion}.\footnote{At the current exchange rate, dividing by seven gives a rough estimate of the dollar equivalents.} Secondly, we find that a plausible range from \numberblank{61\%} to \numberblank{82\%} of the realized gross gains from trade in license plates are lost to transaction costs.
\vspace{0.2in}\\
\noindent Our paper ties in different literatures. One is an older, and mostly theoretical,
literature on the economic impact of rationing on welfare and incentives for illegal trade (\citealp{tobin52, dreze75,stahl85,dye86}).
Our paper also complements a literature within the field of public finance that estimates
the size of the informal economy. Various indirect measures have been proposed, from
monitoring excessive electricity consumption to currency velocity; see \cite{schneider10} for an overview. The interest in this literature however tends to be in the scale of
tax evasion in the macroeconomy. Instead, our interest is in measuring the performance
of a particular black market, which calls for a more tailored empirical approach. 
\vspace{0.2in}\\
\noindent Finally, the market we study is of interest in itself. Our paper complements a
small literature on welfare effects of the recent rationing of car sales in larger Chinese
cities (\citealp{li18, xiao17, tan19, huang19}).
Importantly, this literature has assumed away the existence of a black market. Our
results show that black market trade plausibly ranges from 11\% to 27\% of the quota, and as such cannot be ignored in a thorough economic analysis of car sales in China.
\vspace{0.2in}\\
\noindent We give an overview of the lottery rationing mechanism and the black market in
Section \ref{sec:lottery}. We describe the data in Section \ref{sec:data} and report a standard event study in Section \ref{sec:did}. The optimal transport methods are laid out in Section \ref{sec:inferBounds} along with the estimation
results. We develop and evaluate the market equilibrium model in Section \ref{sec:inferCosts}, including
the estimated gains from trade and transaction costs. We conclude with a brief discussion
in Section \ref{sec:discussion}.

\section{The lottery and the black market for license plates} \label{sec:lottery}
\noindent In late December 2010, the Beijing government announced that, effective January 2011, car license plates would be required to drive without restrictions within the city limits.\footnote{The first public suggestion of a plan to ration car licenses was issued on December 13, 2010. See Section \ref{sec:E} for details about the lottery implementation as well as exact restrictions and eligibility criteria.} A quota of non-transferable license plates, which was set to about 40\% of the previous year's sales, was allocated by lottery. The lottery application process was simple: the pecuniary
costs were low and the application could be completed quickly online. Each applicant
had the same probability of winning a license plate, and there were limits on the number
of applications each household can submit. Immediately following the introduction of the
lottery, the number of newly registered cars in Beijing was reduced to the level of the
quota. 
\vspace{0.2in}\\
\noindent Allocating the license plates by lottery creates gains from trade: rationed prospective
car buyers may find license plate holders or corrupt officials who are willing to sell
theirs. News outlets, including government controlled ones, reported of a black market for
license plates that emerged soon after the introduction of the lottery. Table \ref{tab:news} gives results from a search in Chinese news reports that include mentions of black market transaction
prices. The most frequent mentions of yearly rental prices for a license plate are between
RMB 6\numcomma000 and RMB 12\numcomma000 annually. To provide context, the average monthly wage in
Beijing in 2011 was around RMB 5\numcomma000.  Given an interest rate of about 6\% at the time,
the twenty-year net present value (NPV) of the rental income stream of a license plate
ranges from RMB 73\numcomma000 to RMB 145\numcomma000.
\vspace{0.2in}\\
\noindent The news reports of purchase prices range from RMB 16\numcomma000 to RMB 650\numcomma000. Note that the latter price is for a so-called Jing-A plate, which carries particular prestige, 
is predominantly used by party officials and certain institutions, and is not available via the lottery. According to a \href{https://www.nytimes.com/2016/07/29/world/asia/china-beijing-traffic-pollution.html}{New York Times article}, these seem to be rare
transactions of exceptionally high value (\citealp{guo2016}). The majority of reported
purchase prices, without ownership, are between RMB 70\numcomma000 to RMB 120\numcomma000. The
upper end of that range is on par with the Ford Focus, which was the best-selling car at the time (\citealp{scmp2014}).
\vspace{0.2in}\\
\noindent \href{https://www.reuters.com/article/us-china-carplates/new-car-plate-restrictions-fuel-beijing-black-market-idUSBREA121K120140203}{Reuters} and \href{https://www.nytimes.com/2016/07/29/world/asia/china-beijing-traffic-pollution.html}{New York Times} reported a number of online sites that match license plate holders with buyers (\citealp{shen2014, guo2016}). Car dealerships seem to act as intermediaries. In 2014,
\href{https://www.reuters.com/article/us-china-carplates-idUSBREA121K120140203}{Reuters} quoted a sales representative who
claims that he can provide prospective buyers with license plates (\citealp{shen2014}).\footnote{Wang Shaoyong, sales manager at a Peugeot dealer in Beijing, said his shop provides car buyers with license permits from a partner firm that ``has many car plates registered in its name."} Such reports suggest
that the non-transferability of license plates is leniently enforced.
\vspace{0.2in}\\
\noindent Perhaps more surprisingly, \href{https://www.theatlantic.com/magazine/archive/2017/10/license-plate-marriages/537867/}{The Atlantic} reported that some license holders engage in fake marriages with buyers in order to legally transfer them a license \citep{carlson17}.  The quoted price for such a transaction is RMB 90,000.
\vspace{0.2in}\\
\noindent The market is reportedly supplied by both license plate holders and corrupt officials. According to \href{https://www.scmp.com/comment/insight-opinion/article/1095072/beijing-car-buyers-question-fairness-number-plate-lottery}{South China Morning Post}, officials involved with drafting the lottery
rules have experienced unusual luck in the lottery and there have been accusations of
the lottery being rigged (\citealp{scmp2012}).\footnote{The Beijing News  reported that a record 1.26 million residents competed for fewer than 20,000 plates in a single month.  ``Liu Xuemei" was so lucky to have won two plates on that month.  Cynical car-plate hunters wondered if they should change their name to stand a better chance in the next contest. Eagle-eyed internet users soon discovered Liu Xuemei was the name of the director of the vehicle and driver management department of the Ministry of Public Security. Liu Xuemei, in her 30s, was in charge of drafting rules for vehicle permits.} 
The \href{https://www.nytimes.com/2016/07/29/world/asia/china-beijing-traffic-pollution.html}{New York Times} reported that the head of the Beijing
Department of Transportation was sentenced to life in prison in 2015 for selling Jing-A
license plates, which suggests stricter enforcement. This sentence, however, appears to be part of general secretary Xi Jinping's campaign against corruption amongst party officials, and may not be representative of the overall policy.

\begin{table}[t]\centering
\small
\ra{\usualra}
\begin{threeparttable}
 \caption{\tabtitle{Black market transaction prices in 2011 and 2012 in Chinese news reports}}
    \begin{tabular}{@{}>{\raggedright}p{0.14\linewidth}>{\raggedright}p{0.25\linewidth}>{\centering}p{0.18\linewidth}>{\centering\arraybackslash}p{0.38\linewidth}}
 \toprule
 \tabhead{Article Date} & \tabhead{Newspaper} & \makecell{\tabhead{Rental Price} \\ (\tabunits{RMB 1\numcomma000/mo})} & \makecell{\tabhead{Purchase Price\tnote{$\ast$}} \\ (\tabunits{RMB 1\numcomma000})} \\
 \midrule
 Jul 23, 2011 & \href{http://auto.ce.cn/xwzx/201107/23/t20110723_21010465.shtml}{China Times} & 0.5-2 & 20 (N); 40-90 (N) \\
Sep 18, 2011 & \href{http://auto.sohu.com/?pvid=27112935}{Beijing Times} & 1\tnote{$\dagger$} &  \\
Oct 8, 2011 & \href{http://china.cnr.cn/yaowen/201110/t20111008_508583939.shtml}{China National Radio} & 0.5-1 &  \\
Oct 11, 2011 & \href{http://china.cnr.cn/yaowen/201110/t20111008_508583939.shtml}{China Economic Weekly} & 1 & 80-100 (U) \\
Dec 19, 2011 & \href{http://www.chinanews.com/auto/2011/12-19/3540512_2.shtml}{Beijing News} &  & 85 (U); 130 (U); 75 (U) \\
Feb 18, 2012 & \href{http://style.sina.com.cn/news/b/2012-02-18/072691718.shtml}{Rule of Law Weekend} & & 650\tnote{$\ddagger$} (Y)\:; 260 (Y); 50 (Y). \\
Aug 29, 2012 & \href{http://auto.ce.cn/xwzx/201208/29/t20120829_21241850.shtml}{Beijing Youth Daily} & 0.5-1 & 150 (Y); 30-40 (N)\\
Sep 10, 2012 & \href{http://auto.ce.cn/xwzx/201208/29/t20120829_21241850.shtml}{Beijing Times} & 0.8 & 40 (N)\\
Dec 15, 2012 & \href{http://finance.workercn.cn/c/2012/12/15/121215085904882564398.html}{Workers Daily} & 0.5 & 200+ (Y); 170 (U); 16 (N); 30-50 (N)\\
Dec 19, 2012 & \href{http://epaper.gmw.cn/gmrb/html/2012-12/19/nw.D110000gmrb_20121219_1-05.htm?div=-1}{Guangming Daily} &  & 30 (U)\\
Dec 26, 2012 & \href{http://www.chinanews.com/auto/2012/12-26/4437518.shtml}{ChinaNews.com} &  & 80 (U); 160 (U); 200 (U)\\
 \bottomrule
 \end{tabular} \label{tab:news}
 \begin{tablenotes}
    \item[$\ast$] The letters in the bracket in the column of purchase prices indicate the ownership status: Y = with ownership, N = without ownership, U = ownership unclear. 
    \item[$\dagger$] First year is free, and then the price is RMB 1\numcomma000 per month.
    \item[$\ddagger$] Jing-A is a license plate that is used predominantly by governments and institutions. These license plates command a premium in the market.  These are not available through the lottery.
    \item \hspace{-1.2mm} Sources (by row): \citet{mengmeng11, beijingtimes11, cnr11, beijingnews11, rlw12, xiaohong12, cin12, guangming12, cnn12}
 \end{tablenotes}
\end{threeparttable}
\end{table}
\vspace{0.2in}
\noindent In sum, we find anecdotal evidence of the existence of a material market for license plates. With respect to enforcement, the evidence is mixed, leaning towards lenient. Though a given range of transaction prices is consistently reported across outlets, we find little indication of either the volume of black market trade or of the magnitude of the transaction costs.  The question of the volume of the Beijing black market for licenses seems to remain an open one.  We next turn to the car sales data with which we estimate the size of the market.

\section{Data} \label{sec:data}
We obtained vehicle registration data from \href{http://www.webinsight.cn}{Webinsight Technology and Information
Corporation}, a Chinese marketing research company. The vehicle registration data
have observations at the city-month level on aggregate registration of all vehicle
models available from January 2010 to December 2015 in 35 cities in China. 
The
vehicle models are identified by their unique codes as catalogued by the Motor
Vehicles' Type and Model Designation, published by the National Standard of
People's Republic of China, and various car characteristics are provided. 
\vspace{0.2in}\\
\noindent The vehicle quota data are collected from each city's official publications.\footnote{The official publications of information on allocation mechanisms are from Shanghai International Commodity Auction Co. Ltd., Beijing Passenger Vehicle Quota Administration Office, and Tianjin Information System of Passenger Vehicle Quota Administration.}
The data include the number of applicants, the quota, the average value of winning
bids (in the auction rationing mechanism), and the share of each type of rationing
mechanism if multiple mechanisms are applicable, see Table \ref{tab:rationing}. Table \ref{tab:stat} aggregates
the variables across cities and years.


\begin{table}[t]\centering
\small
\ra{\usualra}
\begin{threeparttable}
 \caption{\tabtitle{Rationing Mechanisms}\tabcaption{}}
 \begin{tabular}{@{}>{\raggedright}p{0.35\linewidth}>{\centering}p{0.15\linewidth}>{\centering}p{0.25\linewidth}>{\centering\arraybackslash}p{0.15\linewidth}}
 \toprule
    & \tabhead{Beijing}  & \tabhead{Tianjin} & \tabhead{Shijiazhuang} \\
    \midrule 
    2015 Per Capita GDP (USD) & 18731 & 13355 & 4576\\
    Rationing Mechanism & Lottery & Auction and Lottery & No Rationing\\
    Announcement Date & Dec 23, 2010 & Dec 15, 2013 & n/a \\
    Implementation Date & Jan 1, 2011 & Dec 16, 2013 & n/a \\
    Average Quota (per mo) & 10\numcomma000 & 8\numcomma000 & n/a \\
    \makecell[l]{Quota Allocation \\ \quad  (electric:lottery:auction)} & 0:1:0 & 1:5:4 & n/a \\
    Vehicles without local plate & restricted & restricted & n/a \\
    \bottomrule
 \end{tabular} \label{tab:rationing}
\end{threeparttable}
\end{table}


\begin{table}[t]\centering
\small
\ra{\usualra}
\begin{threeparttable}
 \caption{\tabtitle{Summary Statistics for MSRP (RMB)}}
 \begin{tabular}{@{}>{\centering}p{0.15\linewidth}>{\centering}p{0.15\linewidth}>{\centering}p{0.15\linewidth}>{\centering}p{0.15\linewidth}>{\centering\arraybackslash}p{0.15\linewidth}}
 \toprule
    \tabhead{Count} & \tabhead{Mean} & \tabhead{Std. Dev.} & \tabhead{Min.} & \tabhead{Max.} \\
 \midrule
    5\numcomma065\numcomma356 & 150\numcomma356 & 100\numcomma919 & 20\numcomma800 & 1\numcomma305\numcomma000 \\
 \bottomrule
 \end{tabular} \label{tab:stat}
\end{threeparttable}
\end{table}

\section{Difference-in-differences analysis} \label{sec:did}
\begin{figure}[H]
  \centering
  \includegraphics[width=\textwidth]{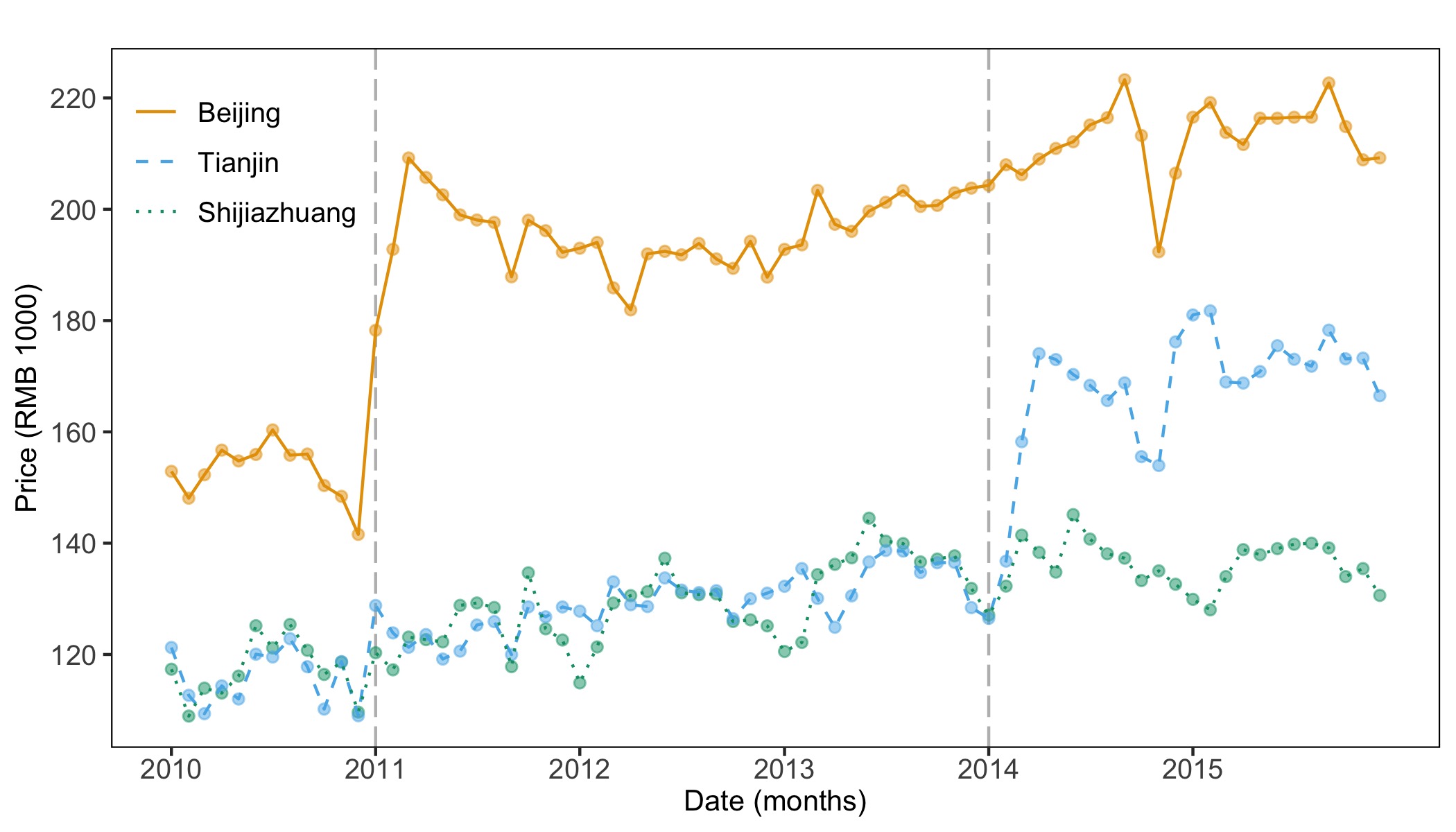}
  \caption{\figtitle{Monthly average car price (RMB 1\numcomma000) for Beijing, Shijiazhuang, and Tianjin.} \figcaption{Beijing introduced the lottery in January 2011, and Tianjin introduced the hybrid auction and lottery mechanism in January 2014.}}
  \label{fig:one}
\end{figure}

\begin{figure}[H]
  \centering
  \includegraphics[width=0.75\textwidth]{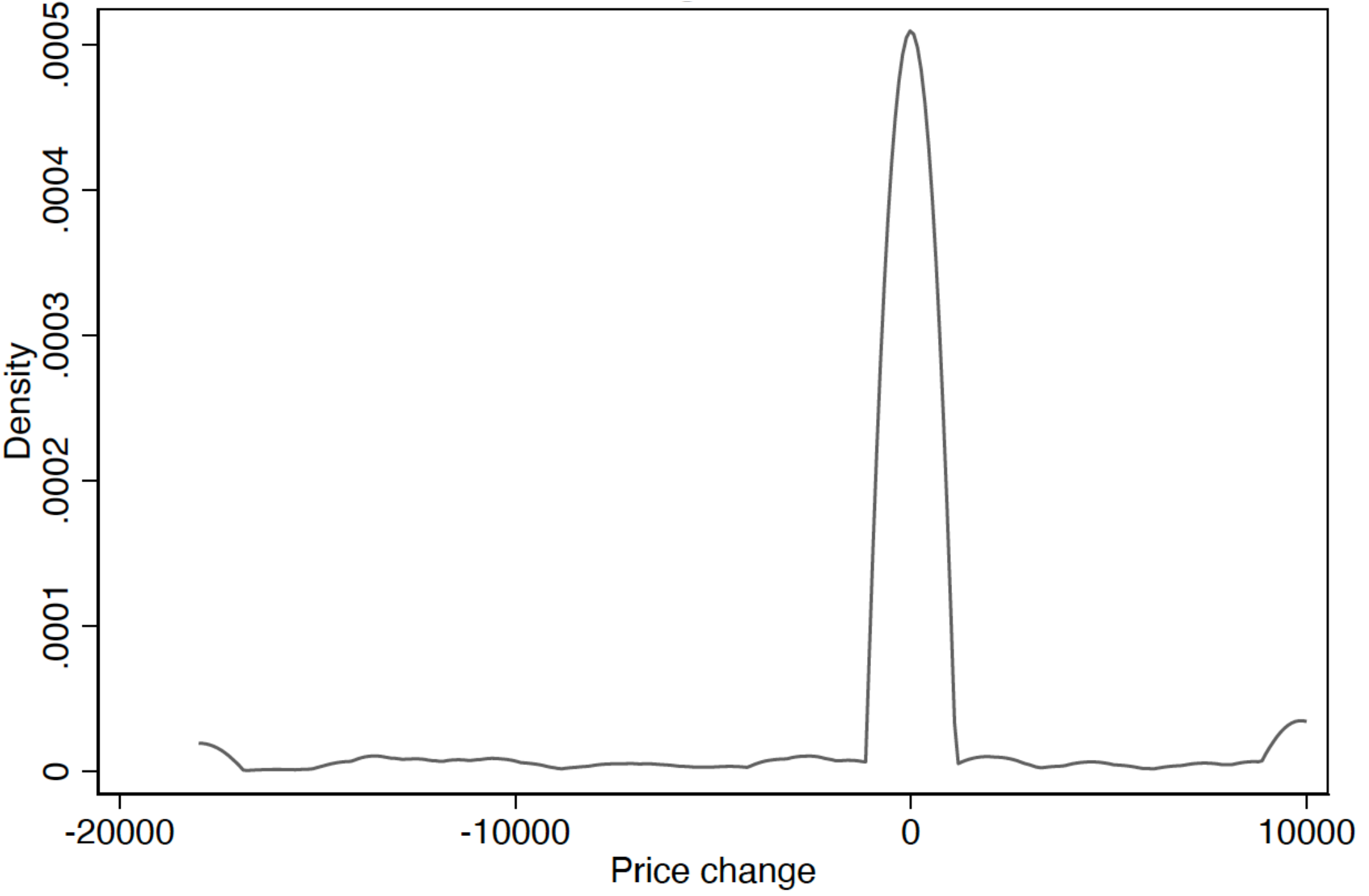}
  \caption{\figtitle{Distribution of price changes between 2010 and 2011.} \figcaption{Nearly all of the density is near 0, suggesting that prices did not change much between 2010 and 2011.}}
  \label{fig:threeprice}
\end{figure}

Figure \ref{fig:one} displays the average car prices in Beijing, Tianjin, and Shijiazhuang on a
monthly level before and after the introduction of the Beijing lottery. Tianjin is the
fourth largest city in China and is about 130 kilometers away from Beijing. With a GDP of about two-thirds of Beijing's GDP, Tianjin experienced
the fastest economic growth of the major Chinese cities in the period covered by the
data. Shijiazhuang, 270 kilometers away from Beijing,  is the capital and the
largest city of North China's Hebei Province.
Although it has about a quarter of the GDP of Beijing and about a tenth of the population, Shijiazhuang has a
similar pre-trend to Tianjin and Beijing. Shijiazhuang had not introduced rationing
by 2016, while Tianjin introduced a hybrid auction/lottery format in January 2014.
The average price in Beijing is seen to increase by \numberblank{24\%} following the introduction
of the lottery. The average prices in Tianjin and Shijiazhuang increased by about
\numberblank{3\%} in the same period. Assuming common trends across the cities, we get a difference-in-differences 
estimate of the effect of the rationing on the average car
prices of about \numberblank{\did}. See Section \ref{sec:A} for the corresponding difference-in-differences
regressions.
\vspace{0.2in}
\noindent The price jump in Beijing could be a supply side price response to the rationing. However, when we
compare the prices of cars that were sold in 2010 with the prices of the same car
models in 2011, we find very little changes in prices.
While, as detailed below, we do not have matched data for the transaction price of each sold car, only its MSRP, we do have access to proprietary data from a market research firm detailing the changes in price of each car model before and after the lottery.  These are presented in Figure \ref{fig:threeprice}.

\vspace{0.2in}

\begin{figure}[H]
  \centering
  \includegraphics[width=\textwidth]{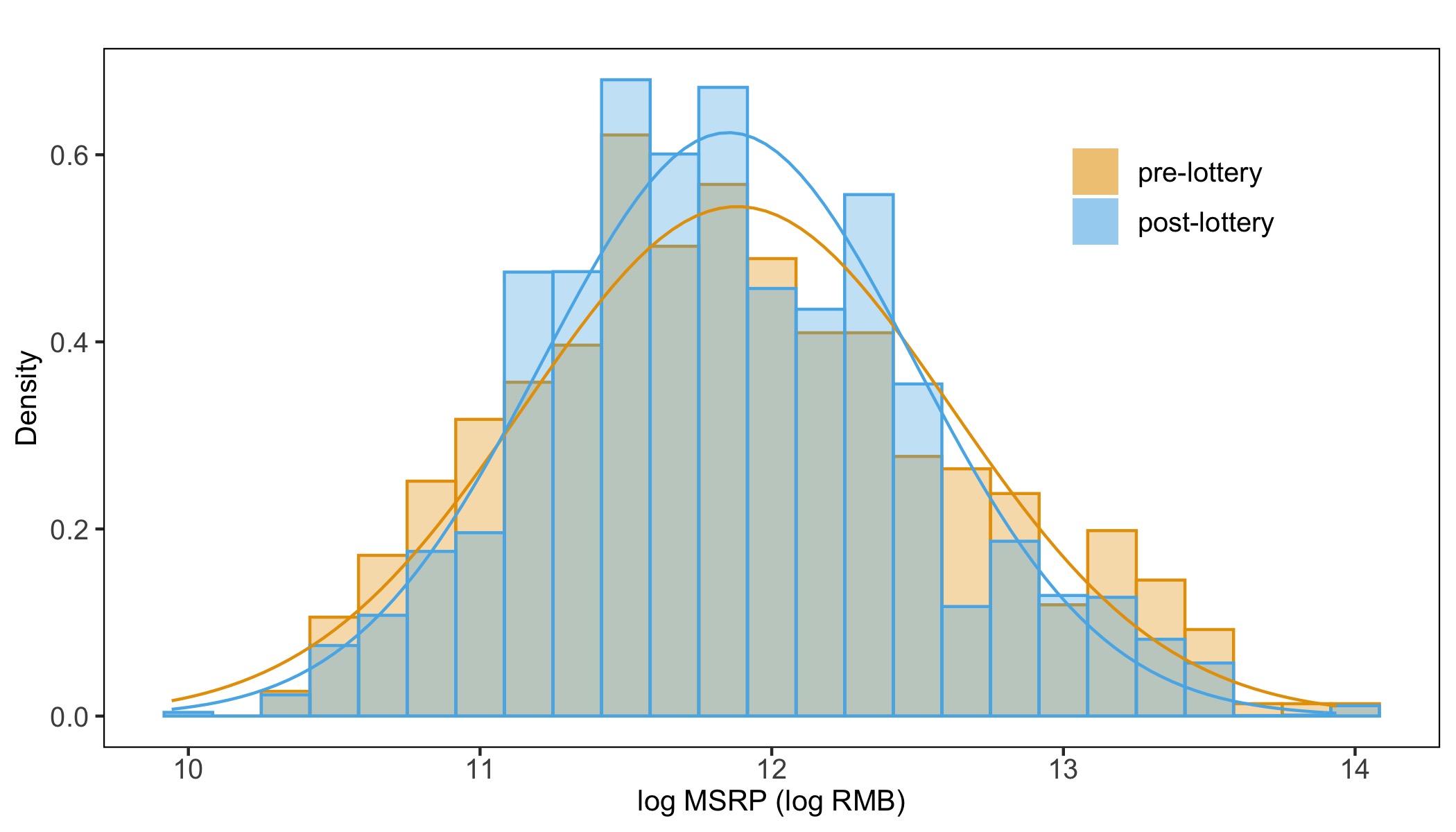}
 \caption{\figtitle{Distribution of Beijing car prices in 2010 (pre-lottery) and new car prices in 2011 (post-lottery).} \figcaption{There is little change in the car prices before the lottery and the car prices of new models after the lottery, irrespective of the number of cars sold at each price. The parameters of the superimposed Gaussian curves are the mean and variance of log MSRP for the respective set of data.}}
  \label{fig:two}
\end{figure}

The lack of supply side price responses is explained by the vertical
restraints that were used in the industry at the time. In 2014 and 2015, following a
large scale price investigation, the National Development and Reform Commission
fined car manufacturers, which included Audi, Chrysler,  Mercedes-Benz,
Dongfeng-Nissan and BMW, a total of more than RMB 2.5 billion for practicing
resale price maintenance for cars and spare parts (see \citealp{cai16}).
While measurement errors in prices, i.e., differences between the measured manufacturers'
suggested retail price (MSRP) and the actual purchase price, is a common concern
in studies of demand for cars in the United States, it is not a threatening empirical issue in the case at hand. Such measurement errors are less of an
issue in the Chinese data since under resale price maintenance, the MSRPs are
enforceable minimum prices.  Indeed, transaction prices are unavailable for large data sets but will typically not deviate from MSRP by a substantial margin.  
\cite{li18} concurs and his analysis likewise relies on this assumptions.  As he explains, ``there could be potential pitfalls in using MSRPs when they are different from the
transaction prices due to promotions,'' however ``different from the promotion-heavy environment in the
U.S., China’s auto market has infrequent promotions from manufactures or dealers and retail
prices are often very close to or the same as MSRPs, a phenomenon commonly seen for luxury
products in China.''
See \cite{xiao17} and references therein for further discussion.
\vspace{0.2in}\\
The bottom line, as we see it, is that not only are the price differences between MSRPs and transaction prices small, they are effectively constant for any given model.
This makes it very unlikely to induce over-counting of sales. 
\vspace{0.2in}\\
Even more to the point, we may directly verify the robustness of the results. We have access to a proprietary data set that is considerably smaller but is less susceptible to measurement error since it contains transaction prices. In particular, the total number of households that were randomly selected to collect the proprietary data in Beijing over the years 2010 and 2011 is 3\numcomma718, whereas the number of households present the Beijing MSRP data during the same time frame is 974\numcomma573.
\vspace{0.2in}\\
Using this proprietary data set, we are able to match a subset of the models in our MSRP dataset with the transaction price, and we analyze this subset using MSRPs and transaction prices separately.
\vspace{0.2in}\\
We find that the absolute difference of the results is quite small, suggesting that  measurement error that arising from using MSRPs instead of transaction prices is not a threat to the validity of the analysis. A longer discussion of this direct robustness check can be found in  Section \ref{sec:msrptp} of the Appendix.
\vspace{0.2in}\\
\noindent Though car dealerships were effectively not allowed to change their prices, \cite{feenstra88} suggests that car dealerships may upgrade the quality of their product lines
in response to the quota. The gain from an increase in the intensive margin can
partially offset the loss from the extensive margin following the rationing. The
increase in the average price in Beijing could therefore be a strategic price response
by the manufacturers to the market contraction. If so, the average prices for the
models that were sold following the rationing, but not before, would be higher than
the models that were sold both before and after the introduction of the lottery.
Figure \ref{fig:two} shows barely noticeable changes in the distribution of the prices of the
models offered before the lottery (pre) and the models that were only offered after
the lottery (post). The post-lottery distribution seems to shift slightly to the left, if
anything. In sum, we find little evidence that the price jump was caused by supply
side responses.
\vspace{0.2in}
\noindent There were other policies in place that may have affected car sales. The Beijing subway system expanded rapidly from 39 stations in 2002 to more than 300 in 2014. In particular, a number of new lines that expanded the network from 228 kilometers to 336 kilometers were opened in December 2010, just before
the rationing went into effect. Public transportation appeals disproportionately to
lower income households who may subsequently select out of the car market,
which would lead to a change in demand that is unrelated to the rationing (\citealp{xie16}). Since the subway expansion in 2010 coincided with the rationing, it is
hard to directly test this alternative account in the data. There were, however, three
other major subway expansions: in December 2012 (70 kilometers), May 2013 (57 kilometers), and
December 2014 (62 kilometers). Figure \ref{fig:one} shows no apparent shift in the Beijing prices at any
one of these expansions. The lack of responses at these later expansions suggests
the effect of the expansion in December 2010 was also modest.  
\vspace{0.2in}\\
\noindent The Beijing price jump may instead be explained by black market trade. If
the Beijing lottery randomly selected car buyers from the population of car buyers,
if there were no illicit trade in license plates, and if income and preferences did not change
right after the lottery, then the sales distribution would not shift with the rationing.
The price jump is instead consistent with a black market that reallocates the license
plates to wealthier households who buy more expensive cars.
\vspace{0.2in}\\
\noindent Figure \ref{fig:one} shows a similar, if smaller, price jump in Tianjin at the beginning
of 2014 when Tianjin introduced license plate rationing. Tianjin
used a lottery/auction hybrid mechanism that allocates 50\% of the license plates 
by lottery, 40\% by auction, and reserves the remainder for electric cars. By design,
the auction, like the black market, allocates license plates to those who have the highest willingness to pay. The Beijing price jump is consistent with a
share of the Beijing license plates being allocated in part by a lottery and in part
by a price-based mechanism.
\vspace{0.2in}\\
\noindent Though the data show price jumps following the rationing in both Beijing
and Tianjin, it is not obvious that price-based allocation mechanisms, be
it a black market or an auction, would shift sales towards more expensive cars.
On the one hand, the price of a license plate serves as a tax on a car purchase
that leads a given buyer to substitute towards less expensive cars. On the other
hand, price-based allocation mechanisms select wealthier buyers that purchase
more expensive cars. The net effect could be that these mechanisms shift sales
towards less expensive cars. \cite{tan19} and \cite{xiao17}, however,
document increasing expenditures in Chinese cities that use lottery/auction hybrids.
In Shanghai, which has used auctions to allocate license plates since 1996, there is
a strong positive correlation between auction prices and car expenditures, so the
net effect has empirically proved to be positive.\footnote{The average Shanghai auction price in 2010 was RMB 39\numcomma000 and increased to RMB 81\numcomma000 in 2015, while the average car price increased from RMB 173\numcomma000 to RMB 215\numcomma000 over the same period.} 
We will remark, however, that our estimation methods do not require the sales to shift towards more expensive cars.

%
%

\section{Inferring bounds on the volume of trade using optimal transport}\label{sec:inferBounds}
\begin{figure}[H]
  \centering
  \includegraphics[width=\textwidth]{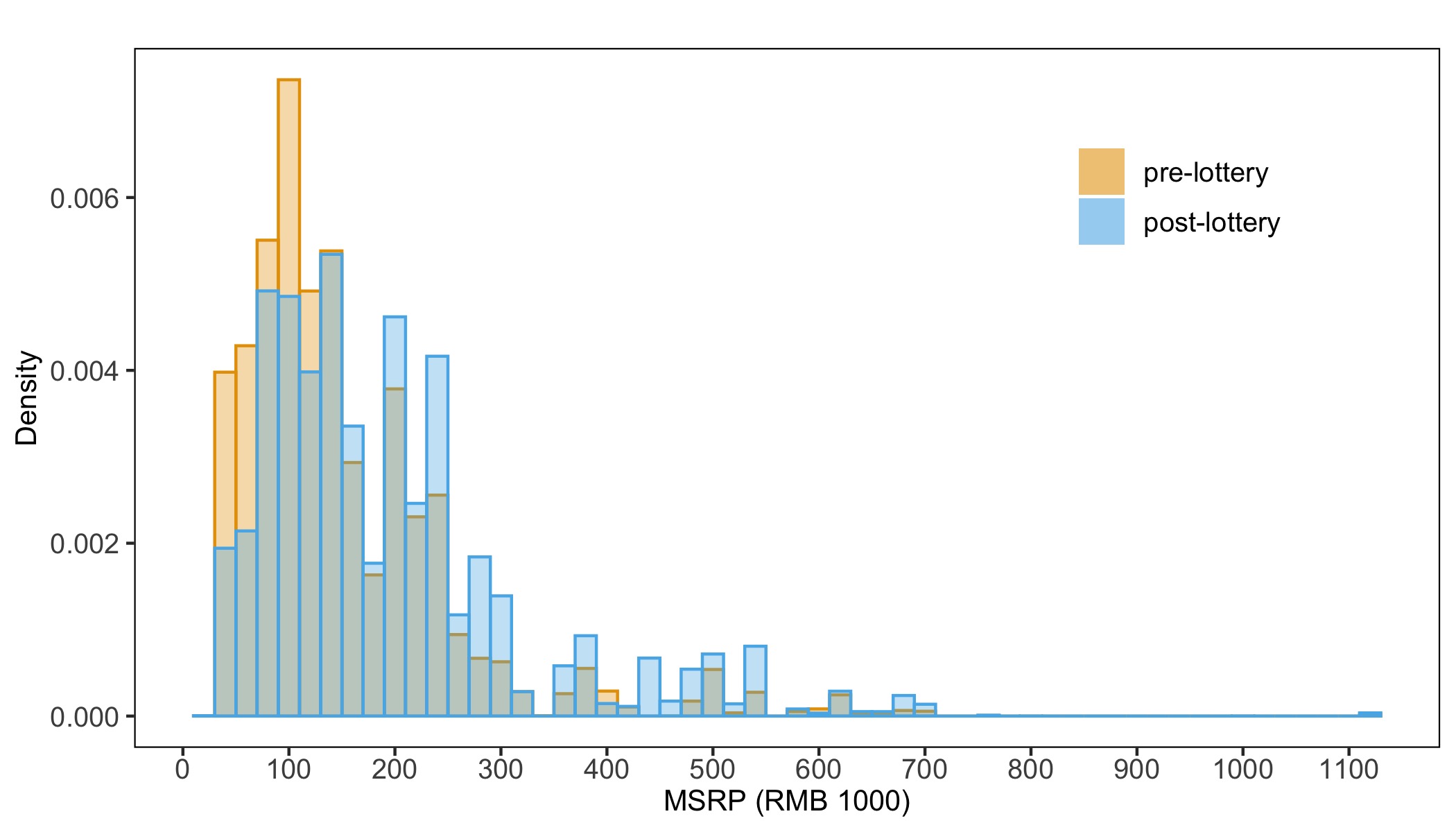}
 \caption{\figtitle{Histograms of Beijing car sales in 2010 (pre-lottery) and 2011 (post-lottery).} \figcaption{Unlike Figure \ref{fig:two}, this plot considers the number of cars sold at each MSRP value.}}
  \label{fig:four}
\end{figure}

The main empirical challenge is that we do not observe transactions in the black
market directly.  
Though a \numberblank{\did} increase in the average price may seem substantial,
one cannot infer the volume of trade from the first moments of the sales distributions.
We instead wish to infer the volume of trade from the observed displacements of
the sales distributions. Figure \ref{fig:four} shows overlaid histograms
of car prices in Beijing in 2010 and 2011. The sales distribution clearly shifts
to the right after the introduction of the lottery. This displacement is consistent
with a black market that reallocates license plates to wealthier households, but,
under assumptions we specify below, is inconsistent with a lottery that randomly
selects car buyers from the population. Our strategy is to infer a lower bound
for the volume of black market trade by quantifying the displacement of the car
sales distributions in Figure \ref{fig:four} using optimal transport
theory and methods.

\noindent Optimal transport has a long history in economics and operations research (see \citealp{kantorovitch58} for an early treatment). 
Furthermore, optimal transport has recently
witnessed renewed interest in economics and applied econometrics, including the analysis of
identification of dynamic discrete-choice models (\citealp{chiong16}), in vector
quantile regression (\citealp{carlier16}), in empirical matching models
(\citealp{galichon18}), and in latent variables (\citealp{arellano2019recovering}).  To the best of our knowledge, this is the first principled use of optimal transport for reduced form analysis in economics, notwithstanding matching applications.  The R package diftrans, implementing the estimators developed in the methodological section, is available on the corresponding author's website.


\vspace{0.2in}
\noindent We develop two estimators. 
First, the \emph{before-and-after estimator} uses displacement, i.e., the shift in the
sales distribution following the rationing in Beijing, in order to estimate the volume of
trade. 
Second, the \emph{difference-in-transports estimator} uses a difference between the displacement in Beijing
and the corresponding displacement in Tianjin. Similar to a standard difference-in-differences estimator, 
this difference in displacements controls for common economic trends. As with standard regression methods, we must specify a cost
function which, in the case of optimal transport, is the distance between points in
the support of the compared probability mass functions.
\vspace{0.2in}\\
\noindent In the following, we refer to a \emph{buyer} as a household that purchases a car
with an illegally purchased or rented license plate. We refer to a \emph{seller} as a
household that won a license plate and then decided to sell or rent it to a
buyer. We refer to the transaction between a buyer and a seller as a \emph{trade}.
\vspace{0.2in}\\
\noindent In the following, it is useful to adopt a potential outcomes notation. Let $\mathbb{P}_{pre}(r,b)$
and $\mathbb{P}_{post}(r,b)$ be the potential pre-lottery and post-lottery population sales distributions in Beijing, respectively. The index $r \in \{0,1\}$ is one when there is rationing
post-lottery, and zero otherwise. The index $b \in \{0,1\}$  is one when there is a black
market post-lottery, and zero otherwise.  
Distinguishing between the rationing and the presence of a black market in the potential outcomes model will allow for a precise comparison between the assumptions needed for both estimators. 
\subsection{Identification Assumptions and Optimal Transport Theory} \label{ssec:bna}
We first provide three assumptions that are sufficient to identify a lower bound for the volume of trade.  The first rules out anticipatory sales effects of the rationing; the pre-lottery sales distribution does not depend on whether there is rationing and/or a black market in the post-lottery period.
\begin{assumption} \label{as:one}
No anticipation: $$\mathbb{P}_{pre}(r,b)=\mathbb{P}_{pre}$$ for $r \in \{0,1\}$ and  $b \in \{0,1\}$.
\end{assumption}
\noindent Assumption \ref{as:one} seems plausible given that there was only a week between the announcement
of the rationing and its implementation.  It is possible that some anticipation occurred in December 2010, the month preceding the inauguration of the lottery.  For robustness, we produce the analysis with and without December 2010 (see Section \ref{sec:v1}).  We find that the results are quite robust; both of our estimates change by at most one percent when we omit December 2010.
\vspace{0.2in}\\
\noindent Next, we assume away time trends that are unrelated to the lottery. Specifically, we assume that the pre-lottery sales distribution is equal to
the potential post-lottery sales distribution when there is no rationing and no black market.
\begin{assumption} \label{as:two}
No time trends: $$\mathbb{P}_{pre}=\mathbb{P}_{post}(0,0).$$
\end{assumption}
\noindent Assumption \ref{as:two} rules out trends in car preferences such as, say, increasing demand
for SUVs. It implies that a household that wins the lottery would purchase a car, in 2011 in a world without rationing, according to the same price distribution as it would in 2010. It
also precludes car dealerships from changing the car prices following the lottery.
This assumption is consistent with
vertical restraints that regulated retail prices in the car market at the time as discussed in Section \ref{sec:lottery}. It is possible to do away with this assumption (see Section \ref{ssec:dit}).  
\vspace{0.2in}\\
\noindent The next assumption concerns the effect of rationing on car
preferences.
\begin{assumption} \label{as:three}
No general equilibrium effects: $$\mathbb{P}_{post}(0,0)=\mathbb{P}_{post}(1,0).$$
\end{assumption}
Assumption \ref{as:three} is an analogue to the common stable unit treatment value assumption, also called no interference assumption, in the program evaluation literature. It forces the potential post-lottery
sales distribution, without rationing and with no black market, to equal
the potential sales distribution post-lottery, with rationing and with no black market.
An example of general equilibrium effects that violates Assumption \ref{as:three} would
be if rationing license plates, which leads to fewer cars on the roads, increased the
willingness-to-pay for a car. We however believe that such general equilibrium effects
are negligible, since the rationing controls the 
flow of new cars to Beijing (on the order of 260\numcomma000 per year), and not the stock (on the order of five million at
the time). The assumption also rules out a direct effect of the rationing on the
car preferences. This assumption is common in the literature on rationing, but it is untestable, and it has been contested in the past. One example is \cite{tobin52},
which entertains the idea that rationing can change tastes over time.
\begin{displayquote}
``\textit{Experience under rationing may alter the consumer's scale of preferences. He
may learn to like pattern of expenditures into which rationing forces him, or
to dislike it even more intensely than if it had not been forced upon him."} (p.
548).
\end{displayquote}

\noindent \cite{shen19} explores the hypothesis that lottery winners experienced a sense
of luck that shifted their preferences towards more expensive car models, which
violates Assumption \ref{as:three}. Neither \cite{shen19}'s hypothesis nor Assumption \ref{as:three}
seems to be testable in our data, and thus must be assessed according to their plausibility.  
\vspace{0.2in}\\
\noindent A potential threat to the validity of Assumption \ref{as:three} is that the rationing can mechanically induce a difference in the observed distributions $\mathbb{P}_{pre}$ and $\mathbb{P}_{post}(1,0)$ by bringing about a change in their composition. Specifically, it is possible that the distribution of cars purchased with a new license and the distribution of cars purchased with an existing license –i.e., one which was previously associated with another car– are different. Because only the former purchases are being rationed, the observed distribution may change due to the fact that its composition has changed. As detailed in Section \ref{ssec:bna_est} and in Section \ref{sec:B} of the Appendix, the magnitude of this effect may be assessed, and it is not a threat to our analysis.
\vspace{0.2in}\\
\noindent Another potential issue deserving discussion is the possibility that households planning to purchase multiple cars, upon winning the lottery, purchase a more expensive car than they otherwise would have, knowing that they will in all likelihood only get to buy one.  We argue that such an effect does not threaten our analysis.  First, the many conditions such buyers must satisfy suggest that they are few and far apart; we speak of prospective car buyers who had at least two fewer cars than they wanted to purchase in the short- or medium-term, were substituting at least one less car for a substantially more expensive car, but were not willing to go purchase it on the black market.  Second, and more importantly, our estimator is specifically robust to this concern.    Indeed, upon inspection of the transport map corresponding to the solution of the optimal transport estimator presented in Section \ref{ssec:bna_est}, one finds that approximately $99\%$ of the trades that account for the estimated total are for more than RMB 100\numcomma000 (which, for instance, is almost half of the 95th percentile household income in China).\footnote{This approximation requires a unique and interpretable solution $\hat{\gamma}$, as defined below.  This obtains by using a ground cost $\mathbf{C}_{i,j}+\lambda|x_{i}-x_{j}|$, for all $i,j$, for small $\lambda$, as it picks out a solution with small trades. We used $\lambda=0.01$.}  This means we would only have as confounders  the unlikely households who are spending an additional amount of at least RMB 100\numcomma000 on the intensive margin because they are constrained on the extensive margin, and are still not willing to transact on the black market.  We consider those to be very unlikely households.
\vspace{0.2in}\\
\noindent If license plates are reallocated by black market trade, then $\mathbb{P}_{post}(1,1)$ may
not equal $\mathbb{P}_{post}(1,0)$. If a buyer purchases a different car than the seller would
have purchased, had he not traded his license, then we can think of that trade as
shifting or transporting mass from $\mathbb{P}_{post}(1,0)$ to $\mathbb{P}_{post}(1,1)$. The smallest number
of such trades required to account for the shift in distributions gives a lower bound
estimate of share of black market trade. It is a lower bound since, for instance, a trade where the
buyer purchases the same car as the seller would have purchased, had he not sold
his license, goes undetected. Since all license plates can be traded without shifting
the sales distribution, the upper bound for the volume of trade is 100\% of the
quota. The counterfactual distribution $\mathbb{P}_{post}(1,0)$ is identified since Assumptions \ref{as:one},\ref{as:two}, and \ref{as:three} imply that
\begin{equation}\label{eq:1} \mathbb{P}_{pre}=\mathbb{P}_{post}(1,0).\end{equation}
\vspace{0.1in}\\
\noindent We could have given condition \eqref{eq:1} directly as an assumption. By distinguishing
between Assumptions \ref{as:two} and \ref{as:three}, and only relaxing Assumption \ref{as:two} in Section \ref{ssec:dit} where we develop a difference-in-transports estimator, we get a clear comparison of
sufficient conditions for identification.
\vspace{0.2in}\\
\noindent Under these assumptions, the minimum amount of mass we need to transport between the (observable) distributions $\mathbb{P}_{pre}$ and $\mathbb{P}_{post}(1,1)$ estimates a lower
bound for the volume of trade. For ease of exposition, we drop the potential outcome
arguments from $\mathbb{P}_{post}(1,1)$ and, abusing notation, let $\mathbb{P}_j(x)$ for $j \in \{pre, post\}$ and $x \in \mathcal{X}$ denote the probability mass functions in the rest of this section, where $\mathcal{X}$ is the support of car prices. If we observed the population
distributions $\mathbb{P}_{pre}$ and $\mathbb{P}_{post}$ directly, we could compute the desired lower bound via
the following oracle problem,
%
%
\begin{mini}
{\gamma \in \Gamma}
{\int c(x_0,x_1) \gamma(x_0,x_1)dx_0dx_1}
{\label{eq:2}}
{OT(\mathbb{P}_{pre}, \mathbb{P}_{post}) = }
\addConstraint{\int\gamma(x_0,x_1)dx_1}{= \mathbb{P}_{pre}(x_0),\quad}{\textnormal{for all } x_0 \in \mathcal{X}} 
\addConstraint{\int \gamma(x_0,x_1)dx_0}{= \mathbb{P}_{post}(x_1),\quad}{\textnormal{for all } x_1 \in \mathcal{X}.} 
\end{mini}
\vspace{0.2in}\\ 
\noindent The optimal transport cost has two components. The first is $c(x_0,x_1)$, the cost of
transport between point $x_0$ in $\mathbb{P}_{pre}$ and point $x_1$ in $\mathbb{P}_{post}$. The second is 
$\gamma(x_0,x_1)$, the amount of mass that is transported between point $x_0$ in $\mathbb{P}_{pre}$ and point $x_1$ in $\mathbb{P}_{post}$. The optimization is over $\gamma \in \Gamma$, where $\Gamma$ is the set of all bivariate probability distributions. The constraints ensure that the marginal distributions exactly equate
the observed distributions  $\mathbb{P}_{pre}(x)$ and  $\mathbb{P}_{post}(x)$. We choose the cost specification
\begin{equation}\label{eq:3}
c(x_0,x_1)=\mathbbm{1}(|x_1-x_0|>0).
\end{equation}
\vspace{0.2in}\\
\noindent This function assigns the same cost for transport between any pair of points $x_0$, $x_1$
when $x_0\neq x_1$, and zero cost when $x_0=x_1$. The objective function of the optimal
transport problem therefore exactly represents a volume of trade that transforms the pre-lottery distribution
into the post-lottery distribution. The optimal transport cost $OT$ for the problem in \eqref{eq:2} can be interpreted as \emph{the smallest volume of trade that is consistent with the data}. Our estimand is then
\begin{equation}\label{eq:4}
s(\mathbb{P}_{pre}, \mathbb{P}_{post})=OT(\mathbb{P}_{pre}, \mathbb{P}_{post}).
\end{equation}
\vspace{0.1in}\\
\noindent An example may be instructive. Suppose cars are sold either at a price $p_1$ or a price
$p_2$, so $\mathcal{X} = \{1,2\}$. Pre-lottery, eight cars are sold: six at $p_1$ and two at $p_2$, such
that the population distribution is $\mathbb{P}_{pre}=\frac{1}{8}\lbrack6,2\rbrack$. Post-lottery, the quota is set to
four cars. One is sold at $p_1$ and three at $p_2$, such that the population distribution is
$\mathbb{P}_{post} = \frac{1}{4} \lbrack1,3\rbrack$. It is immediately clear that at least two license plates must be traded to rationalize the shift in the sales distribution. One solution $\gamma$ that satisfies the constraints in \eqref{eq:2} is
\begin{equation}
\gamma=\frac{1}{4}
\begin{pmatrix}
1 & 2\\
0 & 1
\end{pmatrix}\nonumber.
\end{equation}
\vspace{0.1in}\\
\noindent The diagonal terms imply no transport and hence no cost. We can therefore
calculate the optimal transport by summing all off-diagonal terms of $\gamma$. This gives
$OT(\mathbb{P}_{pre}, \mathbb{P}_{post}) = 50\%$ as a lower bound on the share of illegal trade. This example
assumed away sampling variation in the empirical distributions. We account for
sampling variation in Section \ref{ssec:bna_est} below.
\vspace{0.2in}\\
\noindent There may be more than one $\gamma$ that solves \eqref{eq:2}. Our parameter of interest is
however not $\gamma$, but $OT(\mathbb{P}_{pre}, \mathbb{P}_{post})$, which is clearly unique. Multiplicity of $\gamma$ solutions therefore has no implications for the identification of a lower bound for the volume of trade.
\vspace{0.2in}\\
\noindent It is tempting to visualise, in Figure \ref{fig:four}, the displacement as the area of the two histograms that is non-overlapping.  This turns out to be correct; it represents the total variation distance between the two distributions, which obtains as the dual representation of the optimal transport distance when the cost is as specified in \eqref{eq:3}.
\vspace{0.2in}\\
\noindent The estimand in \eqref{eq:4} does not require that buyers purchase more expensive
cars than the sellers would have purchased if they could not trade, i.e., that the
sales distribution shifts to the right; it only requires that they purchase cars of different prices so that the trades are observed. Any
displacement of the sales distributions is counted as trade.

\subsection{Before-and-after estimator} \label{ssec:bna_est}
\begin{figure}[H]
  \centering
  \includegraphics[width=0.95\textwidth]{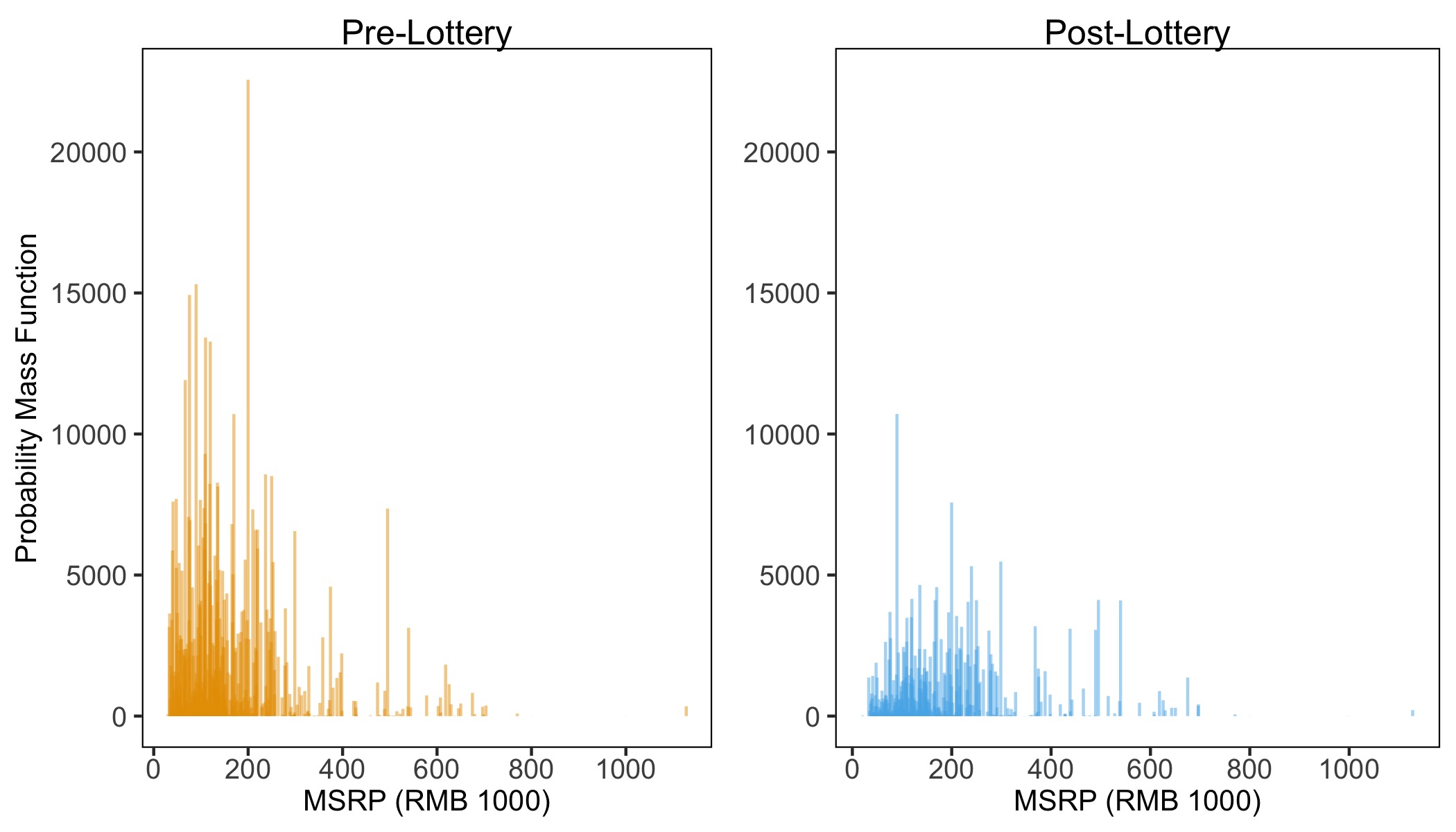}
 \caption{\figtitle{Exact empirical distributions of Beijing car sales in 2010 (pre-lottery) and 2011 (post-lottery).} \figcaption{These are labeled $\hat{\mathbb{P}}_{pre, n_{pre}}$ and $\hat{\mathbb{P}}_{post, n_{post}}$, respectively.}}
  \label{fig:five}
\end{figure}

\begin{figure}[H]
  \centering
  \includegraphics[width=\textwidth]{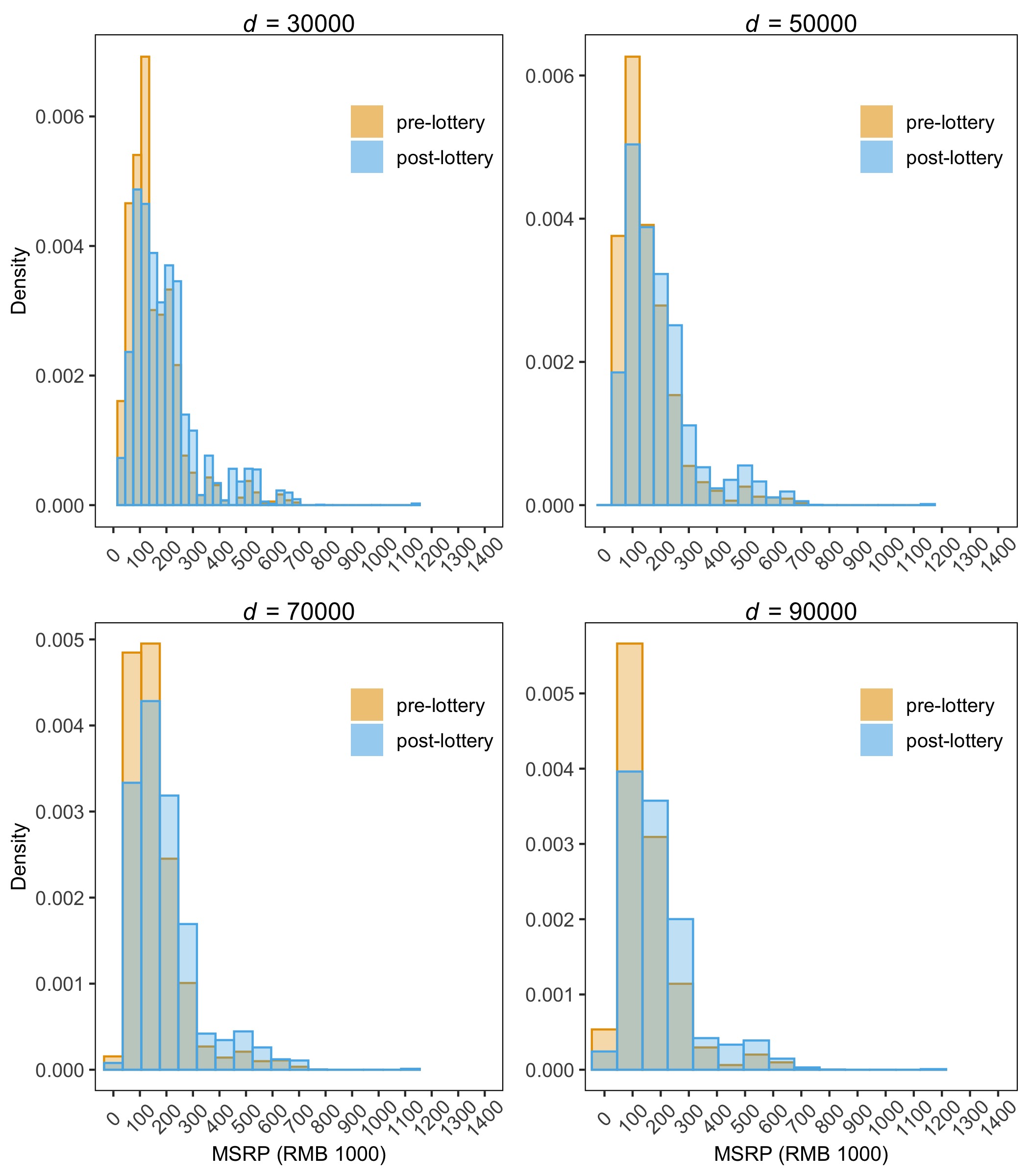}
 \caption{\figtitle{Smoothed empirical distributions of Beijing car sales in 2010 (pre-lottery) and 2011 (post-lottery) at candidate values of $d$.} \figcaption{The width $d$ of the bins in each of these plots is different, approximating the different levels of smoothing applied to Figure \ref{fig:five}.}}
  \label{fig:six}
\end{figure}

We need to account for the fact that we have access to the sample distributions, and
not the population distributions. The sample pre-lottery distribution is $\hat{\mathbb{P}}_{pre, n_{pre}}$ ,
where $n_{pre}$ is the number of observations, and $\hat{\mathbb{P}}_{post, n_{post}}$ and $n_{post}$ are defined analogously for the post-lottery period.\footnote{The pre-lottery period is the year 2010, and the post-lottery period is limited to be the year 2011 unless otherwise specified.} We can approximate the population problem in \eqref{eq:2} with its sample analog,
\begin{mini}
{\gamma \in \Gamma}
{\sum_{i,j}\gamma_{i,j}\mathbf{C}_{i,j}}
{\label{eq:5}}
{OT(\hat{\mathbb{P}}_{pre,n_{pre}}, \hat{\mathbb{P}}_{post,n_{post}}) = }
\addConstraint{\sum_{i}\gamma_{i,j}}{= \hat{\mathbb{P}}_{pre,n_{pre}}(j),\quad}{\textnormal{for all } j} 
\addConstraint{\sum_{j}\gamma_{i,j}}{= \hat{\mathbb{P}}_{post,n_{post}}(i),\quad}{\textnormal{for all } i} 
\addConstraint{\gamma_{i,j}}{\geq 0,\quad \textnormal{for all } i, j,}
\end{mini}
\noindent where $\mathbf{C}=\mathbf{11}^T-diag(\mathbf{1})$. The observed sample distributions $\hat{\mathbb{P}}_{pre,n_{pre}}$ and $\hat{\mathbb{P}}_{post,n_{post}}$
are plotted in Figure \ref{fig:five}. However, we cannot directly apply the program in \eqref{eq:5} to $\hat{\mathbb{P}}_{pre,n_{pre}}$ and $\hat{\mathbb{P}}_{post,n_{post}}$ because sampling uncertainty  will inflate the transport estimate. Even in the null case of no trade in license plates, i.e., if both $\hat{\mathbb{P}}_{pre,n_{pre}}$ and $\hat{\mathbb{P}}_{post,n_{post}}$ were composed of draws from the same distribution $\mathbb{P}_{pre}$, sampling variation would lead to differences in the realized distributions.
\vspace{0.2in}
\noindent An effective and transparent approach to smooth out sampling variation while maintaining interpretability is to ignore small moves: we can attribute zero cost to transport mass up to some small distance $d\geq0$, where the distance is measured in RMB. We want to choose $d$ large enough to not confound sampling variation with trades but small enough to detect as many trades as possible.  An approximate but immediate tool for visualizing the smoothing implied by ignoring differences of up to $d$ RMB is the histogram of car prices evaluated at different values of $d$ as a bin width, see Figure \ref{fig:six}. 
\vspace{0.2in}\\
\noindent To select an appropriate tuning parameter $d$ so as to smooth out the sampling variation, we use as an approximation the variation in transport
cost between empirical distributions with observed sample sizes $n_{pre}$ and $n_{post}$, where both  empirical
distributions are drawn from the population distribution of car buyers, $\mathbb{P}_{pre}$. We select $d$ such that the transport cost in the placebo problem is
less than the precision level at which we report the transport cost.  We report the percentage cost up to its first decimal, and thus require a placebo cost of less than 0.05 percent.  That way, sampling variation can be safely ignored when interpreting point estimates.  We rely on the mean placebo transport cost, but inspection of the quantiles reveals a rule of thumb relying on the later would produce a similar estimate.

\newpage

\noindent To implement this smoothing criterion, we replace the cost matrix $\mathbf{C}$ with a
thresholded version
\begin{equation} 
\mathbf{C}_{i,j}(d)=\mathbbm{1}(|x_i-x_j|>d) \nonumber,
\end{equation}
where $x_i$ is the $i^{th}$ entry of $\mathcal{X}$. We then solve the sample optimal transport problem
%
\begin{mini}
{\gamma \in \Gamma}
{\sum_{i,j}\gamma_{i,j}\mathbf{C}_{i,j}(d)}
{\label{eq:6}}
{OT(\hat{\mathbb{P}}_{pre,n_{pre}}, \hat{\mathbb{P}}_{post,n_{post}}) = }
\addConstraint{\sum_{i}\gamma_{i,j}}{= \hat{\mathbb{P}}_{pre,n_{pre}}(j),\quad}{\textnormal{for all } j} 
\addConstraint{\sum_{j}\gamma_{i,j}}{= \hat{\mathbb{P}}_{post,n_{post}}(i),\quad}{\textnormal{for all } i} 
\addConstraint{\gamma_{i,j}}{\geq 0,\quad\textnormal{for all } i, j.}
\end{mini}
\vspace{0.1in}\\
\noindent The placebo transport cost
is 
the expected transport cost when both the pre-lottery and post-lottery empirical distributions are drawn from the same population of car buyers.  Because the population distributions are not observed, we use the empirical distribution as a plug-in, and look at the average transport cost when drawing an empirical distribution from the same plug-in distribution.
Specifically,
$$\hat{s}_{placebo}(d) = \mathbb{E}_{\hat{\mathbb{P}}_{pre}} \lbrack OT_d(\tilde{\mathbb{P}}_{n_{pre}}, \tilde{\mathbb{P}}_{n_{post}})\rbrack.$$

\noindent This simple and transparent approach is in line with our intuitive notion of sampling uncertainty.  We select a $d$ such that in expectation, or with prescribed probability, only a controlled quantity of detected trades will in fact be attributable to sampling variation.  This controlled quantity may be chosen to be negligible   in terms of the degree of precision of the reported output.  Trades corresponding to changes in price of less than the threshold $d$ RMB will be ignored, which will attenuate the lower bound.  That is again in line with our philosophy of trading off a tighter but less reliable bound for a more conservative approach delivering a credible lower bound on the volume of trade. 

\begin{figure}[t]
  \centering
  \includegraphics[width=\textwidth]{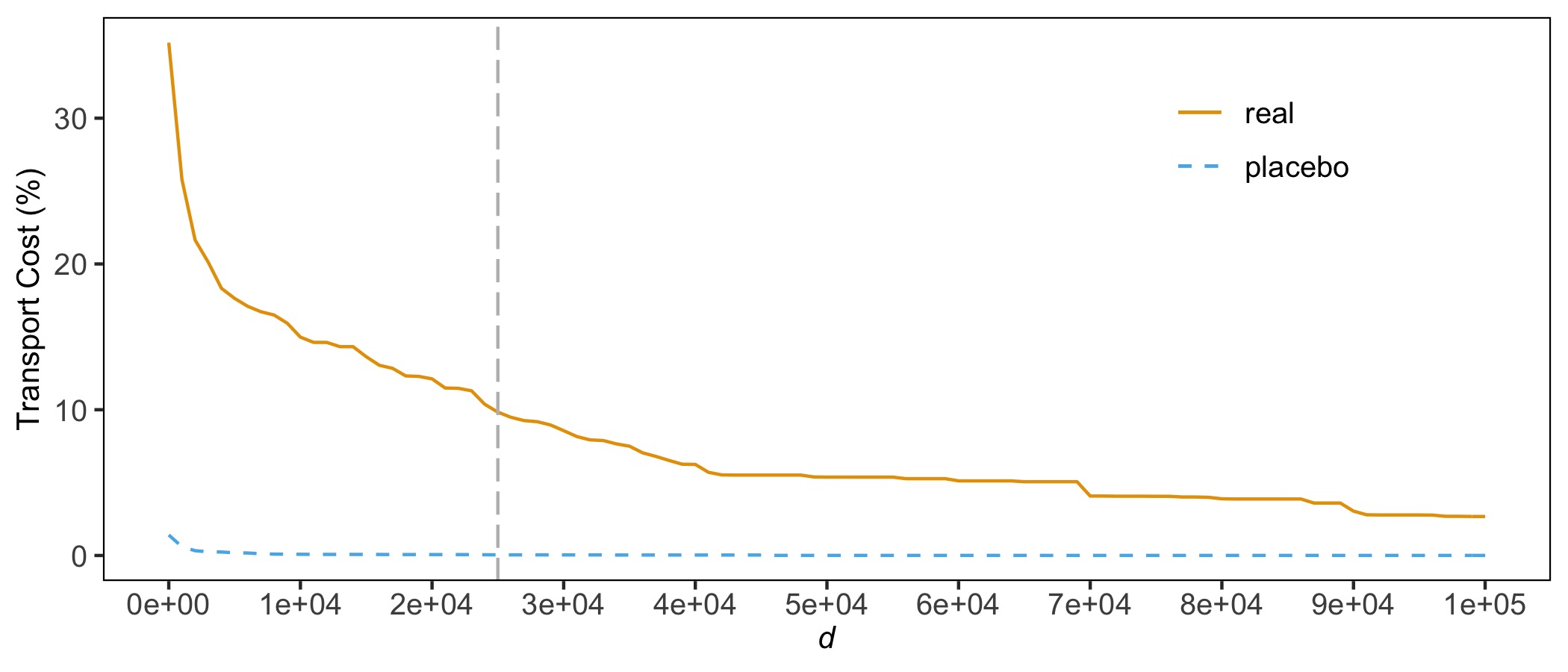}
 \caption{\figtitle{Real and placebo costs as a function of $d$.} \figcaption{The placebo cost in Beijing at each $d$ is the mean of 500 simulated values of $OT_d(\tilde{\mathbb{P}}_{pre, n_{pre}}, \tilde{\mathbb{P}}_{pre, n_{post}})$, while the real cost is $OT_d = (\hat{\mathbb{P}}_{pre}, \hat{\mathbb{P}}_{post}).$  The vertical line at $d=25\numcomma000$ is the smallest bandwidth for which the placebo cost is less than 0.05\%.}}
  \label{fig:seven}
\end{figure}

\vspace{0.2in}
\noindent  We next probe the sensitivity of the estimates over a range of values of $d$'s.
Figure \ref{fig:seven} plots $\hat{s}(d)=OT_d(\hat{\mathbb{P}}_{pre,n_{pre}}, \hat{\mathbb{P}}_{post,n_{post}})$ and $\hat{s}_{placebo}(d)$ against $d$. While $\hat{s}(d)$ estimates the lower bound for the market share, 
$\hat{s}_{placebo}(d)$ estimates the sampling uncertainty as a function of $d$.
\vspace{0.2in}\\
\noindent  According to our rule of thumb, the results in Table \ref{tab:costs} suggest using $d=25\numcomma000$. The placebo transport cost \numberblank{$\hat{s}_{placebo}(25\numcomma000)$} is
negligible at the one decimal accuracy level. We therefore
choose \numberblank{$\hat{s}(25\numcomma000)= 9.8\%$} as our preferred before-and-after estimate. 

\begin{table}[b]\centering
\small
\ra{0.7\usualra}
\begin{threeparttable}
 \caption{\tabtitle{Beijing 2010 placebo transport costs}}
 \begin{tabular}{@{}>{\raggedright}p{0.1\linewidth}>{\centering}p{0.08\linewidth}>{\centering}p{0.08\linewidth}>{\centering}p{0.08\linewidth}>{\centering}p{0.08\linewidth}>{\centering}p{0.08\linewidth}>{\centering}p{0.08\linewidth}>{\centering}p{0.08  \linewidth}>{\centering\arraybackslash}p{0.08\linewidth}}
 \toprule
    \tabhead{$d$:} & \tabhead{10\numcomma000} & \tabhead{15\numcomma000} & \tabhead{20\numcomma000} & \tabhead{25\numcomma000} & \tabhead{30\numcomma000} & \tabhead{50\numcomma000} & \tabhead{70\numcomma000} & \tabhead{90\numcomma000}\\
 \cmidrule{2-9}
    $\hat{s}(d):$  & 15.0\% & 13.6\% & 12.1\% & 9.8\% & 8.6\% & 5.4\% & 4.1\% & 3.0\% \\
    $\hat{s}_{placebo}(d):$ & 0.08\% & 0.07\% & 0.06\% & 0.04\% & 0.04\% & 0.007\% & 0.004\% & 0.004\% \\
 \bottomrule
 \end{tabular} \label{tab:costs}
\end{threeparttable}
\end{table}

\vspace{0.2in}
\noindent Another criterion for selecting a $d$, which was introduced in the discussion of Assumption \ref{as:three} and is detailed in Section \ref{sec:B} of the Appendix, is that it smooths out the displacement due to the changes in the proportion of first-time and returning buyers. As detailed in the Appendix, we find that $d=10\numcomma000$, the estimated effect is about one percent, and for $d=20\numcomma000$, the estimated effect is less than $0.1$ percent.

\subsection{Difference-in-transports estimator} \label{ssec:dit}
\begin{figure}[H]
  \centering
  \includegraphics[width=\textwidth]{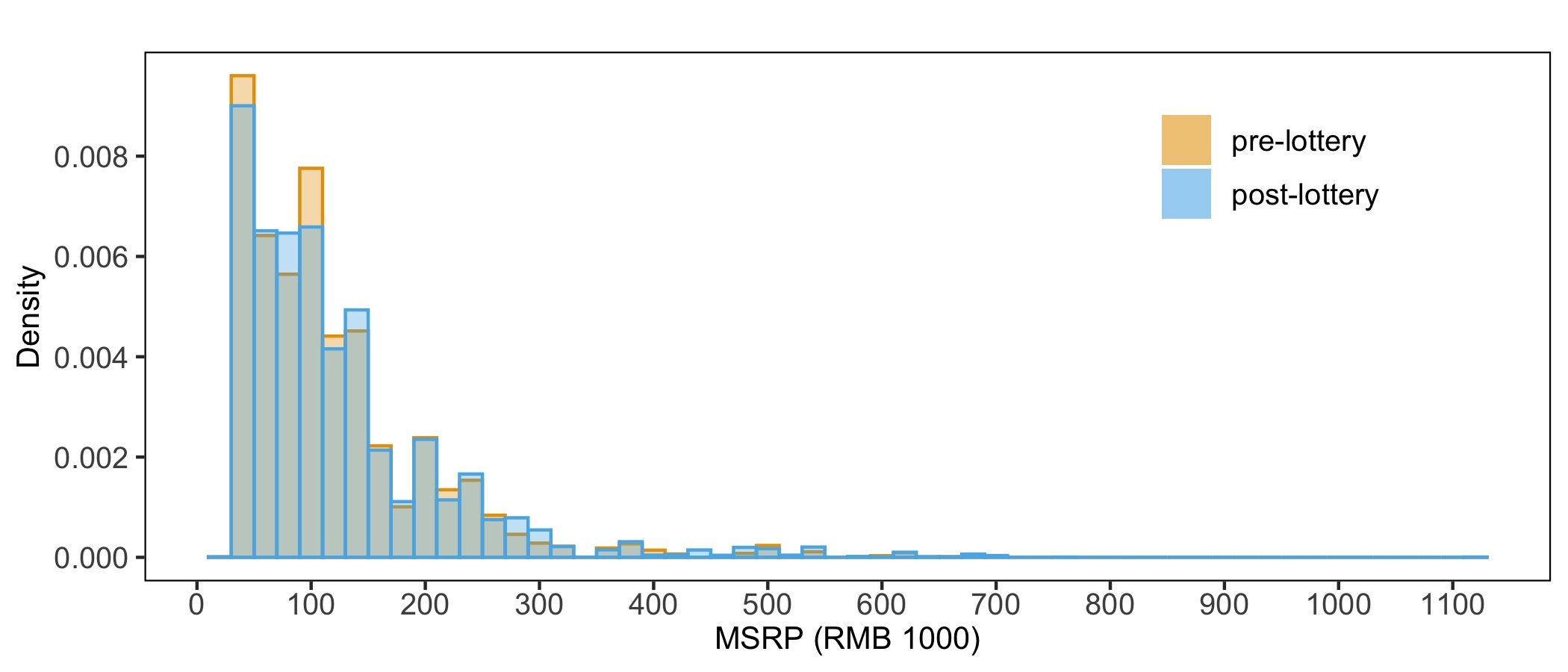}
 \caption{\figtitle{Histograms of Tianjin car sales in 2010 (pre-lottery) and 2011 (post-lottery).} \figcaption{}}
  \label{fig:eight}
\end{figure}

The results in the previous section relied on Assumption \ref{as:two}, which precludes time
trends in the sales distributions. This assumption may be challenged. For instance,
income growth or changes in car preferences could lead the sales distributions to
shift for reasons that are unrelated to black market trade. In Section \ref{sec:did}, we used the
sales data from Tianjin to control for common trends using difference-in-differences
regressions. We draw on the same idea for our second estimator. We want to use
the observable displacement of the Tianjin sales distribution to account for external
factors (changes in income, preferences, etc.) that would have shifted the sales
distribution in Beijing, had it not introduced rationing. We call this analogue to
the standard difference-in-differences estimator a \emph{difference-in-transports} estimator. 
Figure \ref{fig:eight}
shows that the sales distribution in Tianjin, where there was no rationing, may have
shifted slightly to the right. The displacement in Tianjin is clearly less pronounced 
than the corresponding displacement in Beijing, which is seen in Figure \ref{fig:four}.
\vspace{0.2in}
\noindent The target quantity is the displacement between the observed, post-lottery
Beijing sales distribution, with rationing and a black market, and the counterfactual,
post-lottery Beijing sales distribution, with rationing and no black
market,
\begin{equation}
\label{eq:7}
OT(\mathbb{P}_{Beijing, post}(1,0),\mathbb{P}_{Beijing, post}(1,1)).
\end{equation}
\vspace{0.2in}\\
\noindent Its oracle difference-in-transports proxy is given by the transport cost of the
Beijing sales distributions before and after the lottery with rationing and a
black market, net of the transport cost that would have been incurred if there was
rationing, but no black market,
\begin{equation}
\label{eq:8}
OT(\mathbb{P}_{Beijing, pre},\mathbb{P}_{Beijing, post}(1,1))
-
OT(\mathbb{P}_{Beijing, pre},\mathbb{P}_{Beijing, post}(1,0)) 
\end{equation}
\vspace{0.2in}\\
\noindent By the triangle inequality of Proposition 55 in \cite{peyre18}, the population quantity
$s_{dit}$ with $\mathbf{C} = \mathbf{C}(0)$ produces a lower bound on the quantity of interest,
\begin{equation}
\begin{aligned}
\label{eq:9}
&OT(\mathbb{P}_{Beijing, pre},\mathbb{P}_{Beijing, post}(1,1)) 
- OT(\mathbb{P}_{Beijing, pre},\mathbb{P}_{Beijing, post}(1,0)) \\ 
&\leq OT(\mathbb{P}_{Beijing, post}(1,1),\mathbb{P}_{Beijing, post}(1,0)). 
\end{aligned}
\end{equation}
\vspace{0.2in}\\
\noindent This quantity depends on the unobservable  $OT(\mathbb{P}_{Beijing, pre},\mathbb{P}_{Beijing, post}(1,0)) $. In contrast to the before-and-after case, we now relax Assumption \ref{as:two}, thereby allowing the sales distribution in Beijing to shift between the pre- and post-lottery period for reasons other than the lottery, i.e., $\mathbb{P}_{Beijing, post}(0,0)) \neq \mathbb{P}_{pre}$. We instead use shifts in the Tianjin sales distribution to
control for external factors. We replace Assumption \ref{as:two} with the following parallel trends-like assumption.
\begin{assumption}\label{as:four}
Equal displacement:
$$OT(\mathbb{P}_{Beijing, pre},\mathbb{P}_{Beijing, post}(0,0))= 
OT(\mathbb{P}_{Tianjin, pre}(0,0),\mathbb{P}_{Tianjin, post}(0,0)). $$
\end{assumption}
\noindent This assumption requires the observed displacement in Tianjin to equal the corresponding,
counterfactual displacement in Beijing were there to be no rationing and no
black market trade. Assumptions \ref{as:three} and \ref{as:four} together imply
\begin{equation}\label{eq:10}
OT(\mathbb{P}_{Beijing, pre},\mathbb{P}_{Beijing, post}(1,0))= 
OT(\mathbb{P}_{Tianjin, pre},\mathbb{P}_{Tianjin, post}(0,0)). 
\end{equation}
We can therefore write the estimand \eqref{eq:8} in terms of optimal transports between observable distributions
\begin{equation}\label{eq:11}
s_{dit}=
OT(\mathbb{P}_{Beijing, pre},\mathbb{P}_{Beijing, post}(1,1))-
OT(\mathbb{P}_{Tianjin, pre}(0,0),\mathbb{P}_{Tianjin, post}(0,0)), 
\end{equation}
which is the optimal transport analogue of the traditional difference-in-differences estimand. Although
 (\ref{eq:11}) is well estimated by
\begin{equation}
OT_d(\hat{\mathbb{P}}_{Beijing, pre},\hat{\mathbb{P}}_{Beijing, post}(1,1))-
OT_d(\hat{\mathbb{P}}_{Tianjin, pre}(0,0),\hat{\mathbb{P}}_{Tianjin, post}(0,0)) \nonumber
\end{equation}
for a small but non-zero value of $d$, we would like our difference-in-transports estimate
to give a lower bound on $OT_d(\hat{\mathbb{P}}_{Beijing, post}(1,1),\hat{\mathbb{P}}_{Beijing, post}(1,0))$, the sample version of the target quantity, and thus guarantee that the desirable population inequality
 \eqref{eq:9} always holds in sample. We show below that the  estimator 
\begin{equation}\label{eq:12}
\hat{s}_{dit}=
OT_{2d}(\hat{\mathbb{P}}_{Beijing, pre},\hat{\mathbb{P}}_{Beijing, post}(1,1))-
OT_d(\hat{\mathbb{P}}_{Tianjin, pre}(0,0),\hat{\mathbb{P}}_{Tianjin, post}(0,0)) 
\end{equation}
delivers this property.
Under Assumption \ref{as:four}, and for $d$ sufficiently large to smooth out sampling variation, $\hat{s}_{dit}(d)$ is an accurate estimate of
\begin{equation}\label{eq:13}
OT_{2d}(\mathbb{P}_{Beijing, pre},\mathbb{P}_{Beijing, post}(1,1))-
OT_d(\mathbb{P}_{Beijing, pre},\mathbb{P}_{Beijing, post}(1,0)), 
\end{equation}
and lower bounds the sample analogue of the population quantity of
interest.  This can be shown using Theorem \ref{theo1}.  The intuition is that a standard triangle inequality obtains for the smoothed optimal transport, but only if we oversmooth the less-than-or-equal side of the inequality.
\begin{theorem} \label{theo1}
Let $OT_d(\cdot , \cdot)$ be the discrete optimal transport problem described in \eqref{eq:6} with $\mathbf{C} = \mathbf{C}(d)$, and let $\hat{\mathbb{P}}_a$, $\hat{\mathbb{P}}_b$ and $\hat{\mathbb{P}}_c$ be three probability mass functions. Then, the following inequality holds
\begin{equation} \label{eq:14}
OT_{2d}(\hat{\mathbb{P}}_a,\hat{\mathbb{P}}_b)-OT_{d}(\hat{\mathbb{P}}_c,\hat{\mathbb{P}}_b) 
\leq OT_{d}(\hat{\mathbb{P}}_a,\hat{\mathbb{P}}_c).
\end{equation}
\end{theorem}
A proof is given in Section \ref{sec:proof1} of the Appendix.
\vspace{0.2in}
\noindent Theorem \ref{theo1} guarantees that \begin{equation}
OT_{2d}(\hat{\mathbb{P}}_{Beijing, pre},\hat{\mathbb{P}}_{Beijing, post}(1,1))-
OT_d(\hat{\mathbb{P}}_{Beijing, pre},\hat{\mathbb{P}}_{Beijing, post}(1,0)) 
\end{equation} is a lower bound for the sample analog of \eqref{eq:7} for all $d$, and a lower bound for \eqref{eq:7}, the volume of trade, for any
sufficiently large tuning parameter $d$. 

 The most informative lower bound is then given as
$$ 
\hat{s}^*_{dit}=\hat{s}_{dit}(d^*)= \max_{d\geq\underline{d}} 
OT_{2d}(\mathbb{P}_{Beijing, post},\mathbb{P}_{Beijing, pre})-
OT_d(\mathbb{P}_{Tianjin, post},\mathbb{P}_{Tianjin, pre}),
$$
where $\underline{d}$ is a data-dependent minimum value, and may be thought of as the smallest value of $d$ for which equal displacement holds in sample and such that noise is filtered out from the sample analog of the estimand \eqref{eq:7}.\footnote{
The required value for the smoothing parameter $d$ may be characterized by the following bounding argument,
\begin{equation*}
\begin{aligned}
&OT_{2d}(\hat{\mathbb{P}}_{Beijing, post}(1,1),\hat{\mathbb{P}}_{Beijing, pre}) 
- OT_{d}(\hat{\mathbb{P}}_{Tianjing, post}(1,0),\hat{\mathbb{P}}_{Tianjing, pre}) \\ 
&= OT_{2d}(\hat{\mathbb{P}}_{Beijing, post}(1,1),\hat{\mathbb{P}}_{Beijing, pre}) 
- OT_{d}(\hat{\mathbb{P}}_{Beijing, post}(1,0),\hat{\mathbb{P}}_{Beijing, pre}) \\ 
&\le OT_{d}(\hat{\mathbb{P}}_{Beijing, post}(1,1),\hat{\mathbb{P}}_{Beijing, post}(1,0)) \\ 
&\le OT(\mathbb{P}_{Beijing, post}(1,1),\mathbb{P}_{Beijing, post}(1,0)), 
\end{aligned}
\end{equation*}
where the first equality holds (approximately) for $d$ for which the equal displacement holds (approximately) in sample, the first inequality holds by the triangle inequality of Theorem \ref{theo1}, and the last inequality holds for $d$ large enough, as given by Table \ref{tab:costsdit}.}  

\begin{table}[t]\centering
\small
\ra{0.7\usualra}
\begin{threeparttable}
 \caption{\tabtitle{Beijing 2011 placebo transport costs}}
 \begin{tabular}{@{}>{\raggedright}p{0.1\linewidth}>{\centering}p{0.08\linewidth}>{\centering}p{0.08\linewidth}>{\centering}p{0.08\linewidth}>{\centering}p{0.08\linewidth}>{\centering}p{0.08\linewidth}>{\centering}p{0.08\linewidth}>{\centering}p{0.08  \linewidth}>{\centering\arraybackslash}p{0.08\linewidth}}
 \toprule
    \tabhead{$d$:} & \tabhead{4\numcomma000} & \tabhead{7\numcomma000} & \tabhead{10\numcomma000} & \tabhead{15\numcomma000} & \tabhead{20\numcomma000} & \tabhead{25\numcomma000} & \tabhead{30\numcomma000} & \tabhead{35\numcomma000}\\
 \cmidrule{2-9}
    $\hat{s}^{*}_{dit}(d):$  & 9.89\% & 11.87\% & 10.15\% & 6.83\% & 4.99\% & 4.66\% & 4.43\% & 3.40\% \\
    $\hat{s}_{placebo}(d):$ & 0.16\% & 0.05\% & 0.03\% & 0.03\% & 0.03\% & 0.03\% & 0.03\% & 0.02\% \\
 \bottomrule
 \end{tabular} \label{tab:costsdit}
\end{threeparttable}
\end{table}




Remark that an additional virtue of the $2d$ over-correction in the difference-in-transports estimator (motivated above by the in-sample triangle inequality of Theorem \ref{theo1}) is that the subtracted transport cost itself having truncation only at level $d$, the difference likewise over-corrects for the sampling variation, thus producing a lower bound estimate that requires milder sampling variation smoothing.  

\begin{figure}[H]
  \centering
  \includegraphics[width=\textwidth]{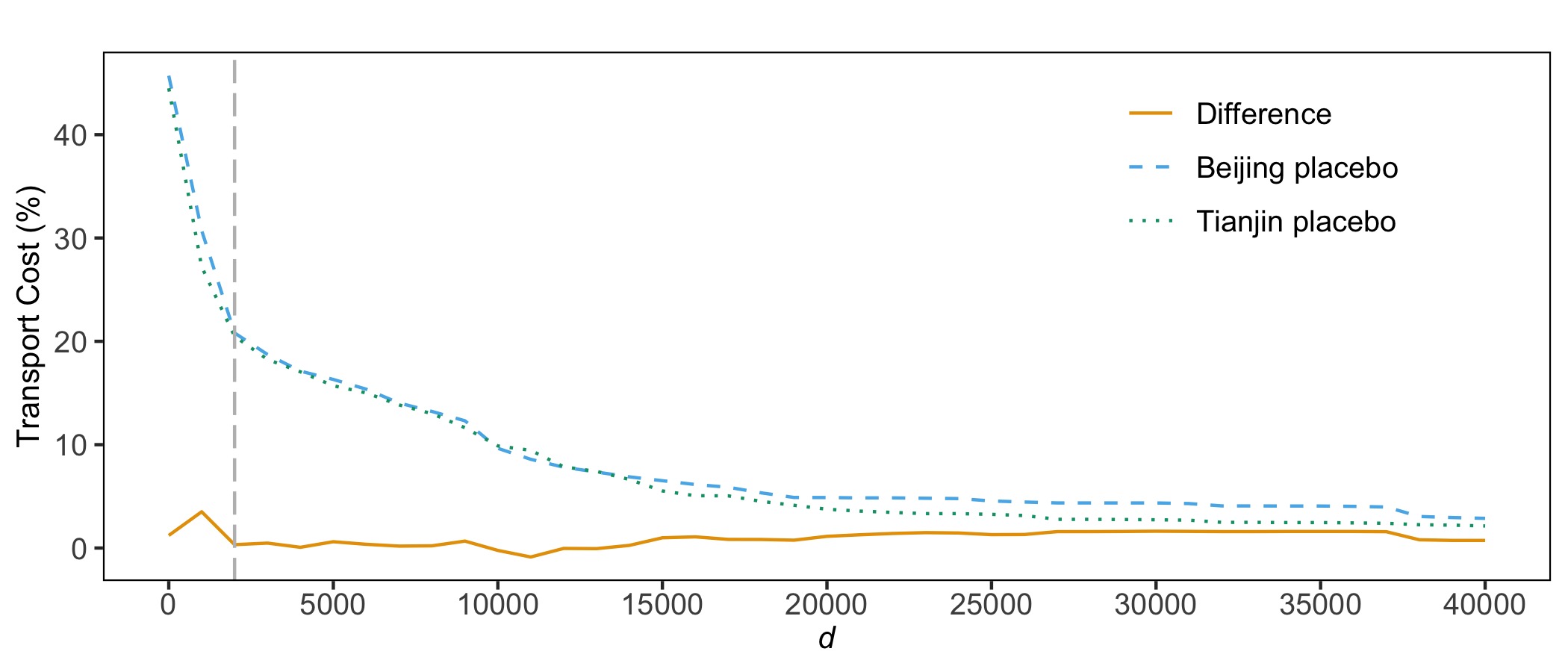}
 \caption{\figtitle{Post-Trends.} \figcaption{The Beijing placebo is given by the optimal transport $OT_{d}\left(\mathbb{P}_{Beijing, 2014}, \mathbb{P}_{Beijing, 2015}\right)$, and the Tianjin placebo is given by the optimal transport $OT_{d}\left(\mathbb{P}_{Tianjin, 2014}, \mathbb{P}_{Tianjin, 2015}\right)$. The difference is $OT_{d}\left(\mathbb{P}_{Beijing, 2014}, \mathbb{P}_{Beijing, 2015}\right)$ - $OT_{d}\left(\mathbb{P}_{Tianjin, 2014}, \mathbb{P}_{Tianjin, 2015}\right)$.}}
  \label{fig:ten}
\end{figure}

We may likewise assess the plausibility of the equal displacement assumption by inspecting trends when both Beijing and Tianjin are in a comparable regime. We inspect displacements in Beijing and Tianjin for 2014 to 2015. In those years, Tianjin had a lottery and an auction while Beijing had a lottery and a black market, with both of them seeing similar average price jumps after the introduction of their own rationing scheme, suggesting they might be comparable. This is confirmed empirically in Figure \ref{fig:ten}, where we see almost equal displacement for $d \ge 2,000$.
While it is more natural to compare trends or displacements when both cities are without rationing since this is the counterfactual we want to account for when differencing transports, we do not observe the Beijing price distribution before 2010, making such a comparison unavailable to us. Observe, from Figure \ref{fig:ten}, that non-trivial parallel trends are captured and differenced out for $d$ up to at least $15,000$.

\begin{figure}[H]
  \centering
  \includegraphics[width=\textwidth]{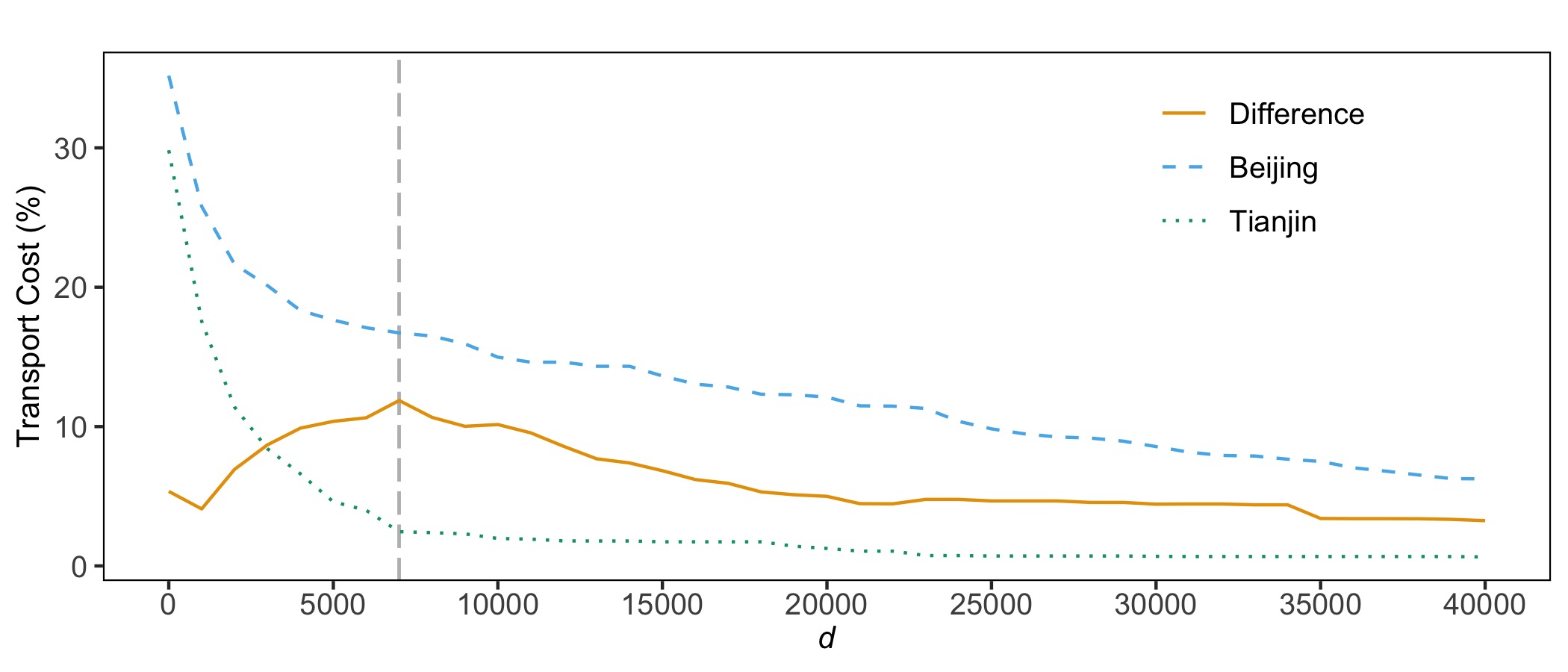}
 \caption{\figtitle{Choosing the most informative admissible $d$.} \figcaption{The Beijing cost is given by the optimal transport $OT_{d}\left(\mathbb{P}_{Beijing, 2010}, \mathbb{P}_{Beijing, 2011}\right)$, and the Tianjin cost is given by the optimal transport $OT_{d}\left(\mathbb{P}_{Tianjin, 2010}, \mathbb{P}_{Tianjin, 2011}\right)$. The difference uses twice the bandwidth to compute Beijing's cost:
$OT_{2d}\left(\mathbb{P}_{Beijing, 2010}, \mathbb{P}_{Beijing, 2011}\right) - OT_{d}\left(\mathbb{P}_{Tianjin, 2010}, \mathbb{P}_{Tianjin, 2011}\right)$; the maximum difference occurs when $d = \dstarcons$, when the transport cost is roughly \ditcons.}}
  \label{fig:eleven}
\end{figure}

\vspace{0.2in}
\noindent The above evidence suggests that noise is filtered out, up to our sensitivity criteria, for $d\ge7,000$, and that equal displacement holds in sample for $d\ge2,000$; we thus pick $\underline{d}=\max \{  2,000, 7,000\}$.
\noindent For the final step, we choose the most informative admissible $d\geq \underline{d}$. The solid yellow
line in Figure \ref{fig:eleven} plots $\hat{s}_{dit}(d)$, the difference between the dotted green line (Tianjin) and
the dashed blue (Beijing) line, for a range of $d$ values.
\vspace{0.2in}
\noindent  The difference-in-transports estimate is seen to be most informative at \numberblank{$d^* = \dstarcons$}, which gives \numberblank{$\hat{s}_{dit}(d^*) = \ditcons$}.  The cost attributable to the change in the composition of the distribution, detailed and calculated in Section \ref{sec:B} of the Appendix, is less than $0.9\%$, and we give $11\%$ as a robust estimate of the volume of the black market.

\section{Inferring transaction costs and prices}\label{sec:inferCosts}
\begin{figure}[H]
  \centering
  \includegraphics[width=0.5\textwidth]{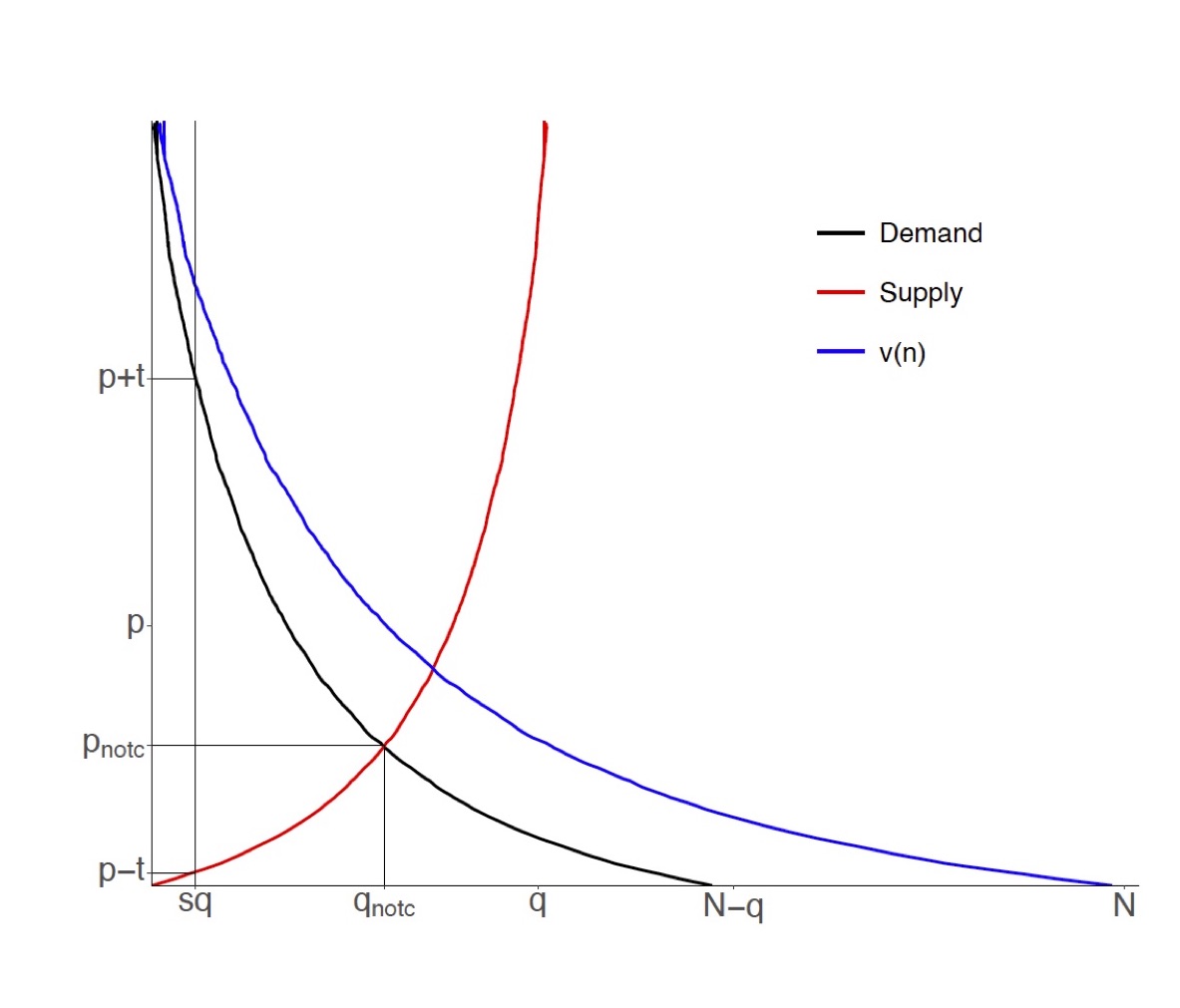}
 \caption{\figtitle{Demand and supply derived from $v(n)$.} \figcaption{The figure is drawn to scale with quantity in 10\numcomma000 and price in RMB 1\numcomma000.}}
  \label{fig:twelve}
\end{figure}

We infer the unobserved transaction costs and prices from a market equilibrium
model that combines our estimated volume of trade with empirical results from the
literature. We derive our model from a common version of the Coase Theorem:
if the initial allocation leaves gains from trade, as is expected when license plates
are allocated by lottery, households will transact until the gains from trade are
exhausted. A market price will form that equates the number of lottery winners
willing to sell their license plates to the number of prospective buyers. 
Transaction costs, which may be both pecuniary and non-pecuniary,
are frictions that reduce the volume of trade. We bound the transaction costs by
treating these as a tax on trade necessary to rationalize the estimated volume of
black market trade.
\vspace{0.2in}\\
\noindent One likely important transaction cost component is potential legal liabilities.
For instance, a new car must be registered in the name of the legal owner of the
license plate, and therefore the formal owner, not the driver,  will be liable for traffic accidents. The fact that trade in license plates
is illegal may also give rise to non-pecuniary transaction costs, such as feelings of
culpability. Search costs are another component: buyers and sellers need to meet
in the market. The reported existence of online market places for license plates,
however, suggests that search costs are small.\footnote{\cite{bhave18} notes that online market places for ticket resale decreases transaction costs and increases the volume of resold tickets in the secondary market.}
\vspace{0.2in}\\
\noindent Potential enforcement, and perhaps moral hazard of the kind associated with
leaving the formal car ownership in someone else's name, are likely the key transaction
cost components. While there may be further transaction cost components, we
do not attempt to distinguish between these. The transaction costs in our model
are thus a simple quantitative measure of the frictions in the black market.
\vspace{0.2in}\\
\noindent We derive a demand and supply curve using the willingness-to-pay for license
plates estimated in \cite{li18}. The estimates are derived by combining a BLP random coefficients discrete choice model, using car registration like we do, with
micro moments from household survey data. These estimates are represented by a
function $v(n)$ with range RMB $\lbrack0,\,280 000\rbrack$.\footnote{
Shanjun Li has generously shared his estimates with us. While \cite{li18}'s
Figure 3 shows positive willingness-to-pay for quantities far beyond the unrationed, pre-lottery market equilibrium,
we require that the marginal willingness-to-pay for a license plate at the unrationed market equilibrium
is zero, i.e., $v(N) = 0$. Our results are, however, not very sensitive to this modification.
}
We take \cite{li18}'s
estimated willingness-to-pay for a license plate $v(n)$ to be known without sampling
variation. It is useful in the following to write $v(n)$ in terms of a cumulative
distribution function $F(v)$, which returns the share of households in the market
with a willingness-to-pay less than $v$, i.e. $F(v) = \frac{1}{N}\sum_{n=1}^N\mathbbm{1}(v(n)\leq v)$. Since $v(n)$
is strictly decreasing, we can recover $F$ from $v$ as
$$\frac{v^{-1}(v)}{N}=1-F(v).$$
We assume that only prospective car buyers enter the lottery.
\begin{assumption} \label{as:five}
The lottery draws q winners with equal probability from the population of N prospective car buyers, whose willingness-to-pay for a license plate is distributed according to the cumulative distribution function F.
\end{assumption}
It follows immediately from Assumption \ref{as:five} that both the $q$ sellers' (winners) and
the $N-q$ buyers' willingness-to-pay are distributed according to the cdf $F$. We set
$q = 260\numcomma000$, the quota in 2011, and the market size to $N = 700\numcomma000$, the total
sales of new cars in 2010 before the lottery. Assumption \ref{as:five} rules out speculators,
which we define to be those who enter the lottery with no intention of buying a car
if they win a license plate. An influx of speculators is consistent with the sharp
decline in the lottery odds, which went from \numberblank{10\%} in the first auction in January
2011, to a monthly average of \numberblank{4\%} for 2011, and dropped further to a monthly
average of \numberblank{2\%} in 2012. We relax Assumption \ref{as:five} and allow for speculators in Section \ref{sec:D} of the Appendix.  
\vspace{0.2in}\\
\noindent We next make an assumption about how prices and transaction costs form.
\begin{assumption} \label{as:six}
Each trade generates transaction costs of $2t$, which are equally borne
by the buyer and the seller. The transaction price p equates demand to supply given
the transaction costs.
\end{assumption}
\noindent A buyer is willing to buy a license plate if $v > p + t$, and a seller is willing to sell a
license plate if $v\leq p-t$. There are $N-q$ rationed prospective buyers that demand
license plates, while there are $q$ lottery winners that supply licenses. This gives the
demand and supply functions
\begin{equation} 
\begin{aligned} \label{eq:16}
&D(p,t)=(N-q)(1-F(p+t)),\\
&S(p,t)=qF(p-t).
\end{aligned}
\end{equation}
Figure \ref{fig:twelve} plots the demand and supply curves derived from \cite{li18}'s estimates.
Suppose first that there are no transaction costs. Then demand equals supply
at price $p_{notc}$ and the quantity is $q_{notc}$. The volume of trade that occurs without
transaction costs $q_{notc}$ is less than the quota $q$ since lottery winners with valuation
in excess of the market clearing transaction price prefer not to trade. Adding
Assumptions \ref{as:five} and \ref{as:six} brings the upper bound of the volume of trade down from
100\% of the fraction to $s_{upper} = \frac{q_{notc}}{q}=\frac{N-q}{N}= 62\%$ of the quota. 
The estimate
of $s_{upper}$ implies that though our lower bound estimate is $\hat{s} = 11\%$ of the quota,
this market can support no more than $\frac{11\%}{62\% }= 18\%$ of trades in equilibrium.
\vspace{0.2in}
\noindent The transaction costs $t$ in Figure \ref{fig:twelve} are seen to drive a wedge between the
supply and demand that lowers the volume of illegal trade to $sq$. The total
transaction costs, summed over buyers and sellers, that rationalize the trade are hence $2tsq$.
\begin{theorem}\label{theo2}
Suppose that $F$ is known, continuous, and strictly increasing. Suppose
that the market clears at transaction price p and transaction cost t for a known volume of $sq$ black market trades. Then the transaction prices and costs are identified.
\end{theorem}
\noindent We can alternatively dispense with the homogeneous transaction cost assumption
and allow buyers and sellers to bear different transaction costs, i.e., $t_{buyer}\neq t_{seller}$.
Though we cannot jointly identify the buyer- and seller-specific transaction costs
along with the transaction price, a pair $\tilde{p}_{buyer} = p + t_{buyer}$ and $\tilde{p}_{seller} = p- t_{seller}$
is identified. This interpretation does not affect the identification of the total
transaction costs, which are $sq(\tilde{p}_{buyer}-\tilde{p}_{seller})$. Since these interpretations
are observationally equivalent, we cannot infer which side of the market bears
the majority of the transaction costs. Hence, we maintain
the homogeneous transaction cost interpretation. The willingness-to-pay $F$, which
we derived from \cite{li18}'s estimates, satisfies the continuity and monotonicity
restrictions.
\vspace{0.2in}\\
\noindent The implied transaction costs and prices at $\hat{s}$ are given in Table \ref{tab:five}, along
with 95\% confidence intervals.\footnote{
We use subsampling to produce draws of $\hat{s}$ for inference on  the share of trade in Table \ref{tab:five}.
}
We show in Section \ref{sec:C} that $\hat{t}$ is an upper bound
for the transaction cost at $\hat{s}$, and that the implied $\tilde{p}_{buyer}$ and $\tilde{p}_{seller}$ at $\hat{s}$ are upper
and lower bound estimates, respectively. In the model, trades take place between
a selection of buyers with particularly high valuations and sellers with particularly
low valuations. This pattern is consistent with the literature on the resale market
of tickets where trades are observed (\citealp{bhave18,leslie13}) and where search costs are relatively small.

\begin{table}[t]\centering
\small
\ra{\usualra}
\caption{\tabtitle{Implied Transaction Prices, Costs, and Net Gains.}}
\begin{threeparttable}
 \begin{tabular}{@{}>{\centering}m{0.3\linewidth}>{\centering}m{0.15\linewidth}>{\centering\arraybackslash}m{0.15\linewidth}}
 \toprule
    \tabhead{Estimates} & \tabhead{at $\hat{s}$ = 11\%} & \tabhead{95\% CI} \\
 \midrule
    $\hat{p}$ & 105 & $\lbrack91,121\rbrack$ \\
    $\hat{t}$ & 100 & $\lbrack83,118\rbrack$ \\
    total transaction costs & 5.7 & $\lbrack3.8,7.3\rbrack$ \\
    net gains from trade & 1.3 & $\lbrack0.5,2.5\rbrack$ \\
 \bottomrule
 \end{tabular} 
 \begin{tablenotes}
    \item Estimates of transaction prices and costs are in RMB 1\numcomma000. Gains from trade and transaction costs are in RMB billon.
 \end{tablenotes}
\end{threeparttable}
\label{tab:five}
\end{table}

\subsection{Transaction costs and net gains from trade} \label{ssec: trsc_costs}
We compare estimates of the total transaction costs and net gains from black market
trade to two benchmarks: one where non-transferability is strictly enforced and one
where there are no transaction costs. The latter can be interpreted as the case with
no enforcement. We consider transaction prices to be transfers between sellers and
buyers that do not affect gross gains from trade (the area between the demand
and the supply curve up to $\widehat{sq}$). Table \ref{tab:five} shows that the gross gains from trade is
RMB 7.0 billion at $\hat{s} = 11\%$, but transaction costs sum up to RMB 5.7 billion. 
The net gains from trade are the gross gains from trade minus the transaction costs.
The lower bound estimate for the net gains from trade is RMB 1.3 billion, while
the upper bound, for the case with no transaction costs at a black market share
$q_{notc} = 62\%$ and price $p_{notc} = \textnormal{RMB } 59\numcomma000$, is RMB 18.8 billion.
\subsection{Further bounds on transaction costs} \label{ssec: further_bounds }
\noindent Under the assumptions of the equilibrium model, the data are consistent with a
volume of trade that ranges from 11\% to 62 \% of the quota. This range implies
transaction costs from zero to RMB 5.7 billion and transaction prices from RMB
59\numcomma000 to RMB 105\numcomma000. We can tighten the bounds with information on the
transaction prices if we are willing to use the news reports from Section \ref{sec:lottery}
as information about transaction prices. Table \ref{tab:six} shows how the implied transaction costs and the volume
of trade vary conditional on a range of known transaction prices. 
 We can therefore
tighten the bounds further if we are willing to use the news reports from Section \ref{sec:lottery}
as informative about transaction prices.
\begin{table}[t]\centering
\small
\ra{\usualra}
\caption{\tabtitle{Transaction Prices and Costs}}
\begin{threeparttable}
 \begin{tabular}{@{}>{\centering}p{0.05\linewidth}>{\centering}p{0.05\linewidth}>{\centering}p{0.05\linewidth}>{\centering}p{0.25\linewidth}>{\centering\arraybackslash}p{0.25\linewidth}}
 \toprule
    \tabhead{$p$} & \tabhead{$t$} & \tabhead{$s$} & \tabhead{Net Gains from Trade} & \tabhead{Share Transaction Costs} \\
 \midrule
    59 & 0 & 62\% & 18.8 & 0\% \\
    57 & 21 & 49\% & 12.8 & 30\% \\
    64 & 42 & 37\% & 8.2 & 49\% \\
    73 & 59 & 27\% & 5.3 & 61\% \\
    105 & 100 & 11\% & 1.3 & 82\% \\
 \bottomrule
 \end{tabular}
 \begin{tablenotes}
    \item The calculations are conditional on $s$ ranging from the estimated lower bound $\hat{s}$ to the upper
bound $s_{upper}$ that the market can support, and assume no speculators. Prices and costs are in
RMB 1\numcomma000. Gains from trade are in RMB billion.
 \end{tablenotes}
\end{threeparttable}
\label{tab:six}
\end{table}
\vspace{0.2in}\\
\noindent Though we found a wide range of reported transaction prices, most are above RMB
70\numcomma000. \cite{li18} estimates that the (counterfactual) market clearing auction price
is about RMB 75\numcomma000 (see its Figure 3). It appears that rentals is the most common
sales format. The most frequently reported rental prices range is from RMB 500 to
1\numcomma000 per month. A twenty-year NPV of RMB 6\numcomma000 in annual rental income is RMB
73\numcomma000, which seems conservative, yet plausible.
\begin{assumption} \label{as:seven}
RMB 73\numcomma000 is a lower bound for the transaction price.
\end{assumption}
\vspace{0.2in}
\noindent Table \ref{tab:six} shows that Assumption \ref{as:seven} tightens the upper bound for $s$ that is consistent
with the data from 62\% to 27\%, and shifts the lower bound for the share of
transaction costs from zero to 61\%. We distinguish the realized gross gains from
trade from the potential gains from trade, which are those that could be realized
in the black market if there were no transaction costs. In our framework, a black market without
transaction costs would theoretically be as efficient as an auction.
Since the potential gains from trade are RMB 18.8 billion, the share of
the realized net gains from trade lies in the range from 7\% to 28\%.
\vspace{0.2in}\\
\noindent Table \ref{tab:six} allows readers who prefer a more conservative lower bound on the
transaction prices to trade weaker assumptions against wider bounds for the
quantities of interest.
\vspace{0.2in}\\
\noindent Table \ref{tab:seven} summarizes the assumptions and the corresponding inferred bounds
for the market performance measures $p$, $s$, $t$, and the share of gains from trade
lost to transaction costs. We started with inferring a lower bound for the volume
of trade using comprehensive car sales data on millions of car sales under weak
assumptions that are common in the program evaluation literature. Under these
assumptions, the car sales data, which are indirectly related to our quantities of
interest, gave an informative lower bound for $s$ but did not deliver useful bounds
on transaction prices and transaction costs. We then imposed increasingly strong
assumptions on black market transactions (Assumptions \ref{as:five}, \ref{as:six}, and \ref{as:seven}). These assumptions are seen
to be powerful in tightening the bounds, in particular Assumption \ref{as:seven}. We note that
this assumption, which serves like a highly informative prior on a variable that is
directly related to our quantities of interest, is supported by only a handful of news 
reports.

\begin{table}[t]\centering
\small
\ra{1.2\usualra}
\begin{threeparttable}
 \caption{\tabtitle{Transaction prices and transaction costs in RMB 1\numcomma000.}}
 \begin{tabular}{@{}>{\centering}m{0.15\linewidth}>{\centering}m{0.15\linewidth}>{\centering}m{0.1\linewidth}>{\centering}m{0.1\linewidth}>{\centering}m{0.1\linewidth}>{\centering\arraybackslash}m{0.20\linewidth}}
 \toprule
    \tabhead{Data} & \tabhead{Assumptions} & \tabhead{$s$} & \tabhead{$p$} & \tabhead{$t$} & \tabhead{Share transaction costs} \\
 \midrule
    car sales & \ref{as:one}, \ref{as:two}, \ref{as:three} & \lbrack14,100\rbrack\% & $\mathbb{R}_+$ & $\mathbb{R}_+$ & \lbrack0,100\rbrack\% \\
    car sales & \ref{as:one}, \ref{as:three}, \ref{as:four} & \lbrack11,100\rbrack\% & $\mathbb{R}_+$ & $\mathbb{R}_+$ & \lbrack0,100\rbrack\% \\
    car sales, F &  \ref{as:one}, \ref{as:three}, \ref{as:four}, \ref{as:five}, \ref{as:six} & \lbrack11,62\rbrack\% & \lbrack59,105\rbrack\ & \lbrack0,100\rbrack\ & \lbrack0,82\rbrack\% \\
    car sales, F, news reports & \ref{as:one}, \ref{as:three}, \ref{as:four}, \ref{as:five}, \ref{as:six}, \ref{as:seven} & \lbrack11,27\rbrack\% & \lbrack73,105\rbrack\ & \lbrack59,100\rbrack\ & \lbrack61,82\rbrack\% \\
 \bottomrule
 \end{tabular}\label{tab:seven}
    \vspace{-3mm}
    \begin{multicols}{2}
        \begin{enumerate}[leftmargin = *, label = ]
            \item \ref{as:one}. No anticipation
            \item \ref{as:two}. No time trends
            \item \ref{as:three}. No general equilibrium
            \item \ref{as:four}. Equal displacement
            \item \ref{as:five}. Market participants
            \item \ref{as:six}. Market equilibrium
            \item \ref{as:seven}. Lowest plausible transaction price
            \item[\vspace{\fill}]
        \end{enumerate}
    \end{multicols}
\end{threeparttable}
\end{table}

\section{Discussion} \label{sec:discussion}
It is puzzling that the Beijing government seems to allow a substantial volume of
black market trade. The government can close the black market if they wish by
allowing for trade. One reason may be that the scale of trade makes the costs of
strict enforcement prohibitive. Another reason may be that some trade takes place
between family members, e.g., a daughter wins a license and lets her parents register
a car in her name. Such trades may be less politically expedient to enforce.
\vspace{0.2in}\\
\noindent Lax enforcement may also strike a balance between allocative efficiency and
equity concerns in a society where only 8\% of the population approves of auctions.
Black market trade may be viewed as a second-best solution that ameliorates
the worst misallocations resulting from the lottery allocations. However, more transparent
mechanisms are available. Other Chinese cities that later introduced
rationing (Guangzhou, Tianjin, Hangzhou, Shenzhen, and Shijiazhuang) chose a
hybrid lottery/auction mechanism which offer a different balance between efficiency and equity. 
\cite{huang19} estimates that the hybrid mechanism in
Guangzhou, which allocates about 50\% of the license plates by lottery and the
other half by auction, preserves 83\% efficiency. Our results give that the Beijing
black market preserves at most \numberblank{30\%} efficiency, down to as little as \numberblank{10\%} efficiency.
\vspace{0.2in}\\
\noindent However, a relatively large black market need not mean that enforcement is lenient.  Indeed, some of the evidence suggests that enforcement is strong. The high willingness-to-pay for license plates in the market creates strong incentives for trade. If interest is
in an equitable allocation, then the size of the transaction costs (up to \numberblank{82\%} of the
gains from trade) suggests that enforcement is in fact quite effective in precluding
illegal trade of high private value.\footnote{
We use enforcement here to mean not only active enforcement 
but also any deterrent effect of codifying the license plates as non-transferable.
}
If interest is instead in an efficient black
market, the size of the transaction costs suggests relaxing enforcement.

\vspace{0.2in}
\noindent Our hope is that the methodological developments detailed herein will find applications and extensions beyond the specific subject matter of this article.   In that respect, because it accounts for trends and self-corrects for sampling variation –i.e., is less sensitive to tuning– we suggest the difference-in-transports estimator as preferable to the before-and-after estimator, when a match is available.

\vspace{0.2in}
\noindent Certainly, the case at hand can be considered at a higher level of generality, where the key observables are sample distributions before and after the time of treatment. While we aimed to count the number of subjects with nonzero treatment effects, a different choice of ground cost could be tailored to produce different estimator.

\vspace{0.2in}
\noindent More sophisticated reduced form applications of optimal transport methods are being considered. For instance, optimal transport presents itself as a natural means with which to compare distributions in bunching type estimation problems (\citealp{pukelis2020estimation}), and to assess covariate balance in program evaluation problems (see \citealp{gracie2020optimal}, \citealp{rosenbaum1989optimal}, and \citealp{rosenbaum2012optimal}). More general methodology could consider accommodating covariates, instruments, and specialized inference.  Relatedly, \citet{gunsilius2020distributional} uses optimal transport as the appropriate notion of distance to extend the synthetic control method to the case in which experimental units correspond to entire distributions, and \citet{arkhangelsky2019dealing} leverages optimal transport theory for identification in specific clustered panel data setups.

\section{Summary} \label{sec:summary}
Black markets may serve as an alternative allocation mechanism to auctions in places
where auctions are either practically or politically infeasible. We adopted a partial
identification approach in the spirit of \cite{manski03} to analyze the performance of
the black market for Beijing license plates. We estimated different bounds for the volume of
trade, the gains from trade, and the transaction costs corresponding to
different and increasingly strong assumptions. We found that the black market plausibly
reallocates between \numberblank{11\%} and \numberblank{37\%} of the rationed license plates and conservatively
estimated the net gains from trade in the black market to be between \numberblank{RMB 1.3 billion} and \numberblank{RMB 5.3 billion}. Yet, the black market realizes between \numberblank{7\%} and
\numberblank{28\%} of the potential gains from trade,  \numberblank{61\%} to \numberblank{82\%} of which are lost to transaction costs. The size of the transaction costs suggests
that enforcement is strong and that the black market realizes modest efficiency
gains compared to hybrid lottery/auction allocation mechanisms that are used in
other large Chinese cities.

\newpage
\hypertarget{scmp2014link}{}\nocite{scmp2014}
\hypertarget{scmp2012link}{}\nocite{scmp2012}
\bibliography{mybib} 

\clearpage

\end{document}


\maketitle

\appendix 
\addcontentsline{toc}{section}{Appendices} 
\vspace{-3em}

\section{Proofs} \label{sec:proof1}

\subsection{Theorem \ref{theo1}}

The proof follows immediately from the argument of Proposition 55 in \cite{peyre18}, and the analogous triangle inequality for the pseudo-distance $\textbf{C}(d)$,
\begin{equation}
\mathbbm{1}(|x_i-x_j|>2d)-\mathbbm{1}(|x_j-x_k|>d) \leq \mathbbm{1}(|x_i-x_k|>d),
\nonumber
\end{equation}
where $x_l$ is the $l^{th}$ entry of $\mathcal{X}$. 
\hfill $\blacksquare$

\subsection{Theorem \ref{theo2}}

For any pair of transaction costs and prices, the valuation of the marginal seller is
$v_{seller} = p-t$ and the valuation of the marginal buyer is $v_{buyer} = p + t$. Together,
we get
\begin{equation}
\begin{aligned}\label{eq:17}
&p=\frac{1}{2}(v_{seller}+v_{buyer})\\
&t=\frac{1}{2}(v_{buyer}-v_{seller})
\end{aligned}
\end{equation}
Equating demand to supply using \eqref{eq:16} at the estimated lower bound of trades
$sq$, the marginal valuations are uniquely recovered by inverting $F^{-1}(s)=v_{seller}$
and $F^{-1}(1-\frac{sq}{N-q})=v_{buyer}$. It follows immediately that $p$ and $t$ are uniquely
determined given $sq$. 
\hfill $\blacksquare$

\section{Additional Figures} \label{sec:addfig}

\begin{figure}[H]
  \centering
  \includegraphics[width=\textwidth]{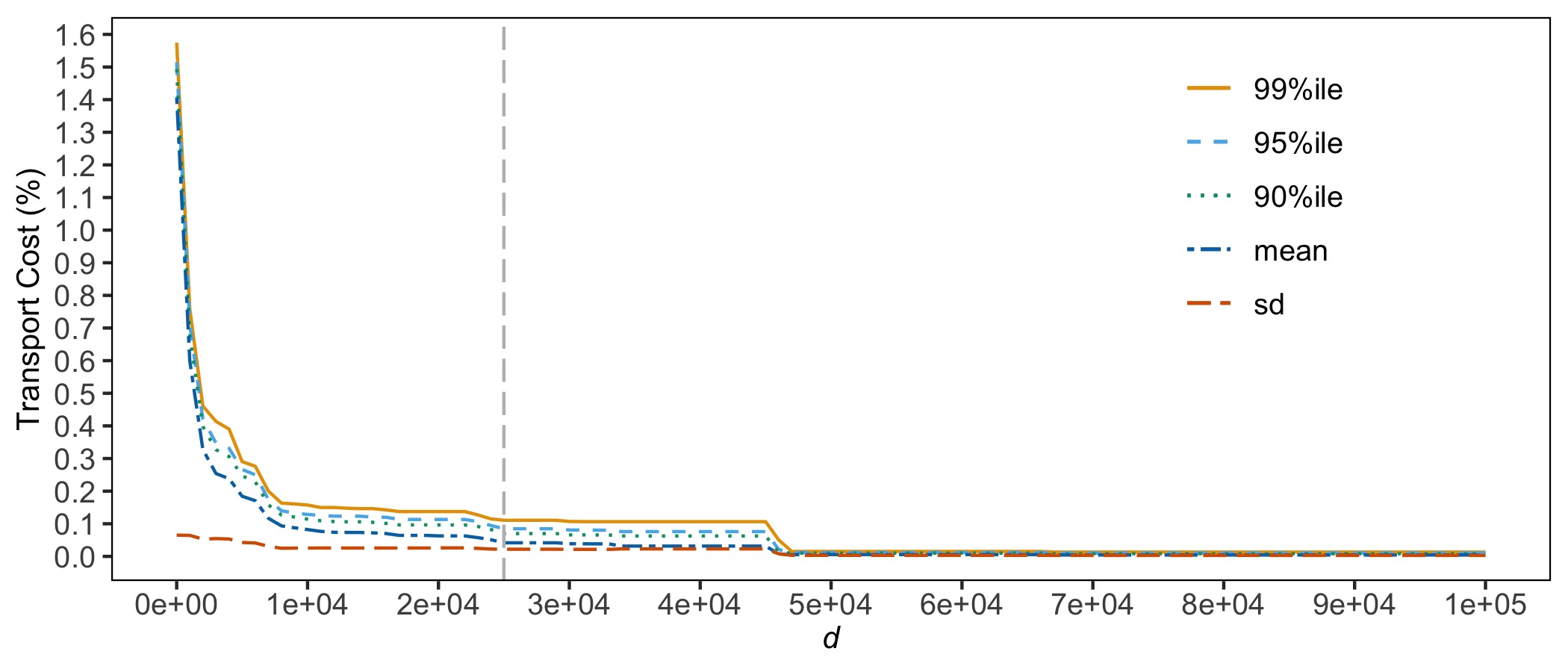}
 \caption{\figtitle{Placebo costs as a function of $d$.} \figcaption{The curves represent the mean, standard deviation, and the quantiles of 500 simulated values of $OT_d(\tilde{\mathbb{P}}_{pre, n_{pre}}, \tilde{\mathbb{P}}_{pre, n_{post}})$ for each value of $d$ in Beijing. The vertical line at $d=25\numcomma000$ is the smallest bandwidth for which the placebo cost is less than 0.05\%.}}
  \label{fig:appfigsix}
\end{figure}

\section{Difference-in-differences regression} \label{sec:A}
After aggregating different car models by city, time period (pre- and post-lottery), and price, we run the following difference-in-differences specifications in logs for three control
groups: Tianjin, Shijiazhuang, and Tianjin and Shijiazhuang combined.  The regression model is
\begin{equation}
p_{j,c,t}=\alpha_0+\alpha_1{Beijing}_{j,c,t}+\alpha_2{post}_{j,c,t}+\alpha_3{Beijing}_{j,c,t}\times{post}_{j,c,t}+\epsilon_{j,c,t}
\end{equation}
where $p_{j,c,t}$ is log of the $j$th price observation in city $c$ in month $t$ and where
${Beijing}_{j,c,t}$ is an indicator that is one for price observations from Beijing and zero
for cities in the relevant control group. The indicator ${post}_{j,c,t}$ is one in all months
after the introduction of the lottery in Beijing, and zero otherwise. Table \ref{tab:eight} shows that
the estimated price jumps, $\hat{\alpha}_3$, are similar, and around \numberblank{21\%} across the specifications.

\begin{table}[H]\centering
\small
\ra{\usualra}
\begin{threeparttable}
 \caption{\tabtitle{Diff-in-diff regression results for different control groups}}
 \begin{tabular}{@{\extracolsep{7pt}}lD{.}{.}{-3} D{.}{.}{-3} D{.}{.}{-3}}
 \toprule
    & \multicolumn{1}{c}{\tabhead{Tianjin}} & \multicolumn{1}{c}{\tabhead{Shijiazhuang}} & \multicolumn{1}{c}{\tabhead{Both}} \\
 \midrule
    Beijing & 0.266^{\phantom{***}} & 0.233^{\phantom{***}} & 0.255^{\phantom{***}} \\ 
      & (0.002) & (0.002) & (0.001) \\
      & & & \\
     post & 0.035 & 0.027 & 0.032 \\
      & (0.002) & (0.003) & (0.001) \\
      & & & \\
     Beijing $\times$ post & 0.208 & 0.215 & 0.210 \\
      & (0.002) & (0.003) & (0.002) \\
      & & & \\
     Constant & 11.493 & 11.527 & 11.505 \\
      & (0.001) & (0.002) & (0.001) \\ 
    & & & \\ 
 \midrule
    Observations & \multicolumn{1}{c}{1,411,305} & \multicolumn{1}{c}{1,200,418} & \multicolumn{1}{c}{1,637,150} \\ 
R$^{2}$ & \multicolumn{1}{c}{0.074} & \multicolumn{1}{c}{0.059} & \multicolumn{1}{c}{0.076} \\ 
 \bottomrule
 \end{tabular} \label{tab:eight} 
\end{threeparttable}
\end{table}

\section{Transport costs due to changes in the composition of sold car prices in Beijing} \label{sec:B}
From \cite{yang14} and the Beijing Municipal
Commission of Transport cited therein, we know the number of licenses emitted on a monthly
basis in Beijing in 2010. We also observe, as detailed in Section
\ref{sec:data}, the monthly sales for each model and price. Therefore, we can estimate
for each month of 2010 the proportion $\phi_{f,t}$ of first-time
buyers and the proportion $\phi_{r,t}$ returning buyers. Specifically,
we estimate
\[
\phi_{f,t}=\frac{\rho\cdot L_{t}}{\sum_{i}q_{i,t}},\ \phi_{r,t}=1-\phi_{f,t},
\]
where $t$ indexes the time period, $q_{i,t}$ is the quantity of cars sold
at price $p_{i}$ in month $t$, and $L_{t}$ is the number of licenses given out in period $t$.
The coefficient $\rho$ is the proportion
of new cars, as opposed to used cars, purchased with a license. We
take this proportion to be $\rho = 0.5$ (\citealp{yang14}).

The monthly estimates are convenient here because the underlying population distributions of preferences of first purchases, $f$, and returning purchases, $r$, are plausibly stable, as is $\rho$, while we get identifying variability from $(\phi_{f,t},\phi_{r,t})$
across $t$.

We may then solve for our best guesses $\hat{f}(\rho)$, $\hat{r}(\rho)$
for the distributions, as a function of $\rho$, by solving
\[
\min_{f,r\in\Lambda}\sum_{i,t}\left(\phi_{f,t}f_{i}+\phi_{r,t}r_{i}-p_{i,t}\right)^{2}.
\]

Given $\hat{f}(\rho)$, $\hat{r}(\rho)$, we can compute
\[
\hat{p}_{\mathrm{pre}}:=\theta_{f,\mathrm{pre}}\hat{f}+\theta_{r,\mathrm{pre}}\hat{r}
\]
as well as
\[
\hat{p}_{\mathrm{post}}:=\theta_{f,\mathrm{post}}\hat{f}+\theta_{r,\mathrm{post}}\hat{r},
\]
 where $\theta_{f,\mathrm{pre}}$ and $\theta_{r,\mathrm{pre}}$ are, respectively, the fraction of car purchases in 2010 for which a license is used for the first time and the fraction for which the costumer used a returning license, i.e., one that has been previously used to purchase another car.  The scalars $\theta_{f,\mathrm{post}}$ and $\theta_{r,\mathrm{post}}$ are defined analogously for 2011.

 Consequently, we can compute before-and-after estimate for a chosen smoothing parameter
$d$,
\[
OT_{d}(\hat{f}(0.5),\hat{r}(0.5)).
\]
 

\begin{table}[H]\centering
\small
\ra{\usualra}
\caption{\tabtitle{Estimated cost due to change in composition}}  \label{tab:tabB}
\begin{threeparttable}
    \begin{tabular}{@{}>{\centering}p{0.15\linewidth}>{\centering\arraybackslash}p{0.15\linewidth}}
    \toprule
      $d$  & $OT_{d}(\hat{f},\hat{r})$ \\
    \midrule
       4\numcomma000  & 2.4\% \\
       8\numcomma000 & 1.8\% \\
       10\numcomma000 & 1.1\% \\
       20\numcomma000 & 0.04\% \\
    \bottomrule
    \end{tabular}
\end{threeparttable}
\end{table}

The correction applies to the before-and-after estimator for tuning $d$ and to the difference-in-transports estimator for the value of the tuning coefficient on the transport cost in Beijing.  For instance, for the difference-in-transports with tuning $d=4\numcomma000$ and smoothing $2d=8\numcomma000$ on the Beijing transport cost, the appropriate correction according to Table \ref{tab:tabB} is $1.8\%$.  

\section{Comparative statics in transaction costs and prices} \label{sec:C}
We derive the comparative statics for the transaction costs and price estimators.
We drop the hat-notation for expositional convenience. Taking derivatives of the
equilibrium conditions in \eqref{eq:17}, we get
$$ 
2\frac{\partial p(s)}{\partial s}=\frac{1}{f(v_{seller})}-\frac{q}{N-q} \frac{1}{f(v_{buyer})}
$$
and
$$ 
2\frac{\partial t(s)}{\partial s}=-\frac{q}{N-q} \frac{1}{f(v_{buyer})}-\frac{1}{f(v_{seller})}.
$$
It is immediately clear that the implied transaction costs decrease with the volume
of trade $s$. The same is not necessarily true for the transaction price, which may
increase or decrease, depending on the shape of $f$. However, both $\tilde{p}_{buyer} = p+t$ and
$\tilde{p}_{seller} = p-t$ are monotonic in $s$. These statics show that, conditional on a lower
bound estimate s, t and $\tilde{p}_{buyer} = p+t$ are upper bound estimates, and $\tilde{p}_{seller} = p-t$
is a lower bound estimate.

\section{Extension to speculators}\label{sec:D}
We now relax Assumption \ref{as:five} and allow for speculators. Define speculators as having
zero willingness-to-pay for a license plate if non-transferability is strictly enforced,
i.e. a speculator will never purchase a car if she wins a license. Speculators have two effects in the market: 
they crowd out car sellers on the supply side (shift the supply
curve down) and they increase the number of car buyers on the demand side (shift
the demand curve out). Suppose that a share $z$ of the license plates are allocated
to speculators, and suppose that a speculator's reservation price is zero. Then the
demand and supply are
\begin{equation}
\begin{aligned}
&D(p,t)=(N-q(1-z))(1-F(p+t)),\\
&S(p,t)=zq+\frac{(s-z)}{s}qF(p-t).
\end{aligned}
\end{equation}
Speculators have a supply curve that is flat from zero to $zq$, and increasing thereafter
to $q$. As $z$ goes to $s$, the supply curve becomes vertical at $q$. On the demand side,
the demand curve shifts out towards $v(n)$ as $z$ goes to one and all prospective car
buyers must turn to the black market for license plates.


\begin{table}[H]\centering
\small
\ra{\usualra}
\caption{\tabtitle{Estimates of transaction prices and costs at \numberblank{$\hat{s} = 11\%$} and $z = 0.11\%$. Prices and costs are in RMB 1\numcomma000.  Gains from trade are in RMB billion.}}
\begin{threeparttable}
 \begin{tabular}{@{}>{\centering}p{0.3\linewidth}>{\centering}p{0.1\linewidth}>{\centering\arraybackslash}p{0.1\linewidth}}
 \toprule
      & \tabhead{Estimate} & \tabhead{95\% CI} \\
 \midrule
    $\hat{p}$ &110 &\lbrack98,125\rbrack\\
    $\hat{t}$ &105 &\lbrack98,125\rbrack\\
    Total transaction costs &6.0 & \lbrack4.0,7.9\rbrack\\
    Net gains from trade &1.1 & \lbrack0.4,2.1\rbrack\\
 \bottomrule
 \end{tabular}
\end{threeparttable}
 \label{tab:ten}
\end{table}

\noindent In Table \ref{tab:ten}, we report results assuming that all 11\% illegal trades are by
speculators. Since the supply curve is now flat from zero to $sq$, all speculators sell
at $\tilde{p}_{seller}=0$. We see that now $\hat{p}=\hat{t}$. At the same time, the demand curve shifts
out to $v(n)$. The outward shift in the demand curve dominates the downward shift
in the supply curve: both transaction prices and costs increase, from 105 to 110 and
from 100 to 105, respectively. The net gains from trade go down by 15\%, largely
due to further taxation of buyers. 

One limit case of interest is when speculators completely crowd out prospective car buyers and there are no transaction costs, i.e., when $z=q$ and $t=0$.  Then our model gives the same allocation as Li (2018)'s counterfactual auction market, where the supply curve is vertical at $q$ and demand is $v(n)$.  It follows that Li (2018)'s analysis applies.\footnote{except for the congestion and pollution externalities, which we have ignored.}


\section{Allocation of Vehicle License Plates} \label{sec:E}

\subsection{Beijing}
According to the Beijing Municipal People's Government data, the number of motor vehicles in Beijing has grown by 81.8\% from 2.58 million in 2005 to 4.69 million in November 2010. To control the number of motor vehicles, the city of Beijing has introduced a set of rules in 2010 (henceforth, `the rules'). 

There have been several revisions to the rules after they were first introduced in 2010. The most recent draft update to the rules was released in June 2020 \citep{rule2020}. The updated rules are expected to be implemented in 2021. In the rest of this section, we focus on the initial version of the rules that was released in December 24, 2010 by the \citet{rule2, rule1}.

\subsubsection{Application Procedure}
To win a license plate, individuals need to participate in a lottery if they want to buy a new or second-hand car, receive a car as a gift, or change a license plate that belongs to another city in China to a Beijing license plate .  

Individuals can participate in the lottery through the following procedure:
\begin{enumerate}
	\item Start an application to obtain an application code.
	\item After the application is approved, the application code will be validated. 
	\item The validated code can be used to participate in the lottery.
\end{enumerate}

Applications can be made either in person or online. There are 16 government offices that accept applications made in person. There are six main steps to apply online:  
\begin{enumerate}
	\item Log in to the application website, create an account and choose the appropriate application category.
	\item Read and accept the rules.
	\item Fill in personal information, such as name, identification number, email address, contact information, etc.
	\item Choose the license category.
	\item Fill in more personal information, such as date of birth, vehicle type, etc.
	\item Check and confirm the information.
\end{enumerate}

\subsubsection{Lottery Mechanism}

The lottery for individuals is held on the 26 of each month with the first taking place on January 2011.  The lottery policy was officially announced  and implemented on December 24, 2010, but had been publicly hinted at as early as December 13, 2010 (\citealp{yang14}).
The non-winning validated codes will remain valid in the same year, and they are automatically allocated to the next lottery. These codes will remain valid until December 31 of the same year in which they are generated. Applicants who win the lottery must complete the vehicle registration process within six months. Otherwise, they will forfeit their rights to a license plate. 

Licenses for alternative fuel vehicles were not included in the initial set of rules. Licenses for these vehicles were introduced in the revised set of rules that was effective in January 1, 2014 \citep{rule2013}. Based on this revision of the rules, alternative fuel vehicles were referred to as the electric vehicles that are listed in the product catalog.  
The quota for the total number of license plates for vehicles was 240,000 in 2011. 
According to the bureau, the lottery is entirely random and was designed by data scientists to ensure fairness. 

For some statistics related to the license plates lottery in Beijing, please refer to Section \ref{sec:bj_stat}. For details on the revisions to the winning rates in the lottery, please refer to Section \ref{sec:bj_win}.

\subsubsection{License Plates for Individuals}
Individuals who do not have a vehicle can participate in the lottery if they fulfill one of the following requirements:
\begin{enumerate}
	\item They hold a Beijing domicile.
	\item They belong to the army or police force in Beijing.
	\item They are citizens from Hong Kong, Macau or Taiwan, or they are from other parts of the world who hold a valid Beijing residence permit and have resided in Beijing for one year.
	\item They are a Chinese citizen who do not have a Beijing domicile but have a Beijing work and stay permit.
	\item They are allowed to live in Beijing and have paid social security and income tax for five consecutive years.
\end{enumerate}

\subsubsection{Transfer of License Plates}
The license plates are non-transferable. If a vehicle is transferred to another person or the vehicle is reported as a scrapped vehicle, the license plate still belongs to the owner. On the other hand, if the vehicle is sent to another person as a gift, then the original owner of the vehicle cannot keep the license plate. 
For property transfers due to divorce or inheritance, judicial order will be followed instead of the rules.

\subsubsection{Application Statistics} \label{sec:bj_stat}

Table \ref{tab:bj_2011} shows the details of the first vehicle license plate lottery for individuals \citep{bj201101,bj201102,bj202002,bj202006}.

\begin{table}[!ht]\centering
\small
\ra{\usualra}
\caption{\tabtitle{Summary Statistics for the Beijing License Lottery in January 2011}}
\begin{threeparttable}
 \begin{tabular}{@{}>{\centering}p{0.4\linewidth}>{\centering}p{0.1\linewidth}>{\centering\arraybackslash}p{0.2\linewidth}}
 \toprule
 \tabhead{Number of Valid Application Codes} & \tabhead{Quota} & \tabhead{Winning Rate} \\
 \midrule
 187\numcomma240 & 17\numcomma600 & 9.39\%  \\
 \bottomrule
 \end{tabular}
\end{threeparttable}  \label{tab:bj_2011}
\end{table}

\subsubsection{Winning Rate} \label{sec:bj_win}

In light of the surge in the number of applications for vehicle license plates, the Beijing Municipal Commission of Transport had implemented several revisions to the lottery mechanism to increase the winning rate for individuals who were unable to win a license plate after many lotteries. In this section, we focus on the Municipal's first attempt in modifying the winning rate \citep{rule2013}. 

In the first attempt to modify the winning rate, the winning rate is higher if an individual has participated in the lottery for 24 times without winning. Depending on the license type, the winning rate is going to be a multiple of the base winning rate of a certain lottery. This is illustrated by Table \ref{tab:lotwin}.

\begin{table}[!ht]\centering
\small
\ra{\usualra}
\caption{\tabtitle{Revised Winning Rate in 2013}}
\begin{threeparttable}
 \begin{tabular}{@{}>{\centering}p{0.45\linewidth}>{\centering}p{0.2\linewidth}>{\centering\arraybackslash}p{0.2\linewidth}}
 \toprule
 \multirow{2}{*}{\tabhead{Number of Lottery Losses}} & 
 \multicolumn{2}{c}{Winning Rate} \\
 \cmidrule{2-3}
  & \tabhead{with C5 License\tnote{$\ast$}} & \tabhead{Other License} \\
 \midrule
 1 -- 24 & $2r$ & $r$ \\
 25 -- 36 & $3r$ & $2r$ \\
 37 -- 48 & $4r$ & $3r$ \\
 49 -- 60 & $5r$ & $4r$ \\
 61 -- 72 & $6r$ & $5r$ \\
$\vdots$ & $\vdots$ & $\vdots$\\
 \bottomrule
 \end{tabular}
\begin{tablenotes}
    \item $r$ refers to the base winning rate in a particular month
    \item[$\ast$] C5 licenses are designated to small passenger cars for people with disabilities
\end{tablenotes}
\end{threeparttable} \label{tab:lotwin}
\end{table}


The 2017 revision \citep{rule2017} to the winning rate uses a similar method as in \citet{rule2013} with a few differences. Namely, the winning rate will be modified once an individual loses the lottery six times instead of 24. Subsequently, the winning rate will be modified after 12 attempts as in Table \ref{tab:lotwin}.

Any contracts or prepayments on vehicle sales that have been reported to the bureau before the implementation of the rules were exempt from the requirements of the rules.

\subsection{Tianjin}

The rationing rules were introduced in Tianjin in order to control the number of vehicles, alleviate the problem of traffic congestion and to improve the air quality of the city. 

The measures to control vehicle purchase and restrict the traffic in Tianjin were first announced in a press conference by the Tianjin Municipal People's Government at 7 p.m. of December 15, 2013 \citep{news}. The controls and restrictions were effective five hours later on December 16, 2013 at midnight.  The Tianjin Municipal People's Government suspended any new vehicle registration and vehicle transfer in Tianjin between December 16, 2013 and January 15, 2014 in order to ensure a smooth transition and preparation for the new rules.

After the rules were announced in the press conference, some 4S stores (stores that sell vehicles in China) had over 70 consumers queued up to purchase vehicles, and the vehicles of some major brands were all sold out before midnight of December 16, 2013 according to a news report by the \citet{newsnbd}. According the same news report, Toyota sold around 2,000 vehicles in five hours following the press conference, which equals to 20\% of the annual vehicle sales in Tianjin.

The first draft of the rules (henceforth, `the rules') was effective for one year. In January 1, 2015, the control rules were updated and remained effective for the next five years. The most recent update to the rules was released in 2019 and implemented on January 1, 2020. This set of rules would remain effective for the next five years.  In the rest of this section, we focus on the initial draft of rules by the \citet{tianjin2013} that was released in December 15, 2013.

\subsubsection{Application Procedure}
In general, the application procedure is as follows:
\begin{enumerate}
	\item Start an application to obtain an application code. The allocation mechanism has to be stated in the application.
	\item  After the application is approved, the application code will be validated.
	\item The validated code can be used to participate in an auction or a lottery. 
\end{enumerate}

The applications can be made in person or online. There are 14 government offices that accept the applications that are made in person. There are five main steps in making the online application for individuals:  
\begin{enumerate}
	\item Log in to the application website, create an account and choose the appropriate application category.
	\item Read and accept the rules.
	\item Fill in the personal information, such as name, identification number, email address, etc., and choose to participate in the lottery or the auction.
	\item Fill in the contact information such as the mobile phone number. An SMS will be sent to the phone number to confirm that the phone number is correct. 
	\item The application code and application form will be generated by the system.
\end{enumerate}

\subsubsection{Allocation Mechanism}
Vehicle license plates are allocated every 12-month cycle. The number of quotas available in each cycle is 100,000. The license plates are allocated monthly. Any unused quota in a cycle cannot be reallocated to the next cycle.  The quotas are allocated according to the ratio of 1:5:4 for different vehicle types as shown in Table \ref{tab:2}. 

\begin{table}[!ht]
	\centering
	\caption{Allocation of License Plates in Tianjin per cycle}

	\begin{tabular}{lll}
		\toprule
		\tabhead{Vehicle type} & \tabhead{Allocation mechanism} & \tabhead{Number} \\
		\midrule
Energy-efficient vehicles &	Lottery &	10,000 \\
Normal vehicles &	Lottery &	50,000 \\
Normal vehicles &	Auction &	40,000 \\
		\bottomrule
	\end{tabular}
	\label{tab:2}
\end{table}

Among the available license plates in each cycle, 88\% of the them are allocated to individuals. The rest are allocated to organizations. 

Each code can be used in either an auction or a lottery. All applications made after the 9th of the month will be allocated to the next month. Applicants who wish to withdraw from the application should do so before the 20th of the month. 
Once an individual wins the auction or lottery, they need to complete the application and use the winning code within six months. Otherwise, the code will be deactivated.

\subsubsection{License Plates for Individuals}

Individuals can only have one validated code and can apply for an application code if they fulfill the following requirements: 
\begin{enumerate}
	\item They reside in Tianjin and fulfill one of the requirements:
	\begin{enumerate}
		\item They hold a Tianjin domicile.
		\item They belong to the army or police force in Tianjin.
		\item They hold a valid personal identification from Hong Kong, Macau or Taiwan, or other parts of the world, and have lived in Tianjin for at least 9 months each year for the past two years.
		\item They do not hold a Tianjin domicile but they have a Tianjin residence permit and have paid for social security for at least 24 months.
	\end{enumerate}
	\item They hold a valid driving license.
	\item They do not have a vehicle registered under their name or the vehicle or they do not hold a license plate that has been cancelled. 
	\item They do not have a valid vehicle license plate or do not have a license plate that is renewable.
	\item They do not have an overdue winning code or did not give up any wining code in the past two years via the car lottery.
\end{enumerate}

\subsubsection{Lottery}
The exact number of license plates available in the lottery is announced by the 9th of each month. The lottery is held on the 26th of each month or the next working day if the 26th is not a working day. 
Like the Beijing lottery, the lotteries are organized every month for individuals.
For some statistics for the license plates lottery in Tianjin, please refer to Section \ref{sec:tj_stat}.

\subsubsection{Auction}
Auctions are organized via the internet and details for each auction is released by the 18th of each month. 

To participate in an auction, applicants need to pay a security deposit of RMB 2\numcomma000 per application code. The reservation price for each license plate is RMB 10\numcomma000. There is no maximum bid in the auction. The auctions for individuals and organizations are held separately.

In each auction, the highest bid wins. If two bidders submit the same bid, the bidder who submits the bid earlier wins. Results for the auctions will be announced within five working days after the auction day. If a winning bidder does not pay for the bid, the security deposit will be forfeited. 

Any unallocated quota for vehicle licenses in a certain month will be reallocated to the next month. 

Some statistics for the license plates auction in Tianjin are given in Section \ref{sec:tj_stat}.

\subsubsection{Application Statistics} \label{sec:tj_stat}
Table \ref{tab:tianjin_auction} summarizes the average price, lowest price and number of license plates available in the auctions for vehicle license plates in the February 2014 (the first auction) and June 2020 (the latest month). Note that the highest bid price for the auctions is not provided by the auction house. The 2014 auction data is obtained from the \citet{tianjin2016aucresult}, while the 2020 auction data is obtained from the \citet{tianjin2020aucresult}.

\begin{table}[!ht]\centering
\small
\ra{\usualra}
\caption{\tabtitle{Vehicle License Plate Auctions in Tianjin in February 2014 and June 2020}}
\begin{threeparttable}
 \begin{tabular}{@{}>{\raggedright}p{0.2\linewidth}>{\centering}p{0.2\linewidth}>{\centering}p{0.2\linewidth}>{\centering\arraybackslash}p{0.2\linewidth}}
 \toprule
 & \tabhead{Average Price} \tabunits{(CNY)} & \tabhead{Lowest Price} \tabunits{(CNY)} & \tabhead{Number of License Plates} \\
 \midrule
 February 2014 & 16\numcomma340 & 10\numcomma000 & 3\numcomma203 \\
 June 2020 & 22\numcomma022 & 20\numcomma500 & 3\numcomma059 \\
 \bottomrule
 \end{tabular}
\end{threeparttable} \label{tab:tianjin_auction}
\end{table}


As shown in Table \ref{tab:tianjin_auction}, the average price for individuals in June 2020 is higher than the corresponding average price in February 2014. In addition, the reservation price for the individual license plates are no longer hit in June 2020. 

On the other hand, the lottery for vehicle license plates are also competitive, but they are a little less competitive than in Beijing. In the first Tianjin vehicle lottery in February 26, 2014, there was a total of 151,139 validated application codes but only 4,450 normal vehicles and 910 renewable vehicles \citep{tianjin2014lotresult}. Table \ref{tab:tianjin_lot} summarizes the details of the lottery in June 2020 \citep{tianjin2020lotresult}. 

\begin{table}[!ht]
	\centering
	\caption{Vehicle License Plate Lottery in Tianjin in June 2020}
	\begin{tabular}{lccc}
		\toprule
\multirow{2}{*}{\bf Category} & \bf Number of valid  	& \bf License plates  & \bf Winning \\
& \bf application codes & \bf allocated & \bf rate \\
\midrule
\multicolumn{4}{l}{\emph{Individuals}}	\\
\hspace{1em} Renewable vehicles	&	3,406	&	3,406	&	100.00\%\\
\hspace{1em} Normal vehicles	&	809,957	&	8,747	&	1.08\%\\[0.5em]
\multicolumn{4}{l}{\emph{Organizations}}\\	
\hspace{1em} Renewable vehicles	&	21&21	&	100.00\%\\
\hspace{1em} Normal vehicles	&	9,996	&	500	&	5.00\%\\
\bottomrule
	\end{tabular}
	\label{tab:tianjin_lot}
\end{table}

Although no official data are available by the Tianjin Municipal Transportation Commission on the breakdown between individuals and organizations in the 2014 lottery data for Tianjin, the first lottery for license plates in Tianjin is much less competitive than the most recent one; the number of valid application codes in June 2020 is 544.78\% more than the number of valid application codes in February 2014.

\section{Robustness to Marginal Data}\label{sec:F}

The Beijing lottery began in January 2011. It is possible that anticipation occurred as early as mid-December 2010 (\citealp{yang14}). This motivates us to drop the month of December from the control year 2010 and repeat the analysis.  Lottery winners do not immediately purchase a vehicle upon obtaining their license, and distributional differences may arise due to the sales only ``getting started" with rationed license when there is the quota-regulated flow of them but no stock.  For that reason, we also consider the analysis when, in addition to omitting December 2010, we omit January and February 2011.

Our results are largely unchanged, suggesting that anticipation of or response to the restrictions does not drive our results.  To summarize, when omitting only December 2010, the difference-in-transports estimate smoothing both differenced transport costs by $d$ is $13.5\%$, and the conservative difference-in-transports estimate smoothing the Beijing transport cost by $2d$ is $10.9\%$.  When omitting December 2010, January 2011 and February 2011, the difference-in-transports estimate smoothing both differenced transport costs by $d$ is $14.2\%$, and the difference-in-transports estimate smoothing the Beijing transport cost by $2d$ is $11.6\%$.

Section \ref{sec:v1} replicates the figures and tables of the main text when omitting December 2010.  Section \ref{sec:v2} replicates the figures and tables when omitting December 2010, January 2011 and February 2011.  Section \ref{sec:resultssummary} collects the estimates from the main and the two alternative specifications.

\subsection{Excluding December 2010}\label{sec:v1}

\begin{figure}[H]
  \centering
  \includegraphics[width=\textwidth]{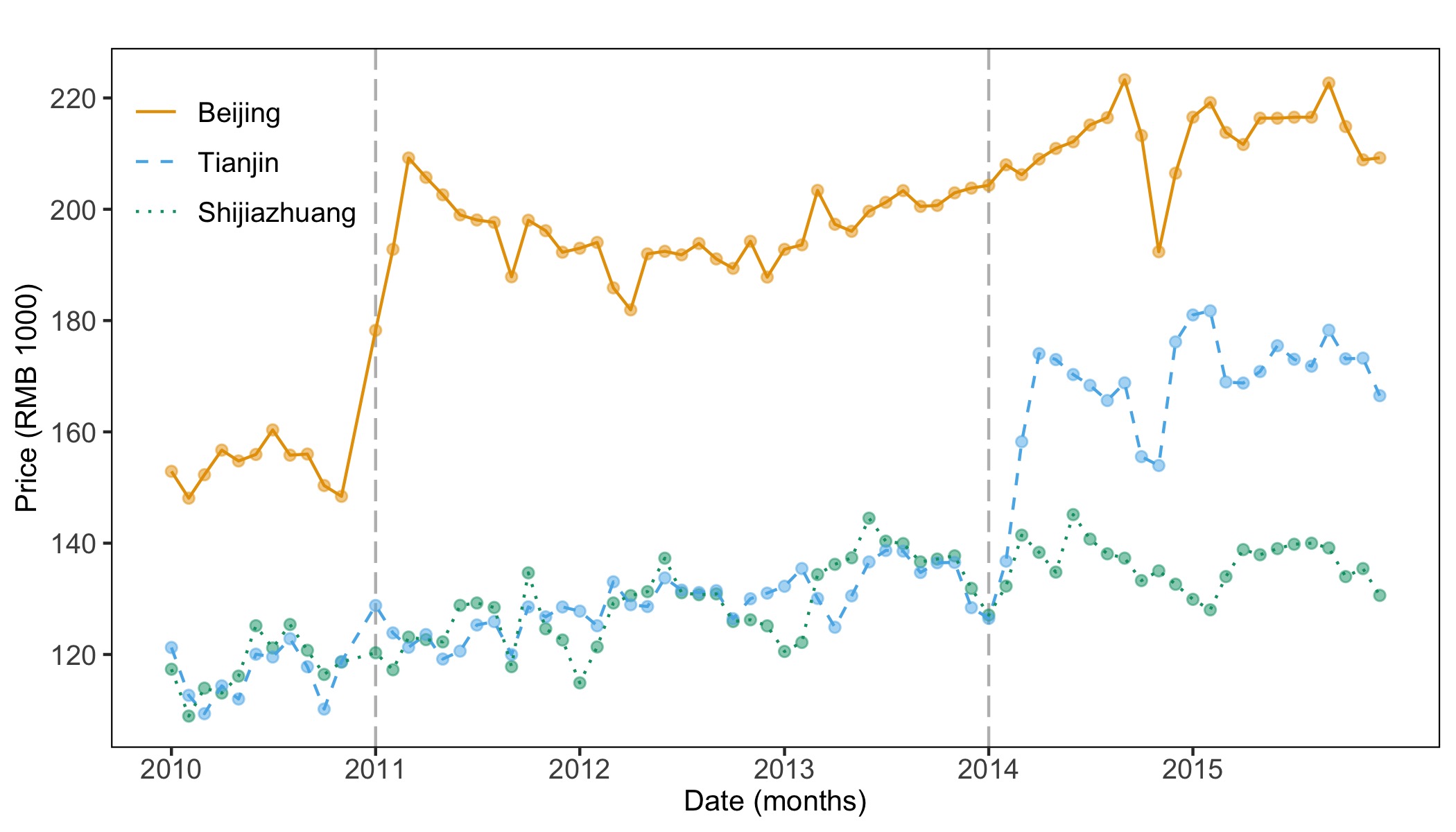}
  \caption{\figtitle{Monthly average swtprice (RMB 1\numcomma000) for Beijing, Shijiazhuang, and Tianjin.} 
  \figcaption{This figure is similar to Figure \ref{fig:one}.}}
  \label{fig:v1-one}
\end{figure}

\begin{figure}[H]
  \centering
  \includegraphics[width=\textwidth]{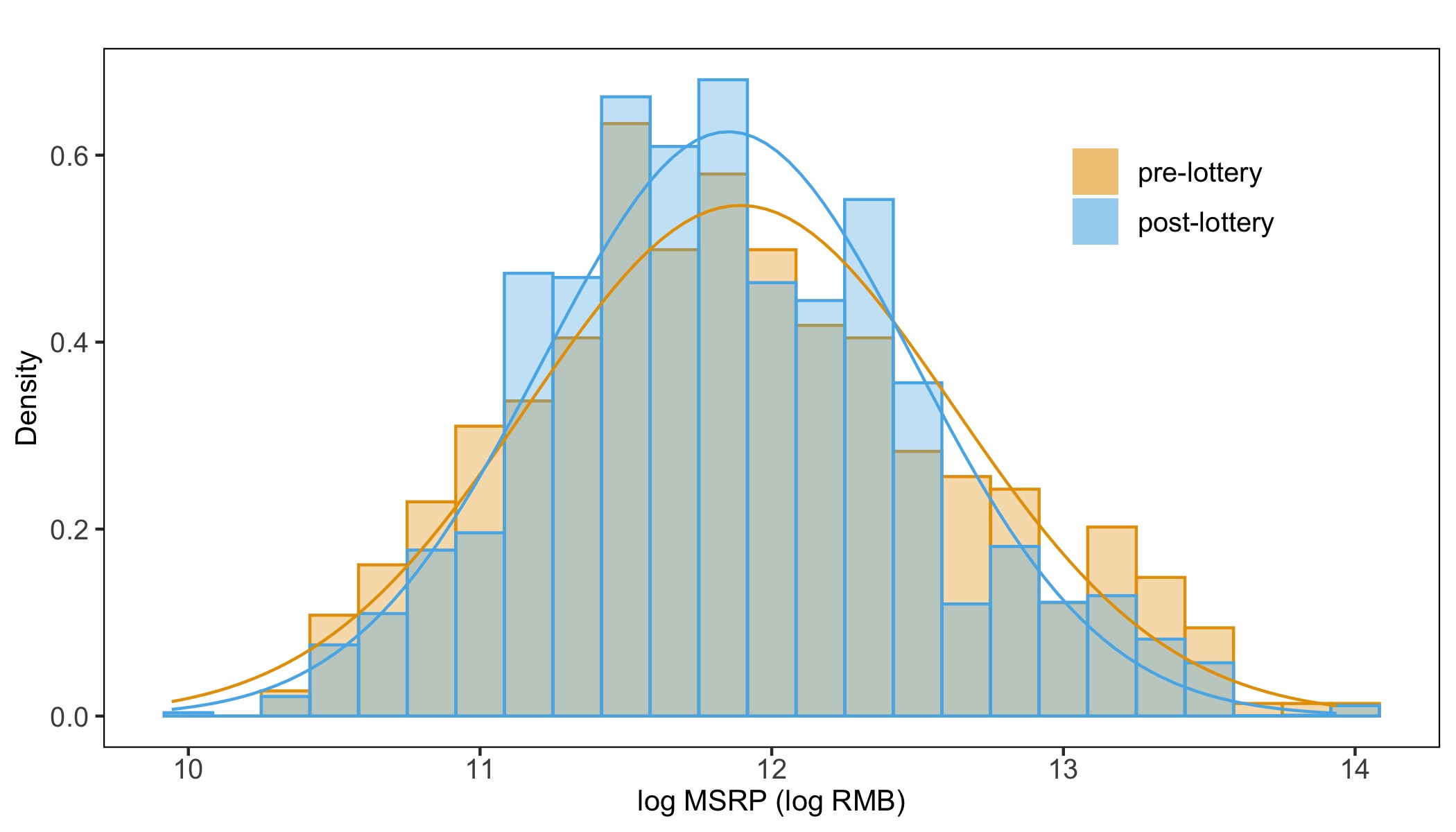}
  \caption{\figtitle{Distribution of Beijing car prices in 2010 (pre-lottery) and new car prices in 2011 (post-lottery).} 
  \figcaption{This figure is similar to Figure \ref{fig:two}.}}
  \label{fig:v1-two}
\end{figure}

\begin{figure}[H]
    \centering
    \includegraphics[width=\textwidth]{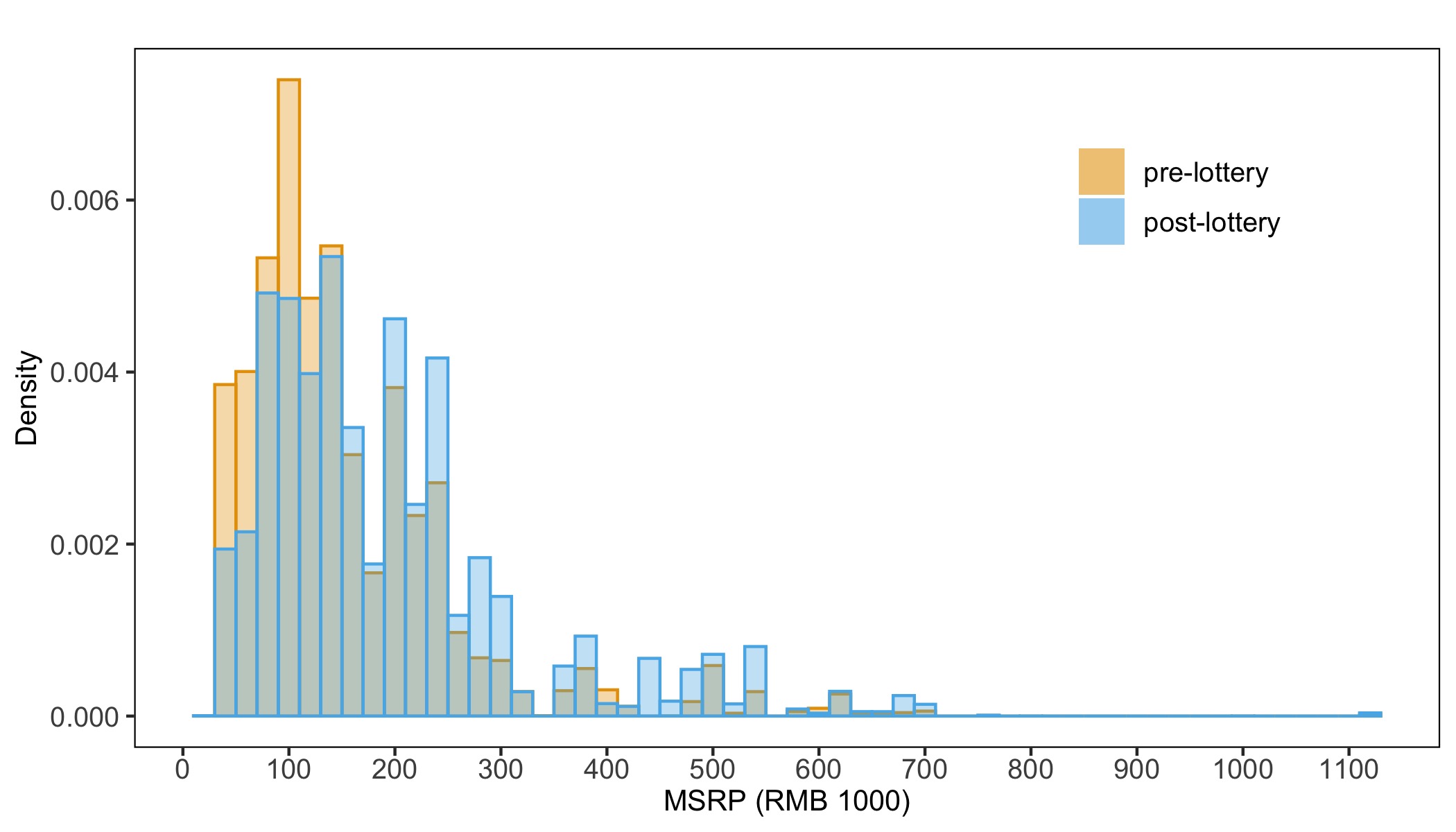}
    \caption{\figtitle{Smoothed empirical distributions of Beijing car sales in 2010 (pre-lottery) and 2011 (post-lottery).} 
    \figcaption{This figure is similar to Figure \ref{fig:four}.}}
    \label{fig:v1-four}
\end{figure}

\begin{figure}[H]
    \centering
    \includegraphics[width=\textwidth]{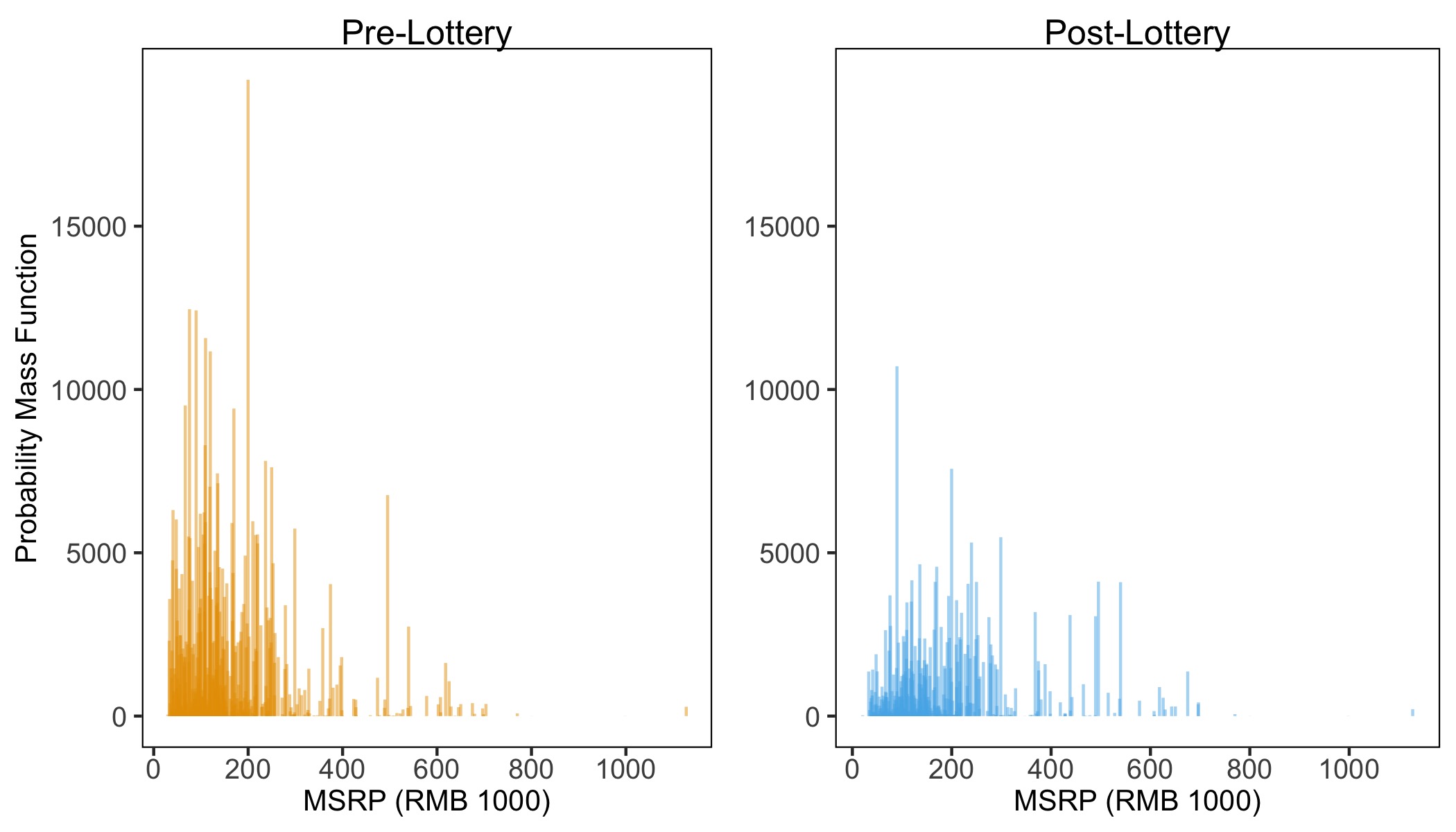}
    \caption{\figtitle{Exact empirical distributions of Beijing car sales in 2010 (pre-lottery) and 2011 (post-lottery).} 
    \figcaption{This figure is similar to Figure \ref{fig:five}.}}
    \label{fig:v1-five}
\end{figure}

\begin{figure}[H]
    \centering
    \includegraphics[width=\textwidth]{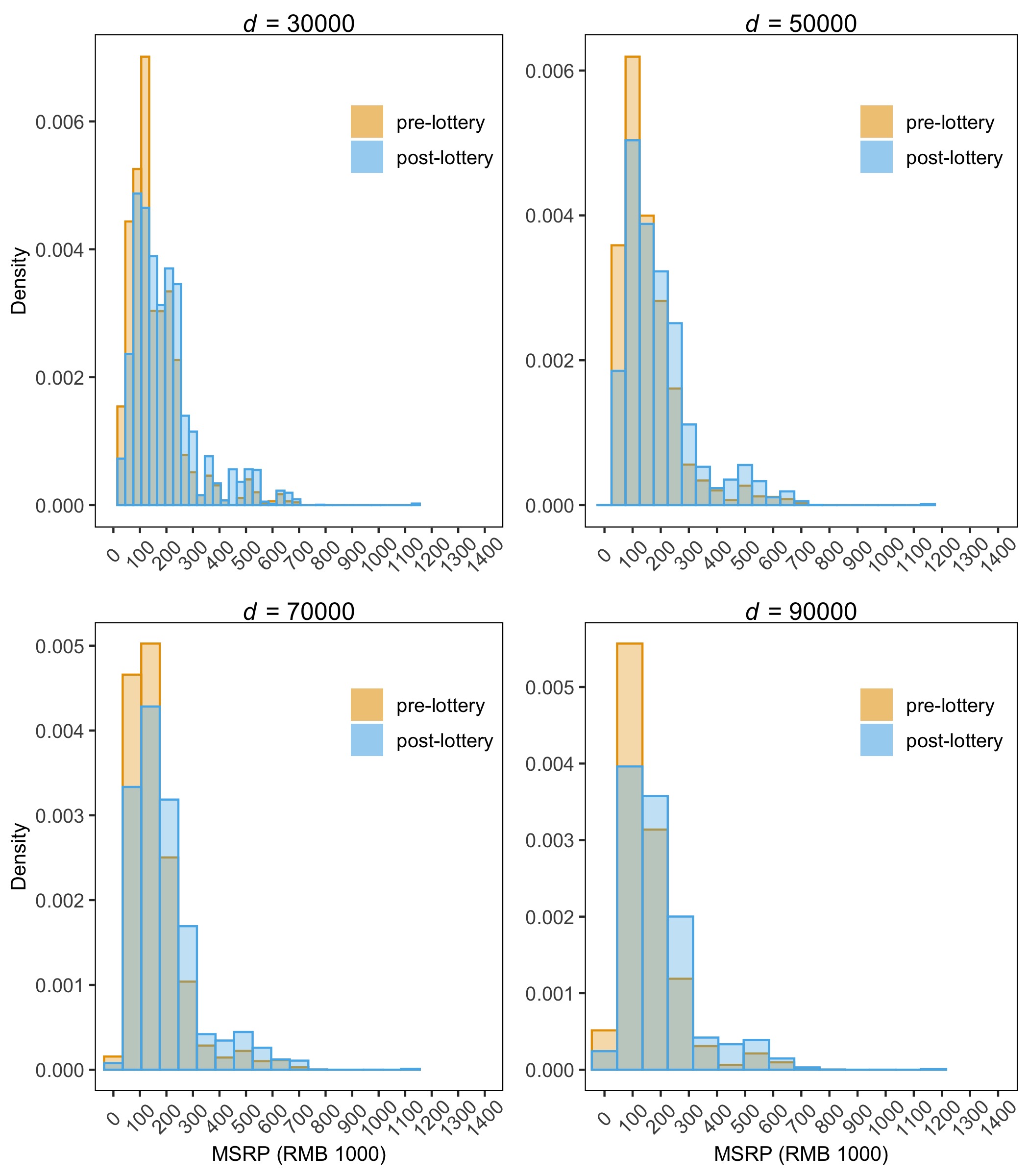}
    \caption{\figtitle{Smoothed empirical distributions of Beijing car sales in 2010 (pre-lottery) and 2011 (post-lottery) at candidate values of $d$.} 
    \figcaption{This figure is similar to Figure \ref{fig:six}.}}
    \label{fig:v1-six}
\end{figure}

\begin{figure}[H]
    \centering
    \includegraphics[width=\textwidth]{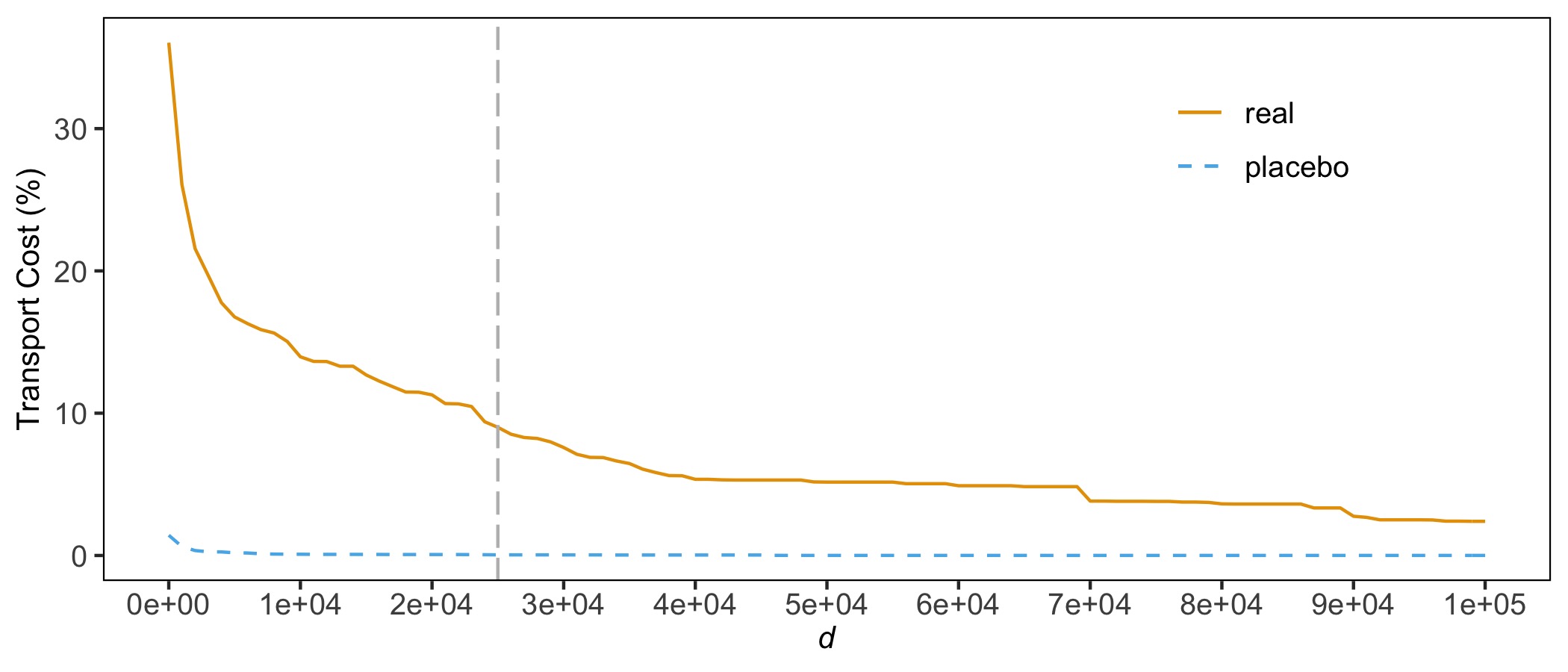}
    \caption{\figtitle{Real and placebo costs as a function of $d$.} 
    \figcaption{This figure is similar to Figure \ref{fig:seven}.}}
    \label{fig:v1-seven}
\end{figure}

\begin{figure}[H]
    \centering
    \includegraphics[width=\textwidth]{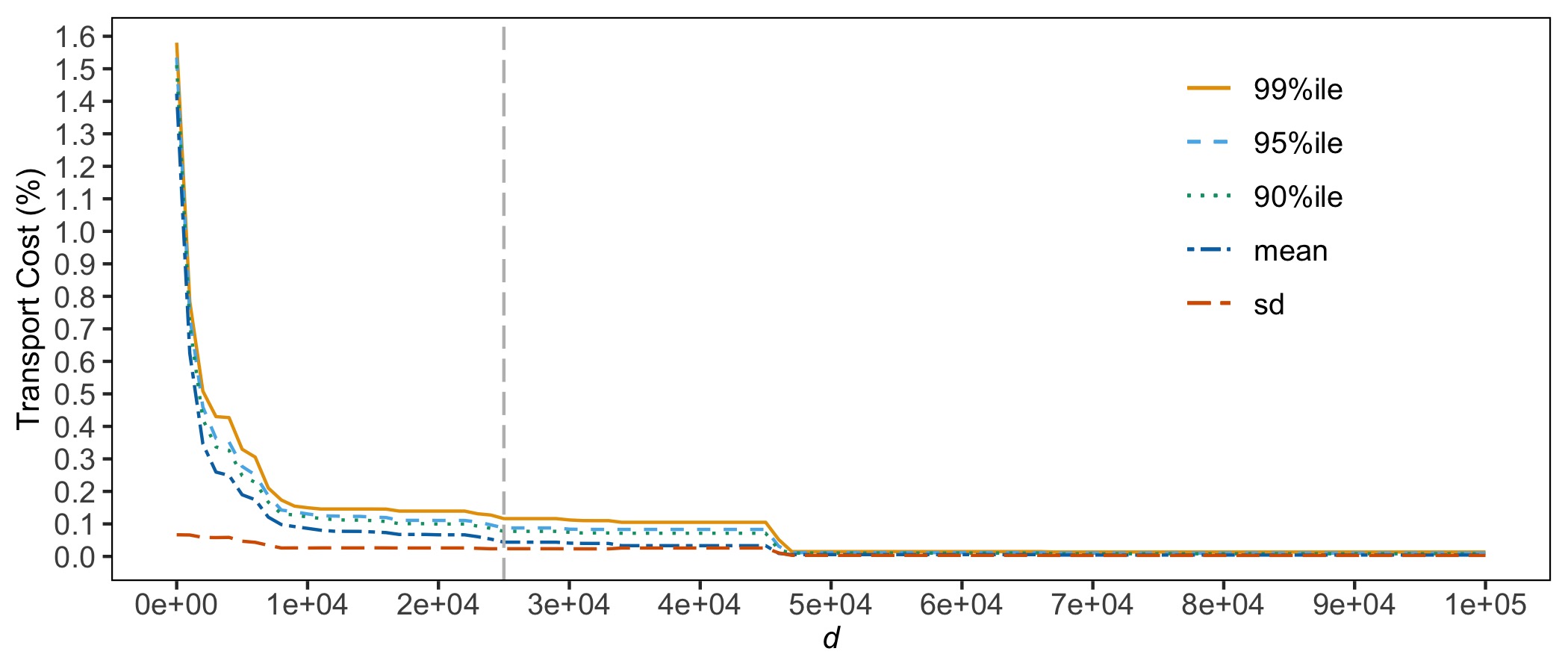}
    \caption{\figtitle{Placebo costs as a function of $d$.} 
    \figcaption{This figure is similar to Figure \ref{fig:appfigsix} in Appendix.}}
    \label{fig:v1-appfigsix}
\end{figure}

\begin{figure}[H]
    \centering
    \includegraphics[width=\textwidth]{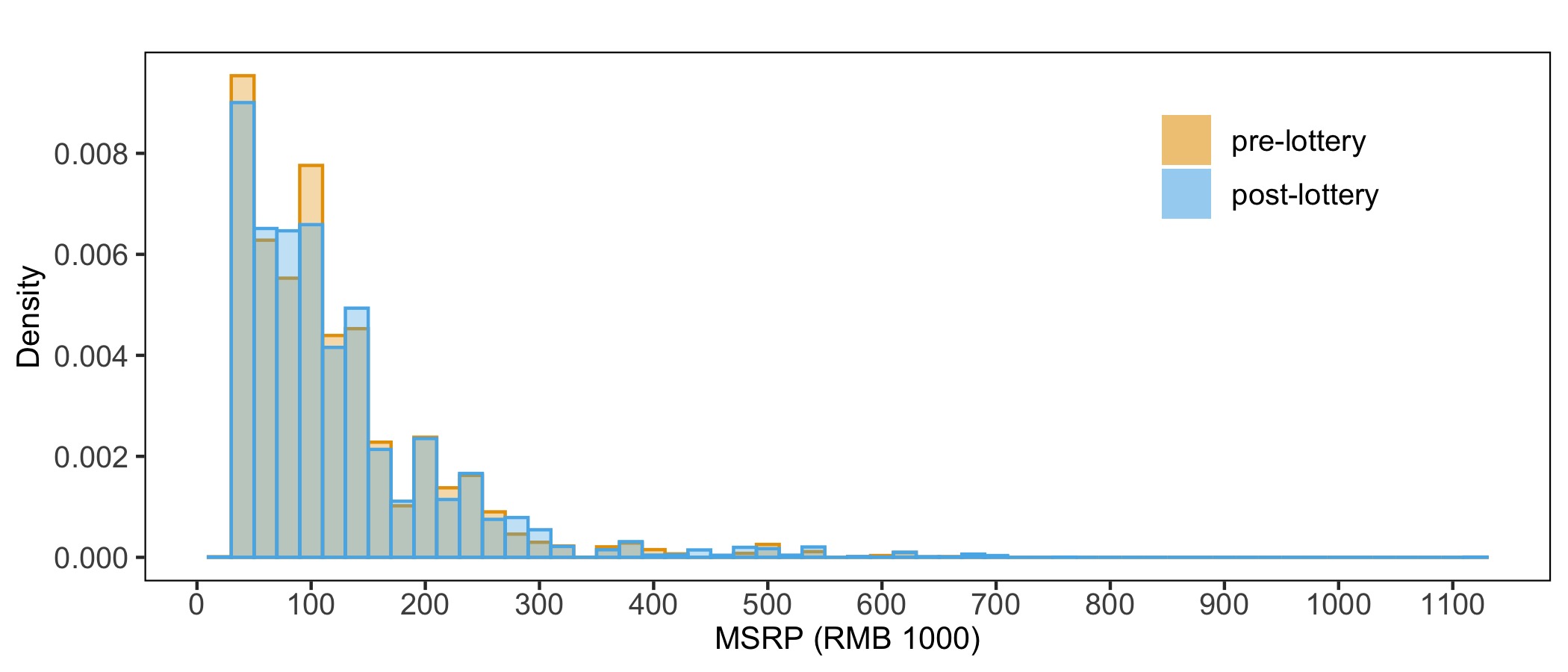}
    \caption{\figtitle{Smoothed empirical distributions of Tianjin car sales in 2010 (pre-lottery) and 2011 (post-lottery).} 
    \figcaption{This figure is similar to Figure \ref{fig:eight}.}}
    \label{fig:v1-eight}
\end{figure}


\begin{figure}[H]
    \centering
    \includegraphics[width=\textwidth]{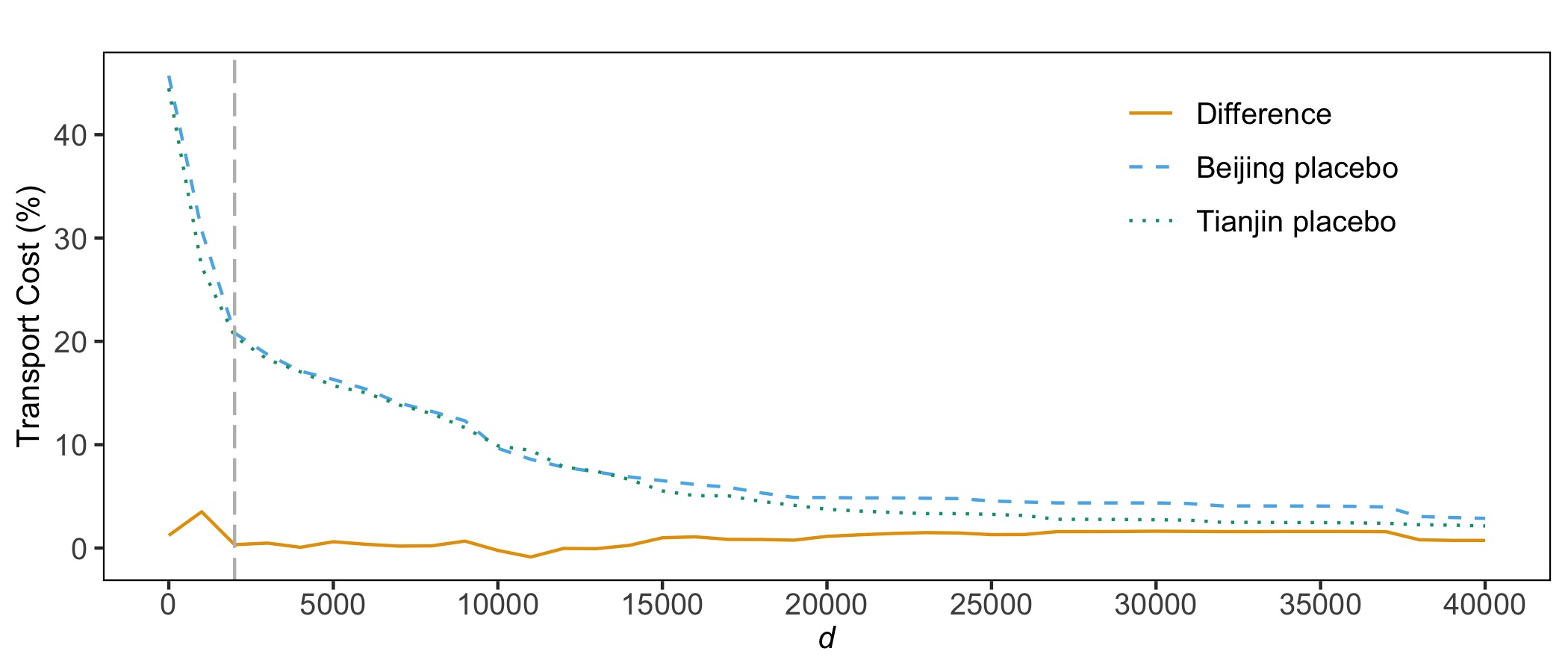}
    \caption{\figtitle{Post-Trends.} 
    \figcaption{This figure is similar to Figure \ref{fig:ten}.}}
    \label{fig:v1-ten}
\end{figure}

\begin{figure}[H]
    \centering
    \includegraphics[width=\textwidth]{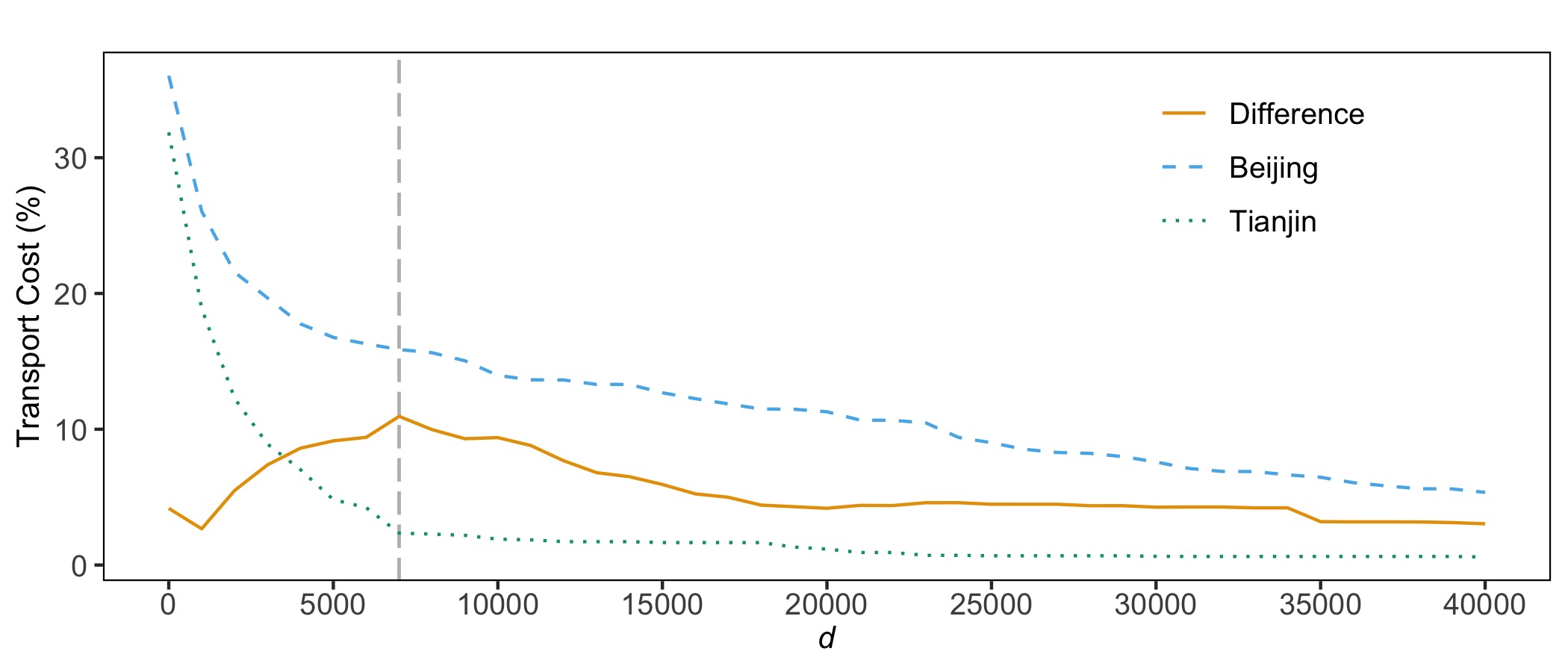}
    \caption{\figtitle{Choosing the most informative admissible $d$.}
    \figcaption{This figure is similar to Figure \ref{fig:eleven}.}}
    \label{fig:v1-eleven}
\end{figure}


\begin{table}[H]\centering
\small
\ra{\usualra}
\begin{threeparttable}
 \caption{\tabtitle{Diff-in-diff regression results for different control groups.} \tabcaption{This table is similar to Table \ref{tab:eight}.}} 
 \begin{tabular}{@{\extracolsep{10pt}}lD{.}{.}{-3} D{.}{.}{-3} D{.}{.}{-3}}
 \toprule
    & \multicolumn{1}{c}{\tabhead{Tianjin}} & \multicolumn{1}{c}{\tabhead{Shijiazhuang}} & \multicolumn{1}{c}{\tabhead{Both}} \\
 \midrule
    Beijing & 0.273 & 0.235 & 0.260 \\
  & (0.002) & (0.002) & (0.001) \\
  & & & \\
 post & 0.026 & 0.014 & 0.022 \\
  & (0.002) & (0.003) & (0.002) \\
  & & & \\
 Beijing $\times$ post & 0.201 & 0.213 & 0.205 \\
  & (0.002) & (0.003) & (0.002) \\
  & & & \\
 Constant & 11.502 & 11.540 & 11.515 \\
  & (0.001) & (0.002) & (0.001) \\ 
    & & & \\ 
 \midrule
    Observations & \multicolumn{1}{c}{1,275,829} & \multicolumn{1}{c}{1,078,516} & \multicolumn{1}{c}{1,486,251} \\ 
R$^{2}$ & \multicolumn{1}{c}{0.078} & \multicolumn{1}{c}{0.061} & \multicolumn{1}{c}{0.079} \\ 
 \bottomrule
 \end{tabular} \label{tab:v1-eight} 
\end{threeparttable}
\end{table}

\subsection{Excluding December 2010, January 2011, and February 2011}\label{sec:v2}

\begin{figure}[H]
  \centering
  \includegraphics[width=\textwidth]{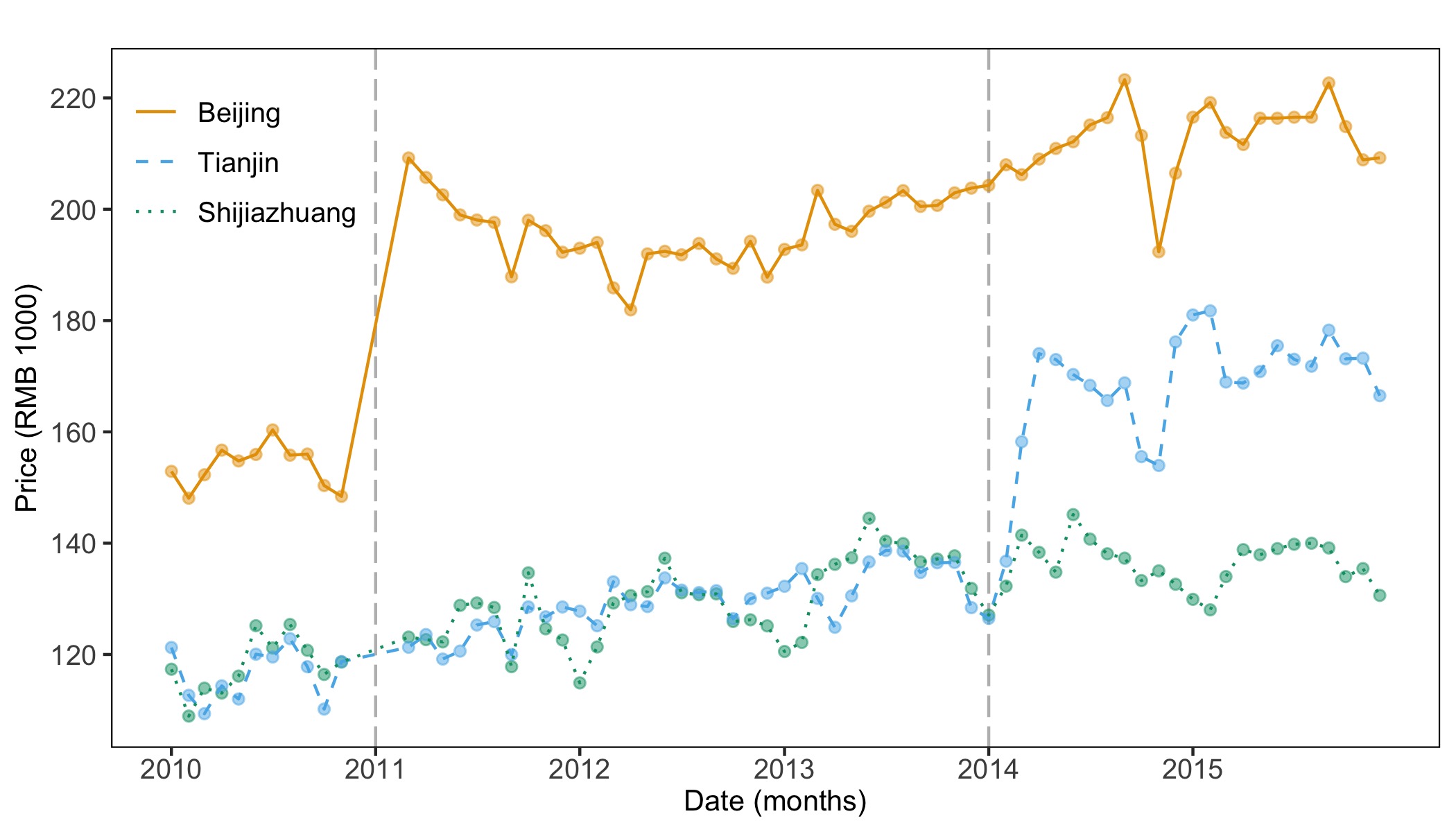}
  \caption{\figtitle{Monthly average swtprice (RMB 1\numcomma000) for Beijing, Shijiazhuang, and Tianjin.} 
  \figcaption{This figure is similar to Figure \ref{fig:one}.}}
  \label{fig:v2-one}
\end{figure}

\begin{figure}[H]
  \centering
  \includegraphics[width=\textwidth]{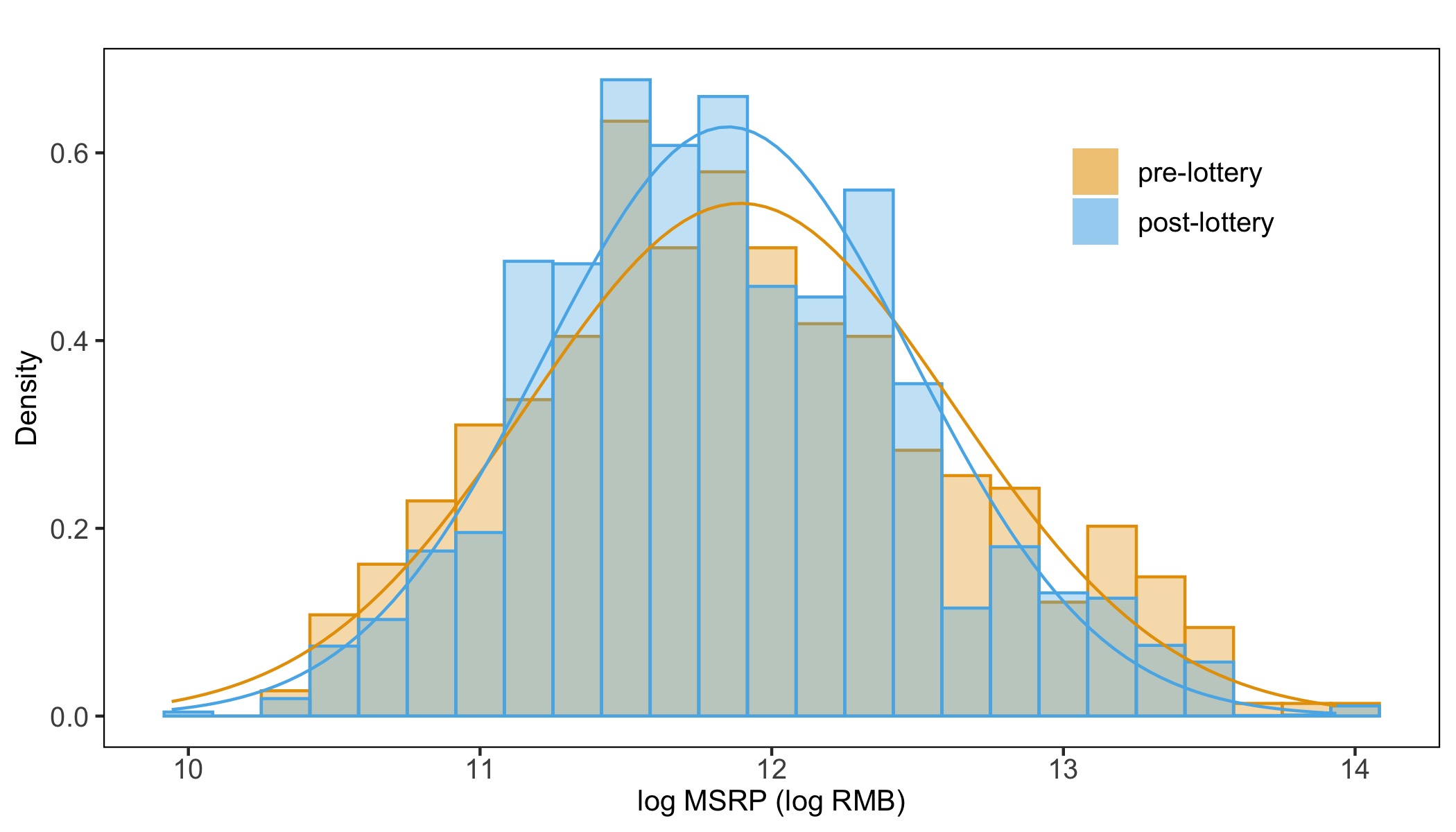}
  \caption{\figtitle{Distribution of Beijing car prices in 2010 (pre-lottery) and new car prices in 2011 (post-lottery).} 
  \figcaption{This figure is similar to Figure \ref{fig:two}.}}
  \label{fig:v2-two}
\end{figure}

\begin{figure}[H]
    \centering
    \includegraphics[width=\textwidth]{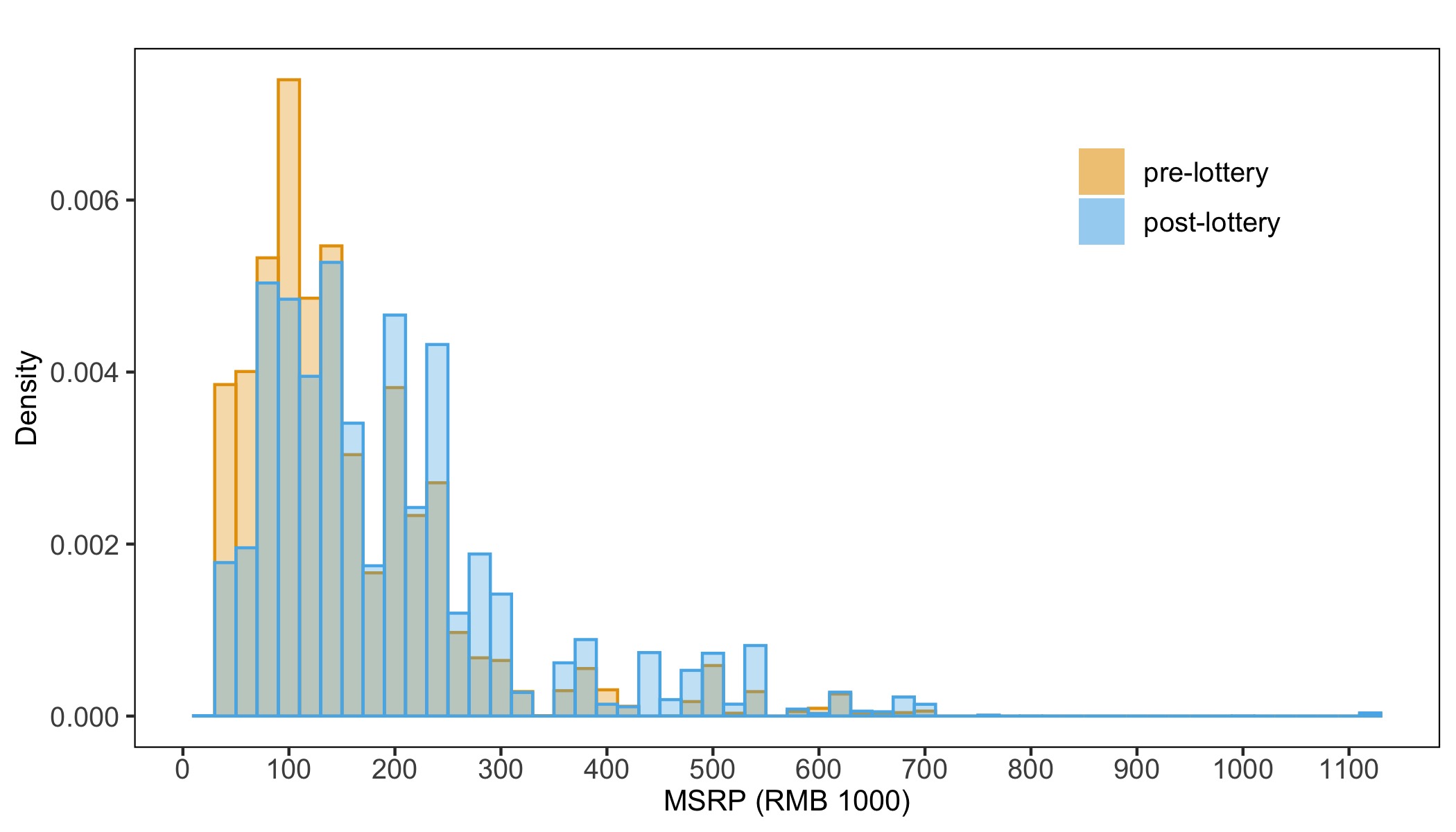}
    \caption{\figtitle{Smoothed empirical distributions of Beijing car sales in 2010 (pre-lottery) and 2011 (post-lottery).} 
    \figcaption{This figure is similar to Figure \ref{fig:four}.}}
    \label{fig:v2-four}
\end{figure}

\begin{figure}[H]
    \centering
    \includegraphics[width=\textwidth]{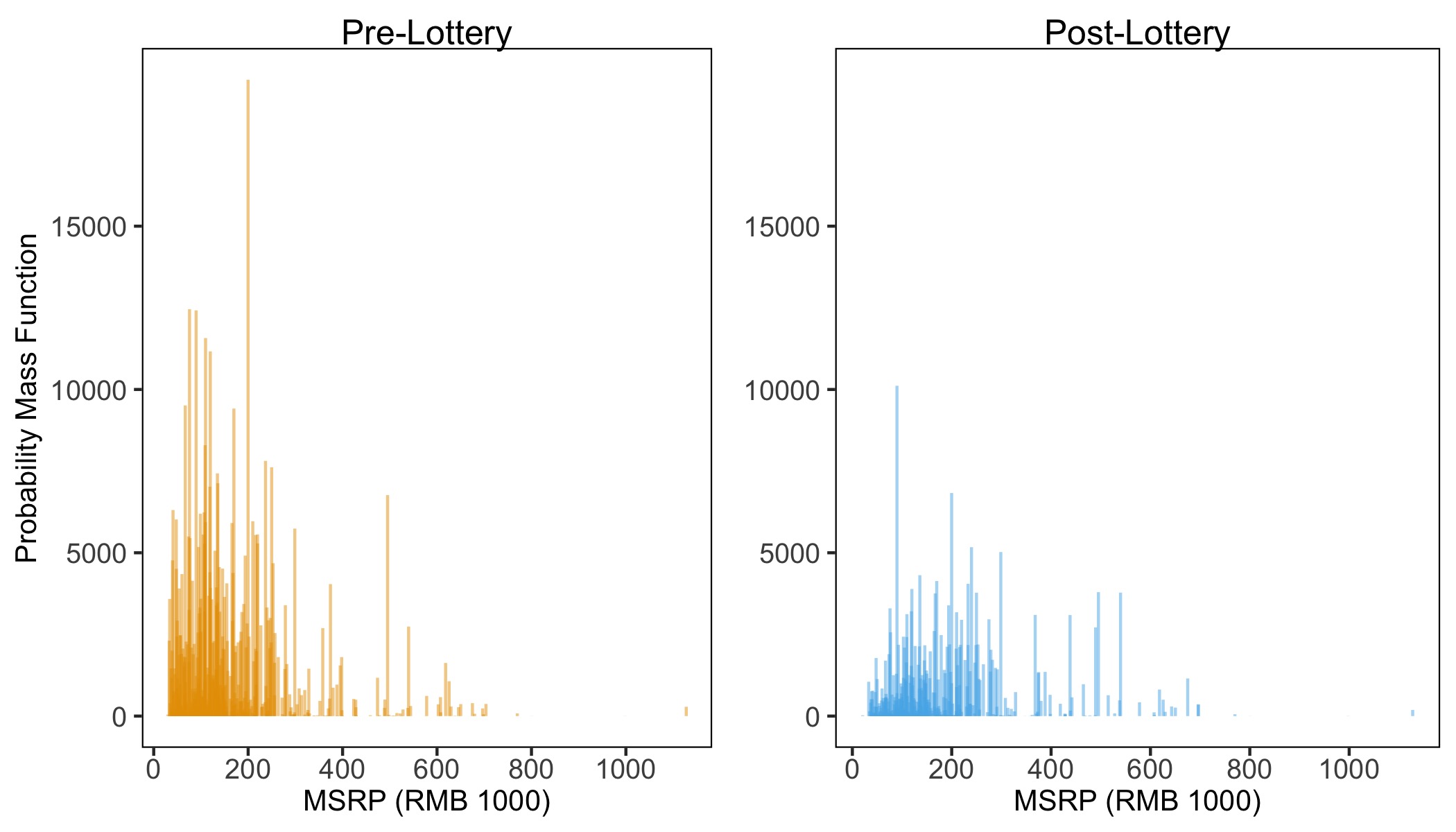}
    \caption{\figtitle{Exact empirical distributions of Beijing car sales in 2010 (pre-lottery) and 2011 (post-lottery).} 
    \figcaption{This figure is similar to Figure \ref{fig:five}.}}
    \label{fig:v2-five}
\end{figure}

\begin{figure}[H]
    \centering
    \includegraphics[width=\textwidth]{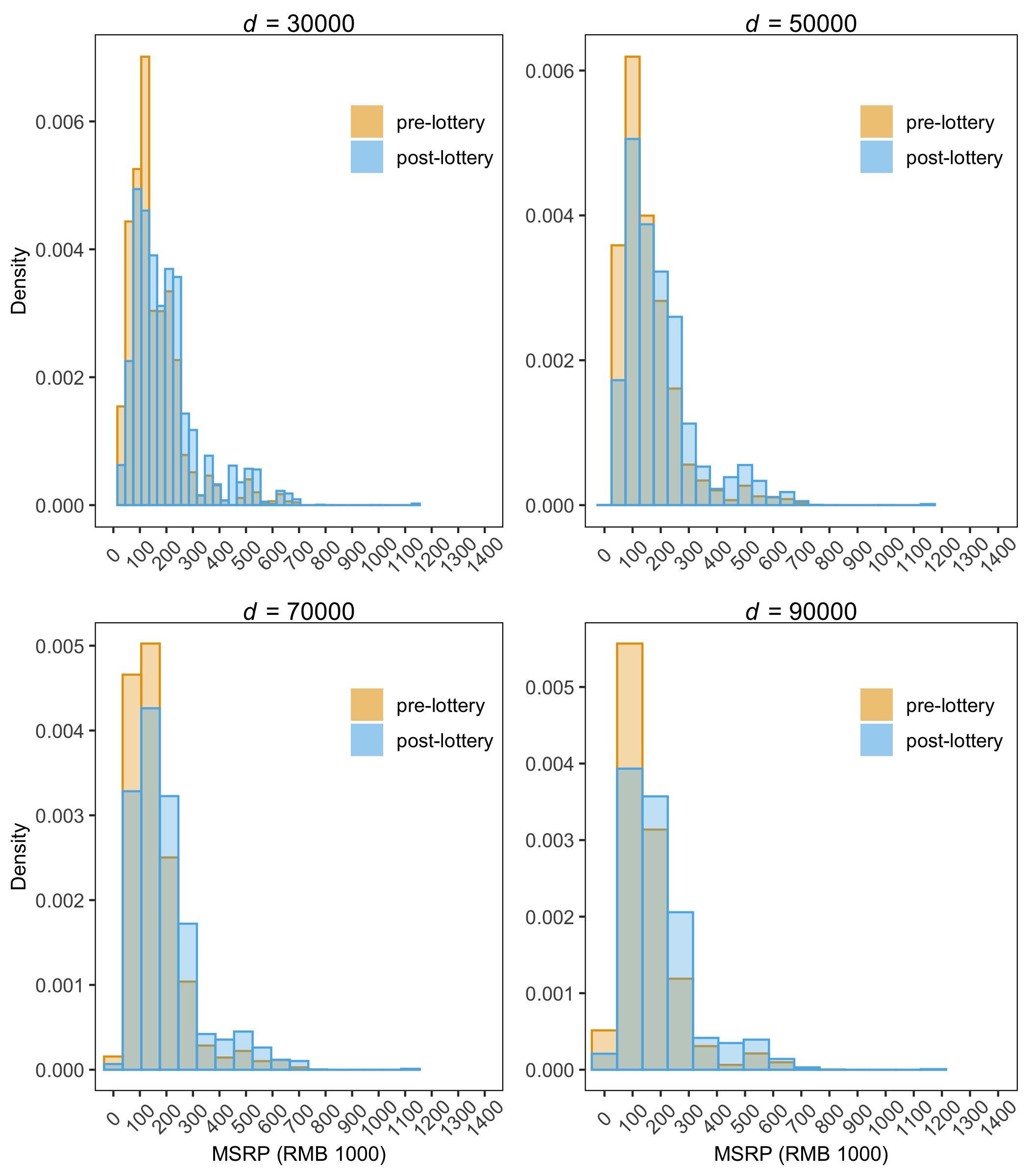}
    \caption{\figtitle{Smoothed empirical distributions of Beijing car sales in 2010 (pre-lottery) and 2011 (post-lottery) at candidate values of $d$.} 
    \figcaption{This figure is similar to Figure \ref{fig:six}.}}
    \label{fig:v2-six}
\end{figure}

\begin{figure}[H]
    \centering
    \includegraphics[width=\textwidth]{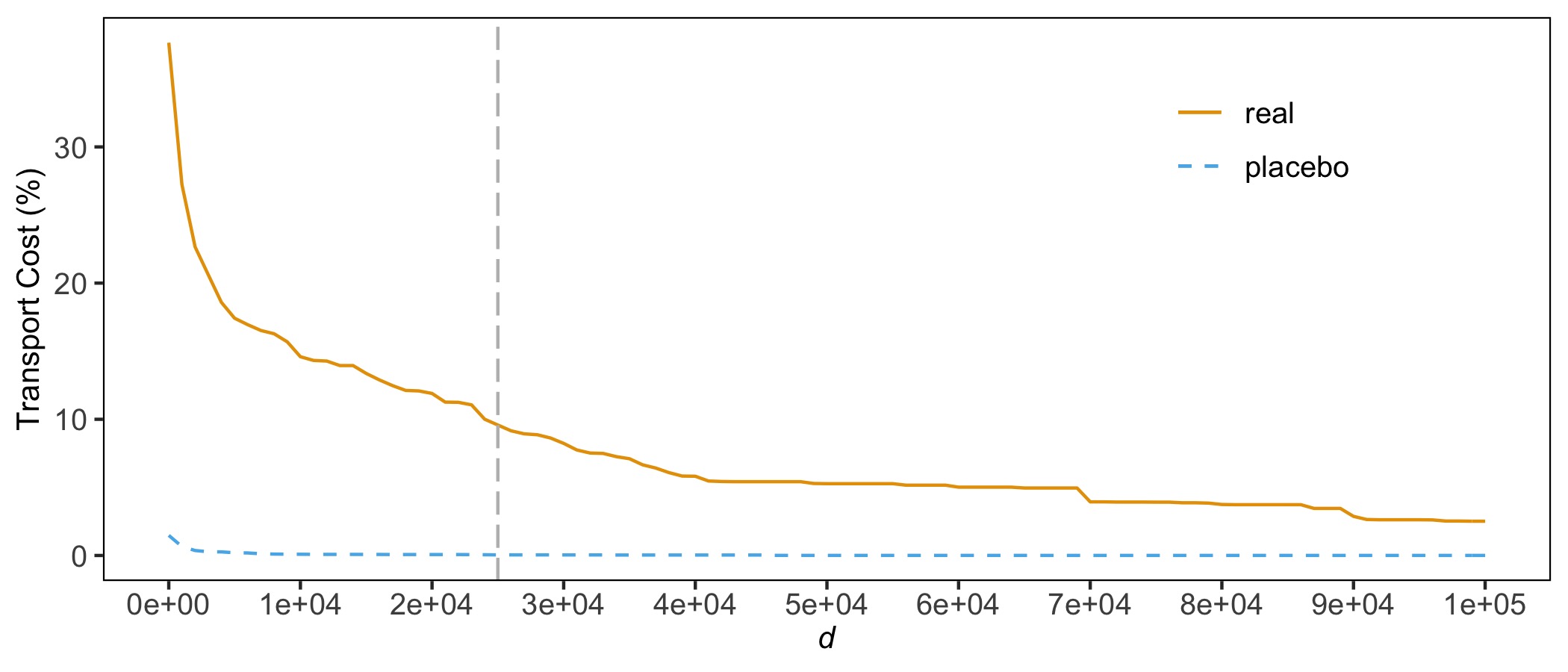}
    \caption{\figtitle{Real and placebo costs as a function of $d$.} 
    \figcaption{This figure is similar to Figure \ref{fig:seven}.}}
    \label{fig:v2-seven}
\end{figure}

\begin{figure}[H]
    \centering
    \includegraphics[width=\textwidth]{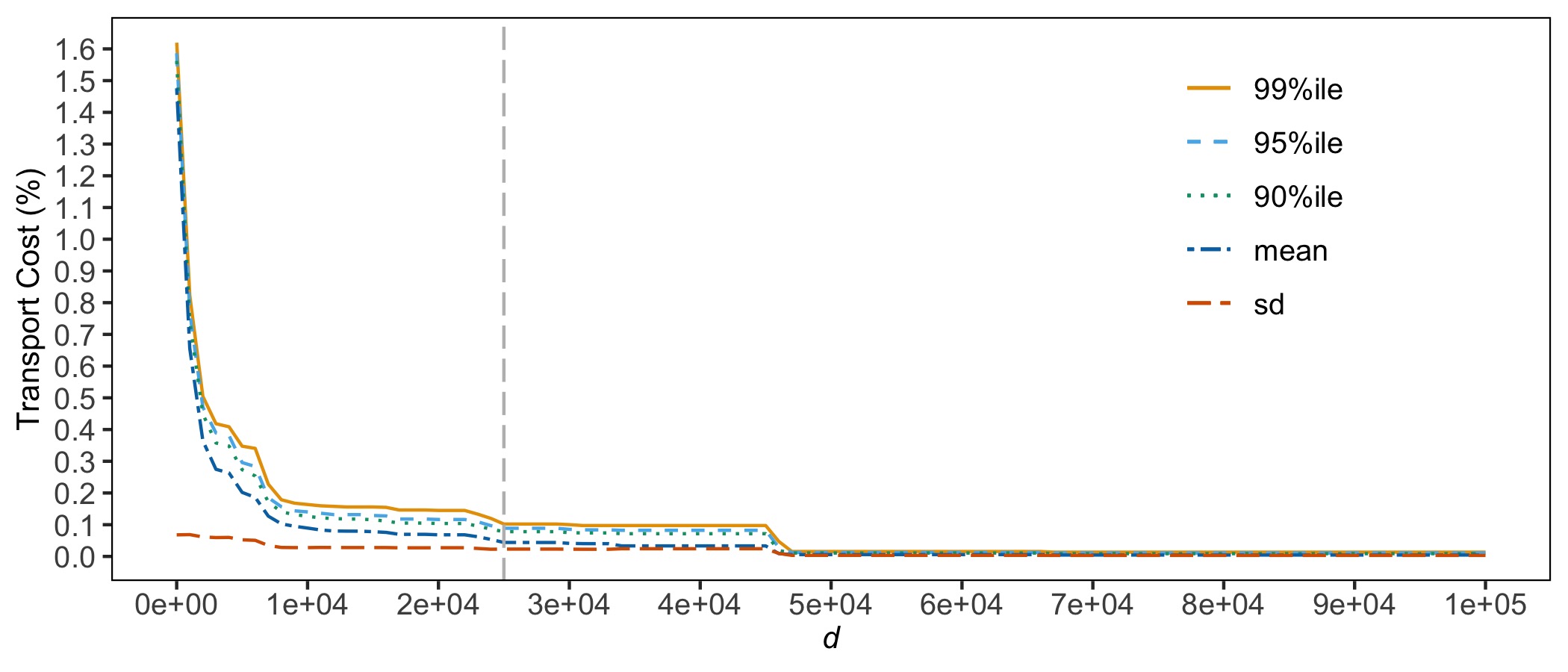}
    \caption{\figtitle{Placebo costs as a function of $d$.} 
    \figcaption{This figure is similar to Figure \ref{fig:appfigsix} in Appendix.}}
    \label{fig:v2-appfigsix}
\end{figure}

\begin{figure}[H]
    \centering
    \includegraphics[width=\textwidth]{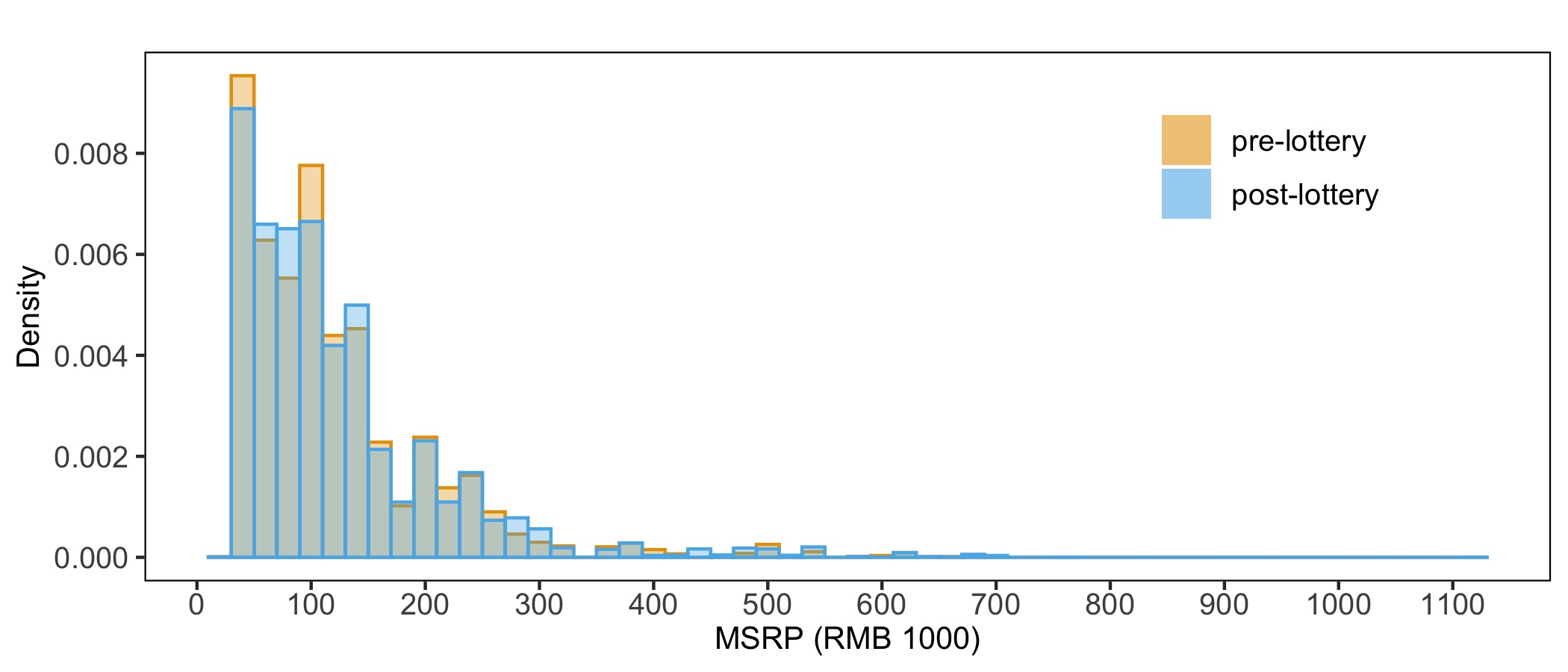}
    \caption{\figtitle{Smoothed empirical distributions of Tianjin car sales in 2010 (pre-lottery) and 2011 (post-lottery).} 
    \figcaption{This figure is similar to Figure \ref{fig:eight}.}}
    \label{fig:v2-eight}
\end{figure}


\begin{figure}[H]
    \centering
    \includegraphics[width=\textwidth]{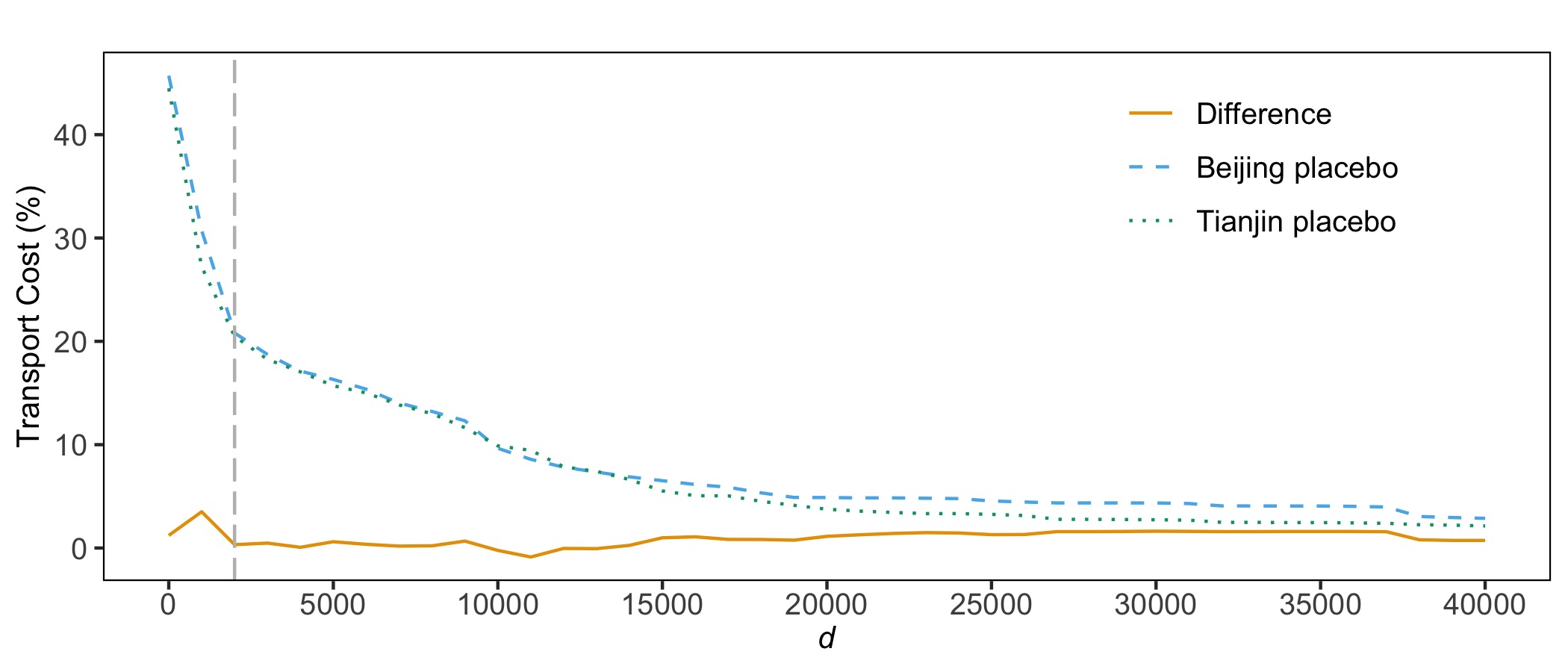}
    \caption{\figtitle{Post-Trends.} 
    \figcaption{This figure is similar to Figure \ref{fig:ten}.}}
    \label{fig:v2-ten}
\end{figure}

\begin{figure}[H]
    \centering
    \includegraphics[width=\textwidth]{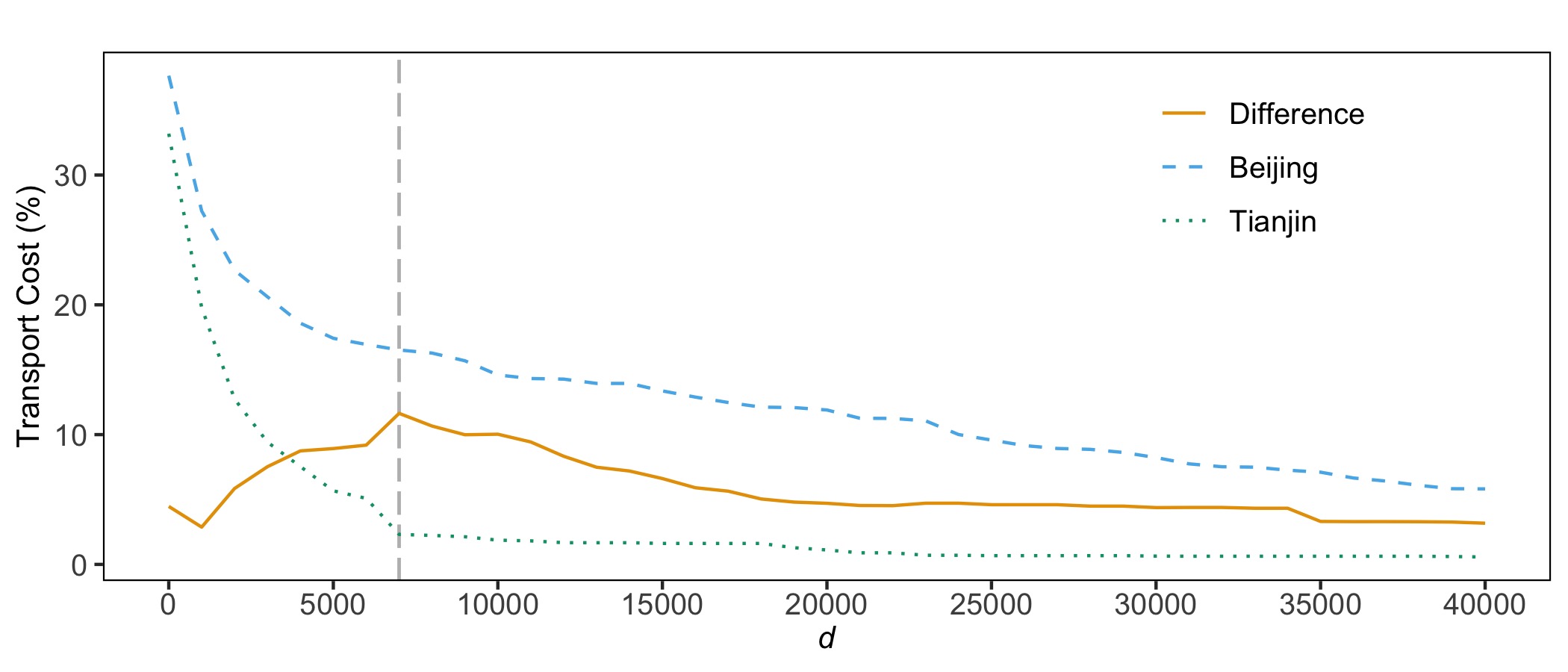}
    \caption{\figtitle{Choosing the most informative admissible $d$.}
    \figcaption{This figure is similar to Figure \ref{fig:eleven}.}}
    \label{fig:v2-eleven}
\end{figure}


\begin{table}[H]\centering
\small
\ra{\usualra}
\begin{threeparttable}
 \caption{\tabtitle{Diff-in-diff regression results for different control groups.} \tabcaption{This table is similar to Table \ref{tab:eight}.}} 
 \begin{tabular}{@{\extracolsep{10pt}}lD{.}{.}{-3} D{.}{.}{-3} D{.}{.}{-3}}
 \toprule
    & \multicolumn{1}{c}{\tabhead{Tianjin}} & \multicolumn{1}{c}{\tabhead{Shijiazhuang}} & \multicolumn{1}{c}{\tabhead{Both}} \\
 \midrule
    Beijing & 0.273 & 0.235 & 0.260 \\
  & (0.002) & (0.002) & (0.001) \\
  & & & \\
 post & 0.025 & 0.023 & 0.024 \\
  & (0.002) & (0.003) & (0.002) \\
  & & & \\
 Beijing $\times$ post & 0.212 & 0.214 & 0.213 \\
  & (0.002) & (0.003) & (0.002) \\
  & & & \\
 Constant & 11.502 & 11.540 & 11.515 \\
  & (0.001) & (0.002) & (0.001) \\ 
    & & & \\ 
 \midrule
    Observations & \multicolumn{1}{c}{1,220,450} & \multicolumn{1}{c}{1,033,006} & \multicolumn{1}{c}{1,412,183} \\ 
R$^{2}$ & \multicolumn{1}{c}{0.079} & \multicolumn{1}{c}{0.060} & \multicolumn{1}{c}{0.080} \\ 
 \bottomrule
 \end{tabular} \label{tab:v2-eight} 
\end{threeparttable}
\end{table}

\subsection{Summary of Results}\label{sec:resultssummary}


\begin{table}[H]\centering
\small
\ra{\usualra}
\begin{threeparttable}
 \caption{\tabtitle{Results}}
 \begin{tabular}{@{}>{\raggedright}p{0.34\linewidth}>{\centering}p{0.17\linewidth}>{\centering}p{0.15\linewidth}>{\centering\arraybackslash}p{0.15\linewidth}}
 \toprule
    & \tabhead{Original} & \tabhead{Section \ref{sec:v1}} & \tabhead{Section \ref{sec:v2}} \\
 \midrule
      Diff-in-diff (t-stat) & \did\ (103.992) & 20.5\% (97.853) &  21.3\% (98.590)\\
    Before-and-after ($d$) & \ba\ (\dba) &  14\% (10\numcomma000) & 14.6\% (10\numcomma000) \\
    Diff-in-transports ($d$) & \dit\ (\dstar) & 13.5\% (7\numcomma000) & 14.2\% (7\numcomma000) \\
    Conservative diff-in-transports ($d$) & \ditcons\ (\dstarcons) & 10.9\% (7\numcomma000) & 11.6\% (7\numcomma000) \\
 \bottomrule
 \end{tabular}
 \begin{tablenotes}
    \item The diff-in-diff estimate uses both Tianjin and Shijiazhuang as the control group. The other estimates are reported with their corresponding bandwidths. The differences-in-transport reports the transport cost when using $d$ for both Beijing and Tianjin. The conservative analogue uses $2d$ as the bandwidth for Beijing calculations. Section \ref{sec:v1} excludes December 2010 from the analysis. Section \ref{sec:v2} excludes December 2010, January 2011, and February 2011 from the analysis.
 \end{tablenotes}
\end{threeparttable}
\end{table}

\section{Robustness to the use of Manufacturer suggested retail price as a proxy for transaction prices}\label{sec:msrptp}

\begin{figure}[H]
    \centering
    \includegraphics[width=\textwidth]{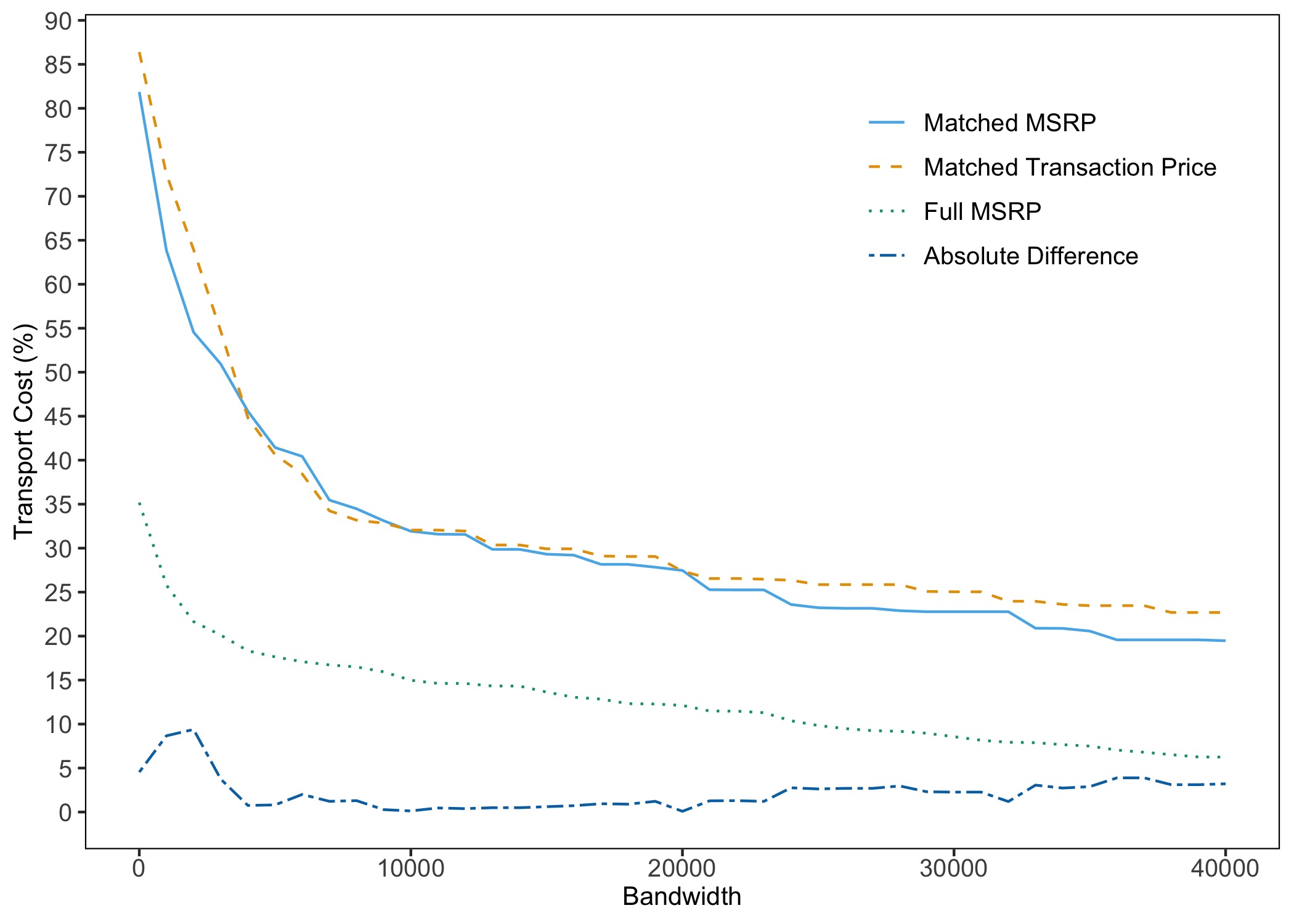}
    \caption{\figtitle{Optimal transport cost using MSRPs and transaction prices, and their difference, for matched data, as well as optimal transport cost using MSRPs for full data.} 
    }
    \label{fig:msrp-vs-tp}
\end{figure}

Focusing our attention on the sales of cars for which we have both the MSRP and transaction price, we compute the transport cost between 2010 and 2011 for bandwidths between 0 and 40,000. As the above figure shows, the transport cost using the transaction prices is very similar to the transport cost using the corresponding MSRPs. The absolute difference of these transport costs does not exceed 5\% beyond a bandwidth of 3000, and for bandwidths between 4,000 and 33,000, the absolute difference is less than 3\%. Thus, on the subset of data for which we have both MSRPs and transaction prices, our results are robust to the choice of MSRP and transaction price. In particular, the before-and-after estimator for $d = 20,000$ is 27.5\% when using transaction prices and 27.4\% when using MSRPs.

Note that because the effective sampling scheme of the proprietary data is different from that of the non-rationed purchases or lottery –corrupted or not– data, the transport cost is different when restricted to the car models which both are in the proprietary data and could be matched to entries in our main data set.

Indeed, this is readily explained.
The proprietary data with information on transaction prices comes from a survey of a small number of people that are supposedly selected at random: 1799 people in Beijing during 2010 and 1919 people in Beijing during 2011. Since very few individuals are sampled, there are fewer distinct models for which we have transaction price data: 401 car models in 2010 and 233 car models in 2011. From this small sample of models, we successfully match a yet slightly smaller subsample: 374 car models in 2010 and 229 car models in 2011.

We may assess the impact of this sampling in practice.   Assuming that successful matches are random selections, we can simulate this process and create a subsample of MSRPs.  If we randomly sample subsets of the size of the proprietary data and then sample uniformly at random a further subset of matched models of size equal to that of the matched sample, we obtain transport costs to the full population distribution of the same order of magnitude as that of the matched sample in MSRP.  The sample selection explains the difference in MSRP transport cost for this matched subsample, and the approximate equivalence for this matched subsample of the MSRP and transaction price transport costs serves as very desirable robustness verification.



%

\newpage
\bibliography{mybib}